1
\documentclass[%
 aip,
 amsmath,amssymb,
 reprint,%
]{revtex4-1}

\usepackage{graphicx}
\usepackage{dcolumn}
\usepackage{bm}

\usepackage[utf8]{inputenc}
\usepackage[T1]{fontenc}
\usepackage{mathptmx}
\usepackage{etoolbox}
\usepackage{color}
 \usepackage[cal=pxtx]{mathalfa}

\makeatletter
\def\@email#1#2{%
 \endgroup
 \patchcmd{\titleblock@produce}
  {\frontmatter@RRAPformat}
  {\frontmatter@RRAPformat{\produce@RRAP{*#1\href{mailto:#2}{#2}}}\frontmatter@RRAPformat}
  {}{}
}%
\makeatother
\begin{document}

\preprint{AIP/123-QED}

\title[Universal scaling of shear thickening transitions]{Universal scaling of shear thickening transitions}
\author{Meera Ramaswamy}
\affiliation{Department of Physics, Cornell University, Ithaca, New York 14853, USA}
\author{Itay Griniasty}
\affiliation{Department of Physics, Cornell University, Ithaca, New York 14853, USA}
\author{Danilo B. Liarte}
\affiliation{Department of Physics, Cornell University, Ithaca, New York 14853, USA}
\author{Abhishek Shetty}
\affiliation{Anton Paar USA, 10215 Timber Ridge Drive, Ashland, Virginia 23005}
\author{Eleni Katifori}
\affiliation{Department of Physics, University of Pennsylvania, Philadelphia, Pennsylvania}
\author{Emanuela Del Gado}
\affiliation{Department of Physics, Georgetown University, Washington DC, USA}
\author{James P Sethna}
\affiliation{Department of Physics, Cornell University, Ithaca, New York 14853, USA}
\author{Bulbul Chakraborty*}
\affiliation{Department of Physics, Brandeis University, Waltham, Massachusetts, USA}
\author{Itai Cohen*}
\affiliation{Department of Physics, Cornell University, Ithaca, New York 14853, USA}
\date{\today}

\begin{abstract}
Nearly all dense suspensions undergo dramatic and abrupt thickening transitions in their flow behaviour when sheared at high stresses. Such transitions occur when the dominant interactions between the suspended particles shift from hydrodynamic to frictional. Here, we interpret abrupt shear thickening as a precursor to a rigidity transition and give a complete theory of the viscosity in terms of a universal crossover scaling function from the frictionless jamming point to a rigidity transition associated with friction, anisotropy, and shear. Strikingly, we find experimentally that for two different systems -- cornstarch in glycerol and silica spheres in glycerol -- the viscosity can be collapsed onto a single universal curve over a wide range of stresses and volume fractions. The collapse reveals two separate scaling regimes, due to a crossover between  frictionless isotropic jamming and frictional shear jamming, with different critical exponents. The material-specific behaviour due to the microscale particle interactions is incorporated into a scaling variable governing the proximity to shear jamming that depends on both stress and volume fraction. This reformulation opens the door to importing the vast theoretical machinery developed to understand equilibrium critical phenomena to elucidate fundamental physical aspects of the shear thickening transition.
\end{abstract}

\maketitle

\section{\label{sec:level1}Introduction}
Suspensions of solid particles in a liquid commonly exhibit shear thickening, an increase in the shear viscosity with the applied stress or strain rate \cite{brown2010generality, wagner2009shear}. This increase in viscosity can span orders of magnitude and even lead to solidification, as illustrated by people running atop vats of cornstarch~\cite{brown2014shear}~\footnote{Search www.youtube.com for walking on cornstarch}. Such strong shear thickening behaviour has been attributed to the change in the nature of particle interactions --  at low stresses, the particle interactions are dominated by the lubrication forces but at high stresses, the particle surfaces are forced closer together and frictional contact interactions dominate \cite{lin2015hydrodynamic, wyart2014discontinuous, cates2014granulation, lootens2005dilatant, seto2013discontinuous, brown2009dynamic, brown2014shear, wyart2014discontinuous}.


Understanding the mechanism underpinning the thickening transition has led to numerous studies focused on altering the system properties to modify the shear thickening behaviour. Commonly, two different strategies have been pursued -- 1) changing the interparticle friction and 2) changing the particle microstructure or suspension packing. Changes to the interparticle friction can be achieved by altering the particle surface roughness~\cite{hsiao2017rheological, hsu2018roughness,hsu2021exploring, hoffman1998explanations, jamali2019alternative, more2020roughness, lootens2005dilatant}, interparticle adhesion~\cite{bourrianne2022tuning}, or hydrogen bonding~\cite{jaishankar2015probing, james2018interparticle, james2019tuning}. The maximum particle packing in the suspensions can be altered by tuning the particle shape~\cite{brown2011shear, james2019controlling, rathee2020role}, size or polydispersity~\cite{guy2020testing, bender1996reversible}, or applying external fields~\cite{ness2018shaken, niu2020tunable, lin2016tunable, sehgal2019using, gillissen2020constitutive, chen2022leveraging, jackson2022designing}. In many of these examples, even a small change in the suspension can result in large changes to the rheology. It has therefore been a major challenge to develop a unified, suspension-independent framework to predict shear thickening.  

A major step towards developing such a framework was proposed by Wyart and Cates \cite{wyart2014discontinuous, cates2014granulation}, who modelled the viscosity as a function of the distance to a stress-dependent jamming volume fraction. As the stress increases, the contacts between the particles become frictional and the jamming volume fraction decreases, resulting in an increase in the viscosity. This theory has been used to fit experimental and simulation data with varying degrees of success~\cite{singh2018constitutive, guy2020testing, rathee2020role, royer2016rheological, lee2020experimental}. Others have proposed similar models to describe shear thickening \cite{singh2018constitutive, gillissen2019constitutive, gillissen2020constitutive}, and more recently, the extent of shear thickening has also been shown to be related to the jamming volume fraction \cite{pradeep2020jamming}, indicating that jamming underpins the shear thickening transition.  Despite these close connections between shear thickening, and jamming -- a well-established equilibrium phase transition -- the relationship between the shear thickening and statistical scaling frameworks remains unexplored. More specifically, it is yet unclear if shear thickening can be described by the proximity to the jamming critical point via universal scaling functions and if jamming with and without shear are associated with the same universality class and scaling exponents. 

In this work, we adopt the idea of scaling to analyze the experimentally observed viscosity of two different shear thickening suspensions and establish the relationship between shear thickening and the associated critical points. Since the advent of the ideas of universality and scaling in equilibrium phase transitions~\cite{cardy1996scaling}, it has been known that scaling analysis provides a litmus test for the presence of critical points in complex phase diagrams.   This approach led to the discovery of non-classical exponents associated with phase transitions, the identification of universality classes, and the development of the renormalization group~\cite{cardy1996scaling}. In contrast to fitting data to functional forms, scaling involves collapsing data over a broad range of multiple control parameters. The advantage of this approach is that it is model-independent and often reveals the underlying physical governing principles, the relevant variables controlling the distance to critical points and the associated scaling functions. Here we propose to use the same machinery to investigate the critical points associated with the shear thickening transition.

More specifically, we pursue the idea that thickening is governed by two different critical points and that a \textit{crossover scaling framework} can be utilized to characterize this transition~\cite{cardy1996scaling}. Crossover scaling was originally
introduced to describe transitions between thermodynamic critical points (e.g., Heisenberg magnets with small uniaxial anisotropy behaving like Ising models), and has become invaluable for describing finite-temperature behaviour induced by quantum critical points~\cite{sachdev2011quantum, sachdev1997theory, sachdev1999universal}, crossovers between universality classes in random matrix theory~\cite{adam2002enhanced}, and fracture and depinning transitions~\cite{chen2015crossover}. Our analysis shows that the same framework provides an excellent unified description of thickening transitions.

\section{\label{sec:level1} Recasting the Wyart and Cates model}
We begin by recasting the Wyart and Cates~\cite{wyart2014discontinuous} model, the current state-of-the-art model used to describe shear thickening, into the framework of crossover scaling. Briefly, the model expresses the viscosity of the suspension in terms of a distance to a stress dependent jamming volume fraction, $\phi_J(\sigma)$, 
\begin{equation}
\label{eq:WC1}
    \eta \sim (\phi_J(\sigma) - \phi)^{-2}
\end{equation}  
As the stress increases, the nature of the contact forces between the particles in the suspension changes from hydrodynamic to frictional, and the jamming volume fraction decreases as: 
\begin{equation}
\label{eq:WC2}
\phi_J(\sigma)  = \phi_0(1 - f(\sigma)) + \phi_\mu f(\sigma). 
\end{equation}
Here, $f(\sigma)$ is the fraction of frictional contacts in the system, $\phi_0$ is the jamming volume fraction in the absence of friction, and $\phi_{\mu}$ is the jamming volume fraction when all the interactions between the particles are frictional ($\phi_\mu<\phi_0$). 
$f(\sigma)$ is a sigmoidal function of the applied stress, with limits of zero at low stress and one at high stress, and Eq.~\ref{eq:WC2} defines a line of critical points at which the viscosity diverges leading to shear-induced jamming at finite $\sigma$. Thus, increasing the stress activates frictional contacts which lowers the jamming volume fraction and increases the viscosity. Despite its simplicity, this model does a remarkable job capturing essential features of the flow behaviours including continuous shear thickening, discontinuous shear thickening, and shear jamming. 

\begin{figure*}[t]
\includegraphics[width=0.8\linewidth]{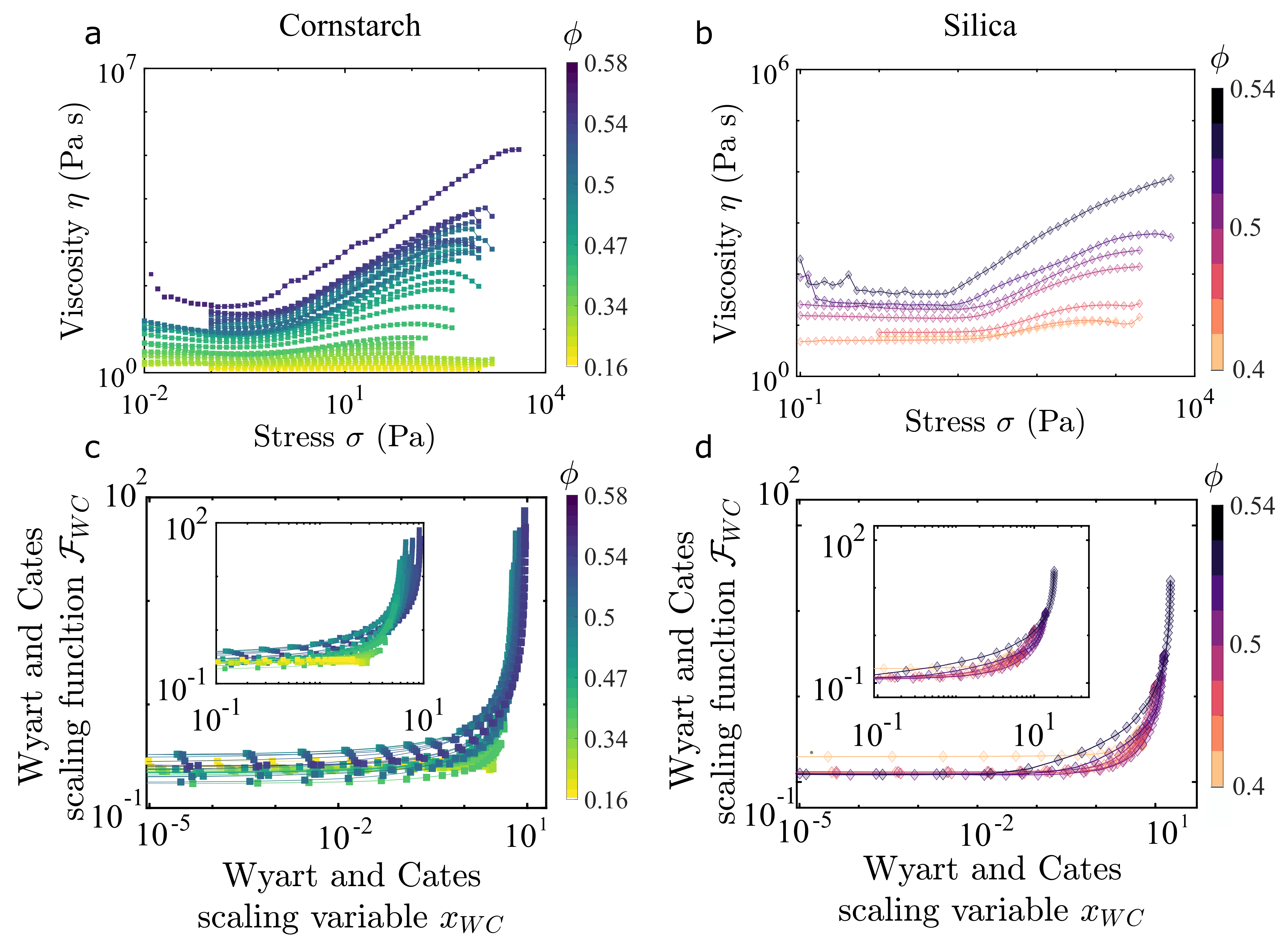}
\caption{\label{fig:FlowCurve} \textbf{Viscosity versus stress measurements and scaling predictions from the Wyart and Cates model}. \textbf{a}\textbf{b}. The viscosity as a function of applied stress for \textbf{a} cornstarch suspensions with volume fractions ranging from 0.16 (yellow) to 0.54 (dark blue) and \textbf{b} silica suspensions with volume fractions ranging from 0.4 (light pink) to 0.52 (black). These suspensions show a range of shear thickening behaviour across volume fractions from weakly shear thickening to discontinuous shear thickening to shear jamming. \textbf{c}, \textbf{d}. Plots of the predicted Wyart and Cates scaling function $\mathcal{F_{WC}}=\eta(\phi_0 - \phi)^2$ versus the Wyart and Cates scaling variable $x_{WC}=f(\sigma)/(\phi_0 - \phi)$ for cornstarch suspensions (\textbf{c}) and silica suspensions (\textbf{d}), showing promising but incomplete scaling collapse. Inset shows the zoom-in of the high $x_{WC}$ region. In the inset of \textbf{c}, the cornstarch scaling function diverges at $\sim$ 10 at low volume fractions, $\sim$ 6 at intermediate volume fractions and $\sim$ 10 at large volume fractions. The silica data in the inset of \textbf{d} diverges at $\sim$ 10 at low volume fractions and $\sim$ 20 at large volume fractions.  
}
\end{figure*}

To recast this model into the crossover scaling framework, we substitute Eq.~\ref{eq:WC2} into Eq.~\ref{eq:WC1} to obtain:
\begin{equation}
\label{eq:WC3}
\eta  \sim  (\phi_0(1 - f(\sigma)) + \phi_\mu f(\sigma) - \phi)^{-2}. 
\end{equation}
Pulling out a factor of $(\phi_0 - \phi)^{-2}$, we find that the viscosity can be expressed as a function that only depends on a specific combination of $f(\sigma)$ and $\phi_0 - \phi$:
\begin{equation}
\label{eq:ScalingEquationWC}
\eta(\phi_0 - \phi)^2  \sim \mathcal{F}_{WC}\left( \frac{f(\sigma)}{\phi_0 - \phi} \right) 
\end{equation}
where the crossover scaling function $\mathcal{F}_{WC}$, specific to the Wyart and Cates model is 
\begin{equation}
\label{eq:ScalingWC}
\mathcal{F}_{WC}  \sim \left( \frac{1}{\phi_0 - \phi_\mu} - \frac{f(\sigma)}{\phi_0 - \phi} \right)^{-2}. 
\end{equation}
At small values of the scaling variable, $x_{WC} = f(\sigma)/(\phi_0 - \phi)$, the scaling function is a constant and the system behaviour is governed by the frictionless jamming critical point $\eta \sim (\phi_0 - \phi)^{-2}$. Crucially, however, the crossover scaling function $\mathcal{F}_{WC}$ has a divergence at $x_c = 1/(\phi_0 - \phi_\mu)$ indicating that as $x_{WC}/x_c \rightarrow 1$ the system is governed by a line of frictional jamming critical points such that $\eta \sim (x_c - x_{WC})^{-2}$. Notably, with this form of the crossover scaling function, the viscosity diverges with exactly the same exponent of $-2$ all along the jamming line.
This recasting of the Wyart and Cates model for the thickening transition in terms of crossover scaling clearly indicates that a major assumption in the model is that frictionless and frictional shear jamming (at nonzero $x_{WC}$) are controlled by the same fixed point and that the only effect of friction is to change the location of the critical point.

Practically, this formulation enables us to move beyond merely fitting the model to viscosity data, and instead attempt a scaling collapse to elucidate the underlying physics controlling this transition. More specifically, Eq.~\ref{eq:ScalingEquationWC} suggests that plotting $\eta(\phi_0 - \phi)^2$ as a function of $f(\sigma)/(\phi_0 - \phi)$ should collapse the viscosity across various stresses and volume fractions onto a universal curve, revealing the scaling function and its singularities. This collapse will also allow us to determine whether thickening is indeed controlled by a unique scaling exponent. 
\begin{figure*}[t]
\includegraphics[width=0.8\linewidth]{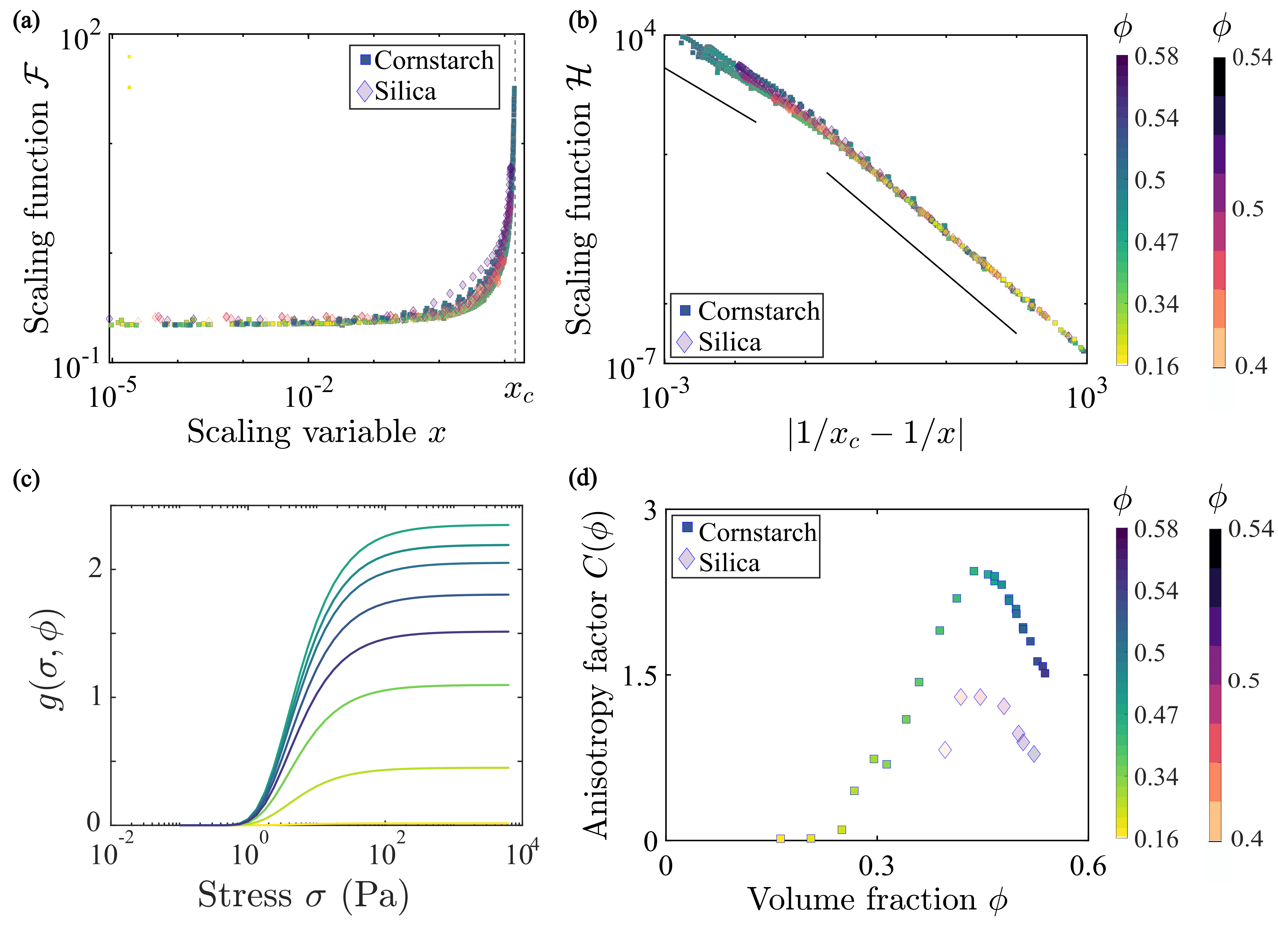}
\caption{\label{fig:Scaling} \textbf{Universal scaling of the suspension viscosity}. \textbf{a}. The scaling function $\mathcal{F} = \eta(\phi_0 - \phi)^2$ as a function of the scaling variable $x = e^{-\sigma^*/\sigma}C(\phi)/(\phi_0 - \phi)$ for all the cornstarch (squares) and silica (diamond) suspensions data. We find that all the data collapse onto a single universal curve that diverges at $x = x_c$. \textbf{b}. The scaling function $
\mathcal{H} = \eta g^2(\sigma, \phi)$ versus $|1/x_c - 1/x|$ for all the cornstarch and silica suspensions data. This way of scaling the data clearly illustrates two distinct regimes - a regime characterized by a power law of -2 at small $x$ and a power law of $\sim -3/2$ at $x \approx x_c$. The solid black lines indicate power laws of $-2$ and $-1.5$ respectively. \textbf{c} The non-linear scaling variable, $g(\sigma, \phi)$ as a function of the stress for a range of volume fractions of the cornstarch suspensions. We note that this parameter has the same sigmoidal shape, now well established for the fraction of frictional contacts, while $C(\phi)$ dramatically affects the overall scale. \textbf{d}. The anisotropy factor $C(\phi)$ as a function of the volume fraction for both silica and cornstarch. We note that $C(\phi)$ is a smooth analytic function as required by theories for scaling variables such as $g(\sigma,\phi)=f(\sigma)C(\phi)$ far from the critical point.} 
\end{figure*}
\section{\label{sec:level1} Experimental Methods}
We test this scaling theory on two different  non-inertial, low Reynolds number systems - a mixture of cornstarch in glycerol and a mixture of hard sphere silica particles in glycerol both of which show shear thickening behaviour but over different ranges of volume fractions. The samples were prepared by weighing out the solutes - cornstarch (Argo) and silica (2~$\mu$m charge stabilized spheres from Angstrom Sphere) and the solvent - glycerol (Sigma-Aldrich). We use glycerol as the solvent in each of these cases because of its low vapour pressure and the ease with which these glycerol-based suspensions can be loaded onto the rheometer. The cornstarch suspensions were used immediately after preparation and the silica suspensions were sonicated for 60 minutes prior to use. The viscosity of the suspension is measured using a stress-controlled method and a parallel plate geometry, with a set gap of 1~mm, on an Anton Paar MCR 702 rheometer. The sample was presheared at a constant stress of 1~Pa for five minutes. The suspension viscosity was then measured by performing a descending stress ramp.

 The viscosity of the cornstarch and silica suspensions of different volume fractions as a function of the stress are shown in Figs.~\ref{fig:FlowCurve}\textbf{a} and~\ref{fig:FlowCurve}\textbf{b} respectively. The low volume fraction data (yellow and light pink data in Figs.~\ref{fig:FlowCurve}\textbf{a} and~\ref{fig:FlowCurve}\textbf{b} respectively) have smaller viscosities and mild shear thickening, or continuous shear thickening (CST). The intermediate volume fraction data (teal and purple data in Figs.~\ref{fig:FlowCurve}\textbf{a} and~\ref{fig:FlowCurve}\textbf{b} respectively) have larger viscosities and show discontinuous shear thickening (DST), where $ d\log(\eta)/d\log(\sigma) \ge 1$.  

\section{\label{sec:level1} Results}
To collapse the viscosity using Eq.~\ref{eq:ScalingWC}, we need to determine the isotropic jamming fraction, $\phi_0$, and the fraction of frictional contacts, $f(\sigma)$. We determine $\phi_0$ from the divergence of the low stress viscosity (see SI for details). We use the form $f(\sigma) = e^{-\sigma^*/\sigma}$ for the fraction of frictional contacts, which is consistent with fits of the Wyart and Cates model in prior literature \cite{guy2015towards, guy2015towards, lee2020experimental}. We have explored a limited set of different expressions for $f(\sigma)$, none of which qualitatively change the results reported here (see SI for results with a different functional form of $f$). By fitting the flow curves in Fig.~\ref{fig:FlowCurve}\textbf{a},\textbf{b} to the Wyart and Cates model, we determine $\sigma^*$. Using the calculated values of $f(\sigma)$, and $\phi_0$, we plot $\mathcal{F}_{WC} = \eta(\phi_0 - \phi)^2$ as a function of $x_{WC} = e^{-\sigma^*/\sigma}/(\phi_0 - \phi)$ in Figs.~\ref{fig:FlowCurve}\textbf{c} and \textbf{d}. While we find that the collapse is promising, at higher volume fractions, the data diverge at different values of $x_{WC}$ (Fig.~\ref{fig:FlowCurve} insets). Since this reformulation of the Wyart and Cates model in the language of crossover scaling assumes a very particular form of the scaling function, $\mathcal{F}_{WC}$, and the scaling variable, $x_{WC}$, we naturally ask whether relaxing these assumptions and using the full machinery of crossover scaling leads to better data collapse and a more accurate description of the observed thickening transitions. 

We find that using a scaling variable where the numerator is both a function of stress and volume fraction; $x = g(\sigma, \phi)/(\phi_0 - \phi)$ dramatically improves the scaling collapse (Fig.~\ref{fig:Scaling}). To simplify the search for the function, $g(\sigma, \phi)$, we assume a product form, $g(\sigma, \phi) = C(\phi)f(\sigma)$. Impressively, this single parameter, $C(\phi)$, for the data set at each volume fraction collapses both the cornstarch and silica suspension data across all measured volume fractions onto a single universal scaling function $\mathcal{F}$ as shown in Fig. \ref{fig:Scaling}\textbf{a}\footnote{We find that the scaling functions for cornstarch silica differ by a multiplicative factor of $\sim$2, which simply reflects a different solvent viscosity.}.

\begin{figure*}[t]
\includegraphics[width=0.8\linewidth]{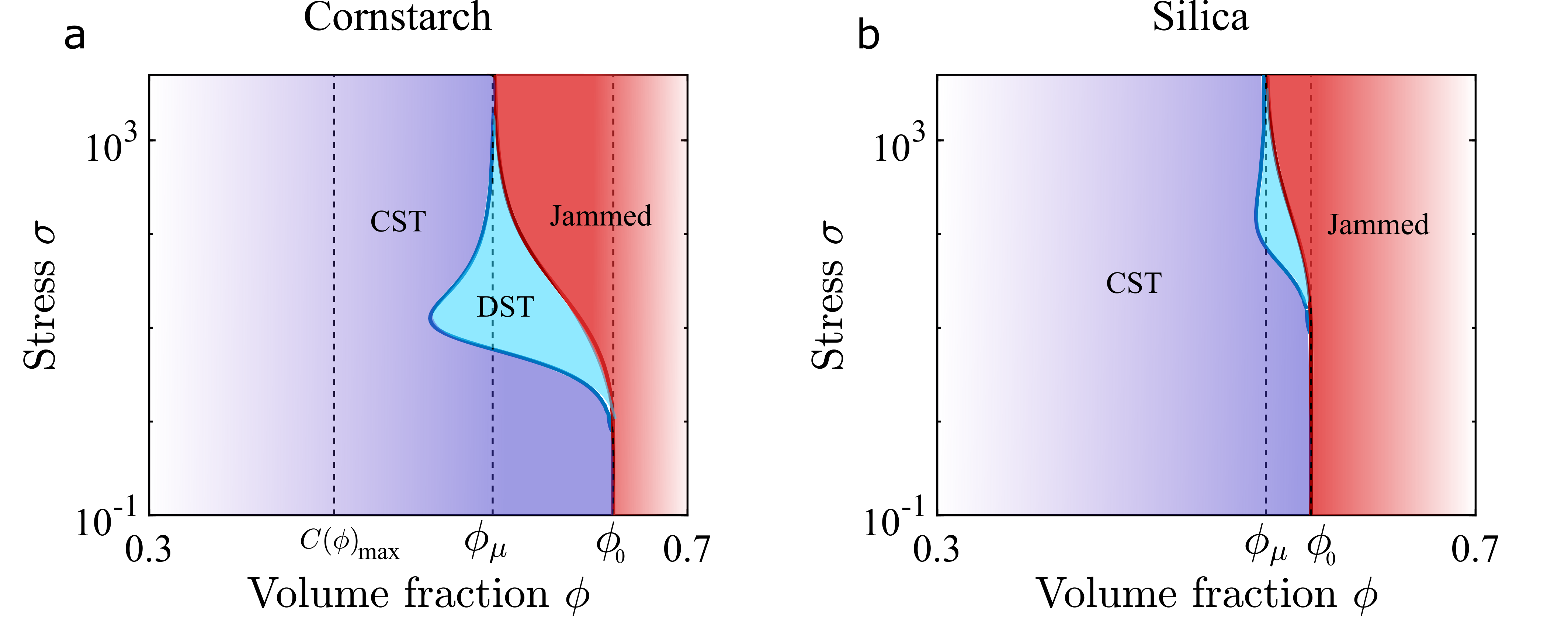}
\caption{\label{fig:PhaseDiagram} \textbf{Phase diagrams for cornstarch and silica suspensions as derived from the scaling analysis}. Three distinct regions are seen in the phase diagrams for the cornstarch \textbf{a} and silica \textbf{b} systems- continuous shear thickening (CST) in purple, discontinuous shear thickening (DST) in blue and a jammed region in red. The shear jamming line (maroon) is determined by $x = x_c$, where $x = e^{-\sigma^*/\sigma}C(\phi)/(\phi_0 - \phi)$ and the DST line (blue) is determined by the condition $d\log\eta/d\log\sigma = 1$. The vertical dotted lines indicate the values of frictionless jamming point, $\phi_0$, and shear jammed point, $\phi_{\mu}$, and the volume fraction at which $C(\phi)$ is maximum.} 
\end{figure*}

The scaling function has several characteristic features that are consistent with the expected behaviour for shear thickening suspensions. At small $x$, $\mathcal{F}$ is a constant, and the divergence in the viscosity is: 
\begin{equation}
    \eta \sim \frac{1}{(\phi_0 - \phi)^2}.
\end{equation}
which is consistent with a number of previous studies \cite{singh2018constitutive, guazzelli2018rheology, morris1999curvilinear, guy2020testing}. As $x$ increases, $\mathcal{F}$ also increases, until it diverges at $x = x_c$, with an exponent $\delta$:
\begin{equation}
    \mathcal{F}(x) \sim \frac{1}{(x_c - x)^\delta},
\end{equation}
with $x_c  = 14.5$. We note that since $C(\phi)$ is a multiplicative factor, there is an overall scale that can be chosen for $x_c$. 
This divergence implies that the viscosity also diverges with the exponent $\delta$. Remarkably, we find that $\delta < 2$ and is significantly different from the exponent observed for frictionless jamming (See SI, Fig. S3). To visualize this change in exponents, we follow Cardy \cite[Sect. 4.2]{cardy1996scaling} and write: 
\begin{equation}
\label{eq:ScalingTheoryCardy}
    \eta g^2(\sigma, \phi) \sim \mathcal{H}\left( |1/x_c - 1/x| \right).
\end{equation}
where $\mathcal{H}$ is a universal scaling function. We then plot $\eta g^2(\sigma, \phi)$ as a function of $|1/x_c - 1/x|$ in Fig.~\ref{fig:Scaling}\textbf{b}. We find excellent collapse over seven orders of magnitude in the scaling variable with two easily distinguishable regimes each characterized by clearly different power-law exponents. At small $x$, far from $x_c$, the behaviour is governed by the frictionless jamming point $\phi_0$ and $\mathcal{H} \sim  |1/x_c - 1/x|^{-2}$. As $x$ approaches $x_c$, we observe a clear change in the value of the exponent from -2, with a crossover between the two regimes at $x/x_c \sim$ 0.1. Our best estimate for the new exponent is -3/2. Importantly, this change in exponent indicates that the crossover underlying shear thickening is between critical points that belong to \textit{different} universality classes. As such, this change in exponent is a remarkable demonstration of frictional shear-jamming being qualitatively different from frictionless jamming at zero shear.

Importantly, the nonlinear scaling variable that drives the suspension towards frictional jamming, $g(\sigma, \phi)$, depends on both the stress and the volume fraction as shown in Fig.~\ref{fig:Scaling}\textbf{c}. $g(\sigma, \phi)$ is sigmoidal in the stress, similar to the fraction of frictional contacts, $f(\sigma)$, in previous works \cite{wyart2014discontinuous, guy2020testing, royer2016rheological}. The volume fraction dependence changes the overall scale and indicates that the fraction of frictional contacts that contribute to and determine the shear viscosity varies with $\phi$. The scaling collapse of the data reveals that this functional dependence, $C(\phi)$, is non-monotonic for both the silica and cornstarch suspensions (Fig.~\ref{fig:Scaling}\textbf{d}). Once these material dependent differences in $C(\phi)$ and $\phi_0$ are accounted for, the collapse is universal. 

Since the scaling form (Eq.~\ref{eq:ScalingTheoryCardy}) is valid for both suspensions at all stresses and volume fractions, we can use it to construct a shear thickening phase diagram as shown in Fig.~\ref{fig:PhaseDiagram}~\footnote{To construct this phase diagram we use an extended data set with data closer to $x_c$. See SI for more details}. Crucially, constructing these phase diagrams is only possible because we have determined the full functional form of the scaling function over seven orders of magnitude in the scaling variable, which is not typical for other analyses. In particular, we plot the shear jamming boundary at $x = x_c$ (border of red region) where the system transitions from a flowing to a jammed state. Since this border is defined by the scaling variable $x_c$, it is independent of the details of the scaling function $\mathcal{F}$(shape, functional form etc.). In addition, the boundary for discontinuous shear thickening is determined from:
\begin{equation}
\frac{d \log(\eta)}{d \log \sigma} = 1,
\label{eq:DST}
\end{equation}
which \textit{does} depend on the form of the scaling function. To obtain this boundary (blue lines), we fit $\mathcal{H}$ to a function consisting of two power law regimes stitched together by a crossover region (See SI), and use this fit to compute the derivatives in Eq.~\ref{eq:DST}. This boundary indicates the transition from continuous to discontinuous shear thickening regimes. Due to the form of the anisotropy factor $C(\phi)$, and differences in $\phi_0$ and $\sigma^*$, the phase diagrams for cornstarch and silica are distinct, and are shown in Figs. \ref{fig:PhaseDiagram}\textbf{a} and \textbf{b} respectively. These phase diagrams are qualitatively similar to those obtained by previous experiments and simulations \cite{peters2016direct, seto2013discontinuous, bi2011jamming} but are generated directly from the scaling collapse of the data. 

\section{\label{sec:level1} Discussion}
\subsection{\label{sec:level2} Altering the scaling variable to change material properties}
A key modification that enabled scaling collapse of the data is the non-linear scaling variable, $g(\sigma, \phi) = f(\sigma)C(\phi)$, that depends on \textit{both} the stress and the volume fraction. Previous studies have interpreted $f(\sigma)$ as the fraction of frictional contacts. If we retain this interpretation, then the inclusion of $C(\phi)$ in $g(\sigma, \phi)$ suggests that only a portion of $f(\sigma)$ contributes to the viscosity divergence. Such a modulation of $f(\sigma)$ is indicative of the role of  force network connectivity or anisotropy in the viscosity divergence consistent with recent simulations \cite{gameiro2020interaction, singh2020shear, gillissen2020constitutive} and models \cite{gillissen2019constitutive, gillissen2020constitutive,baumgarten2019general}. 

Interpreting $g(\sigma, \phi)$ in this manner enables us to envision how we can alter the shear thickening phase diagram. In particular, by modifying material properties corresponding to changes in $\sigma^*$, $\phi_0$, and $C(\phi)$ we can dramatically shift the borders between the jammed and unjammed regimes and by extension the discontinuous versus continuous shear thickening. A change in $\sigma^*$, for example, indicates how easily frictional interactions increase with stress and can be controlled by altering particle roughness~\cite{hsu2018roughness, hsiao2017rheological, hsu2021exploring, hoffman1998explanations, jamali2019alternative, more2020roughness, lootens2005dilatant}, hydrogen bonding~\cite{jaishankar2015probing, james2018interparticle, james2019tuning}, or solvent-particle interactions~\cite{van2021role}. Changes to $\phi_0$ can be generated by modifying particle roughness~\cite{hsu2018roughness, hsiao2017rheological, hsu2021exploring, hoffman1998explanations, jamali2019alternative, more2020roughness, townsend2017frictional}, shape~\cite{brown2011shear, james2019controlling, rathee2020role}, and polydispersity~\cite{guy2020testing, bender1996reversible}. Finally, differences in $C(\phi)$ may result from the constraints governing particle displacements and rotations as well as other perturbations to the flows~\cite{singh2020shear, pradeep2020jamming, o2019liquid, sehgal2019using, lin2016tunable, ness2018shaken, niu2020tunable, white2010extensional, han2018shear}. Thus, differences in these variables directly inform the types of changes that one can use to influence the thickening and jamming behaviours.

\subsection{\label{sec:level2} Exponents related to frictionless and frictional divergence}
The scaling analysis in Fig.~\ref{fig:Scaling}\textbf{b} illustrates two distinct power law regimes, one with an exponent of -2, associated with frictionless isotropic jamming and another with an exponent of -3/2 associated with frictional shear jamming. Measurements of the divergence in the viscosity associated with frictionless and frictional jamming have been reported previously in a wide range of systems. Our -2 scaling result for frictionless isotropic jamming is consistent with a number of previous studies in shear thickening suspensions~\cite{singh2018constitutive, guazzelli2018rheology, morris1999curvilinear, guy2020testing}. 

We note that in pressure driven suspensions there is a claim that the scaling exponent in this regime is -2.85. This apparent discrepancy, however, is a result of an assumption that stress and pressure scale identically close to jamming, which may not hold for dense suspensions. For the expert reader, previous works have shown a scaling of $(\phi_0 - \phi) \sim \mathcal{J}^{\gamma_\phi}$, with $\gamma_\phi = 0.37$ for frictionless particles and $\gamma_\phi = 0.7$ for frictional suspensions~\cite{degiuli2015unified, boyer2011unifying}. Here, $\mathcal{J} = \eta_0 \dot{\gamma} /P$ is the viscous number, $\eta_0$ is the solvent viscosity, $\dot{\gamma}$ is the shear rate and $P$ is the applied pressure~\cite{degiuli2015unified, boyer2011unifying, perrin2019interparticle, perrin2021nonlocal}. Assuming that $P \sim \sigma$ close to jamming, we can invert the equation to obtain $\eta \sim (\phi_0-\phi)^{-1/\gamma_{\phi}} = (\phi_0-\phi)^{-2.85}$ in the frictionless regime, giving an apparent discrepancy in the scaling exponent \cite{degiuli2015unified}. The assumption that $P \sim \sigma$ however, is a conjecture and may not hold for dense suspensions. Indeed, a recent study extending the analysis presented here has shown that the pressure and the shear viscosities are associated with different scaling exponents \cite{malbranche2022scaling, malbranche2023shear}. In these simulations, the shear viscosity exponents are consistent with those presented here and the pressure viscosity exponents are consistent with those presented in the pressure driven systems~\cite{malbranche2022scaling}.

With respect to the exponent of -3/2 for the shear jamming regime, we note that others have previously tried to fit exponents to viscosity versus volume fraction data. Our result is well within the range of previously obtained exponents ~\cite{mari2014shear, guy2015towards, guy2020testing, singh2018constitutive}. Moreover, here, we use scaling collapse of the entire shear thickening transition, using a scaling variable that is a function of stress and the volume fraction to determine both exponents. This approach is more rigorous since multiple measurements corresponding to different combinations of stress and volume fraction are used to determine the value of the universal function $\mathcal{F}$ at each point $x$.

\subsection{\label{sec:level2}Renormalization group flows}
Recasting the non-equilibrium thickening transition as being governed by crossover scaling between two different critical points suggests a renormalization group flow diagram with two fixed points. This picture is directly analogous to that found in equilibrium magnetic systems (e.g. Heisenberg to Ising crossover scaling). We project the flows, obtained from the experimentally determined scaling function, onto the $g(\sigma, \phi)$ and $|\phi_0 - \phi|$ plane. The nonlinear scaling variable,  $g(\sigma, \phi) = f(\sigma) C(\phi)$, is associated with the relevant direction at the frictionless jamming point with the nonlinear terms in $g(\sigma,\phi)$ providing analytic corrections to scaling~\footnote{just as one defines $u_h(P,T)$ as the magnetic-field like perturbation away from the liquid-gas critical point~\cite{cardy1996scaling}. Usually, analytic corrections to scaling are treated perturbatively near criticality. Here we use them to extend our theory to the entire shear-thickening regime.}.
The isotropic jamming fixed point has two relevant directions with flows away from the fixed point. The first is along the $|\phi_0 - \phi|$ axis (black line double arrows) and the second is along the line of critical points on the shear-jamming line (black line single arrow) flowing to the shear jamming fixed point. This critical manifold separates the jammed and the flowing states. The shear jamming fixed point must also have a relevant variable, and the flow along this line determines the exponent $\delta$ (drawn schematically in green line double arrows). Finally, the crossover between the two fixed points can be approximated by the knee in Fig.~\ref{fig:Scaling}\textbf{b} and is depicted in the renormalization group flow by the dashed grey line. This diagram clearly illustrates the deep connection between the shear thickening transition and renormalization group flows that are the hallmark of critical phenomena in thermodynamic systems.

\begin{figure}
\includegraphics[width=1\linewidth]{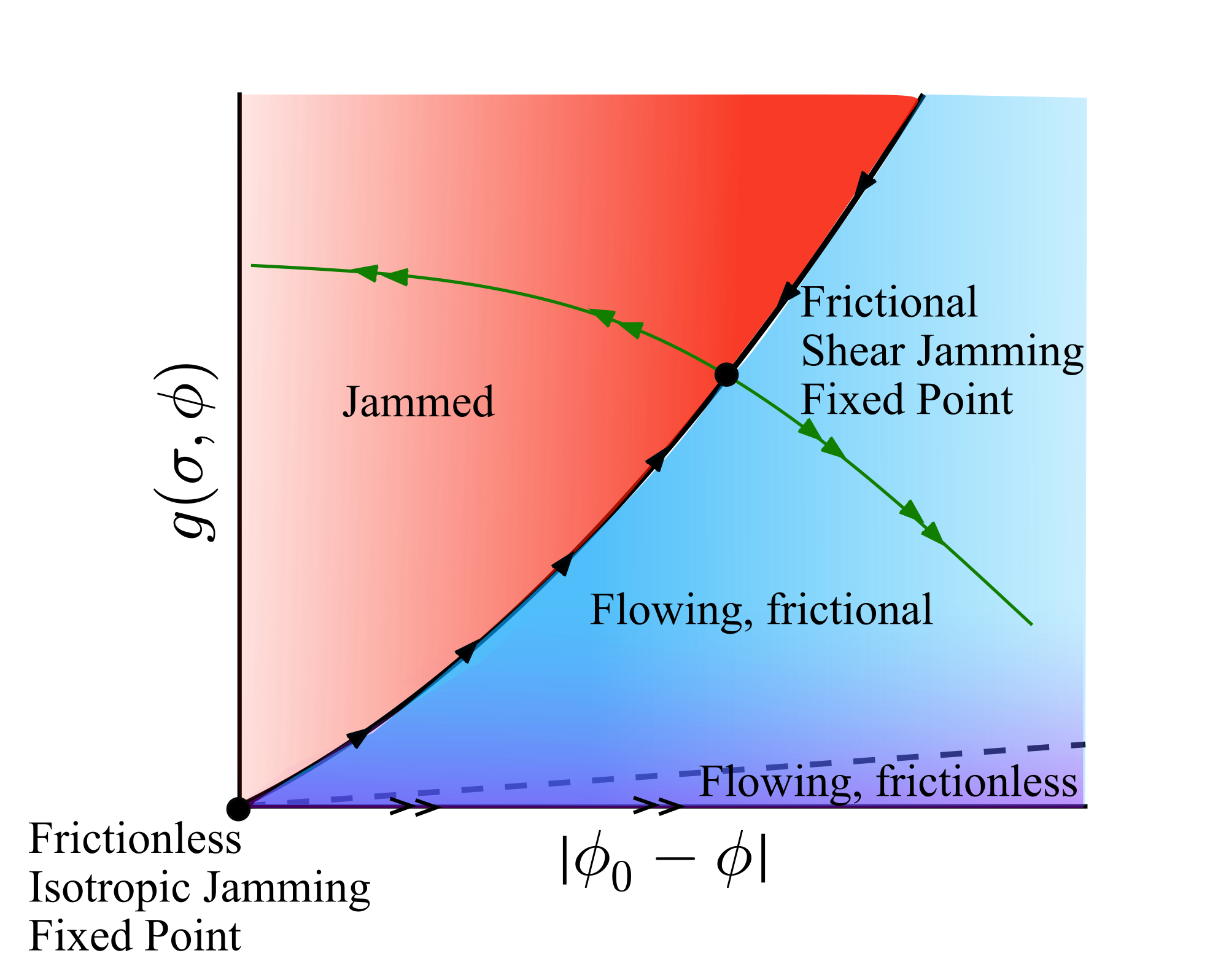}
\caption{\label{fig:RGFlow} \textbf{Renormalization group flow diagram governing the shear thickening transition}. 
We plot the projection of the renormalization group flows onto the $g(\sigma, \phi)$ and $|\phi_0 - \phi|$ plane. The two distinct fixed points are the frictionless isotropic jamming fixed point and the frictional shear jamming fixed point. The black solid line with single arrows is a line of critical points, and the dashed line corresponds to the $x$ value for the knee in Fig.~\ref{fig:Scaling}\textbf{b}. The flows are outward from the isotropic jamming point, which corresponds to a critical exponent of -2 along the $|\phi_0 - \phi|$ axis. The green line is the outward flow line for the shear jammed fixed point, which corresponds to a critical exponent of -3/2 for the viscosity.}
\end{figure}


\section{\label{sec:level2} Conclusions}
Adopting this statistical mechanics framework to describe shear thickening opens several novel avenues for future work. For instance, universal scaling theories and scaling functions such as those presented have previously been used to predict a number of physical properties in equilibrium systems, suggesting that similar approaches may be used in shear thickening. In the well-studied system of Heisenberg and Ising magnets, for example, one can use the crossover scaling function to determine the specific heat, correlation functions, and many other system properties \cite{cardy1996scaling}. Similarly, recent studies have investigated the scaling relations for the jamming transition for frictionless particles, demonstrating that both static and dynamic viscoelastic properties such as the shear stress, bulk modulus, and shear modulus can be predicted by such a scaling formulation \cite{goodrich2016scaling, o2003jamming, goodrich2014jamming, liarte2019jamming, liarte2021scaling, liarte2021universal}. These studies demonstrate the power of scaling theories and universal scaling functions and our work suggests that analogous predictions and theories could be generated for shear thickening and frictional jamming.

\begin{acknowledgments}
We thank Edward Y. X. Ong, Eric M. Schwen, and Stephen J. Thornton for valuable discussions and Anton Paar for use of the MCR 702 rheometer through their VIP academic research program. For this work, IC, IG, and MR were generously supported by  NSF CBET award numbers: 2010118, 1804963, 1509308, and an NSF DMR award number: 1507607. BC was supported by NSF CBET award number 1916877 and NSF DMR award number 2026834. JPS was supported by NSF CBET-2010118 and DMR-1719490. EK was supported by the NSF Award PHY-1554887, the University of Pennsylvania Materials Research Science and Engineering Center (MRSEC) through Award DMR-1720530. EDG was supported by National Science Foundation (NSF DMR-2026842).
\end{acknowledgments}

\nocite{*}
\bibliography{aipsamp}

\begin{thebibliography}{89}%
\makeatletter
\providecommand \@ifxundefined [1]{%
 \@ifx{#1\undefined}
}%
\providecommand \@ifnum [1]{%
 \ifnum #1\expandafter \@firstoftwo
 \else \expandafter \@secondoftwo
 \fi
}%
\providecommand \@ifx [1]{%
 \ifx #1\expandafter \@firstoftwo
 \else \expandafter \@secondoftwo
 \fi
}%
\providecommand \natexlab [1]{#1}%
\providecommand \enquote  [1]{``#1''}%
\providecommand \bibnamefont  [1]{#1}%
\providecommand \bibfnamefont [1]{#1}%
\providecommand \citenamefont [1]{#1}%
\providecommand \href@noop [0]{\@secondoftwo}%
\providecommand \href [0]{\begingroup \@sanitize@url \@href}%
\providecommand \@href[1]{\@@startlink{#1}\@@href}%
\providecommand \@@href[1]{\endgroup#1\@@endlink}%
\providecommand \@sanitize@url [0]{\catcode `\\12\catcode `\$12\catcode
  `\&12\catcode `\#12\catcode `\^12\catcode `\_12\catcode `\%12\relax}%
\providecommand \@@startlink[1]{}%
\providecommand \@@endlink[0]{}%
\providecommand \url  [0]{\begingroup\@sanitize@url \@url }%
\providecommand \@url [1]{\endgroup\@href {#1}{\urlprefix }}%
\providecommand \urlprefix  [0]{URL }%
\providecommand \Eprint [0]{\href }%
\providecommand \doibase [0]{http://dx.doi.org/}%
\providecommand \selectlanguage [0]{\@gobble}%
\providecommand \bibinfo  [0]{\@secondoftwo}%
\providecommand \bibfield  [0]{\@secondoftwo}%
\providecommand \translation [1]{[#1]}%
\providecommand \BibitemOpen [0]{}%
\providecommand \bibitemStop [0]{}%
\providecommand \bibitemNoStop [0]{.\EOS\space}%
\providecommand \EOS [0]{\spacefactor3000\relax}%
\providecommand \BibitemShut  [1]{\csname bibitem#1\endcsname}%
\let\auto@bib@innerbib\@empty
\bibitem [{\citenamefont {Brown}\ \emph {et~al.}(2010)\citenamefont {Brown},
  \citenamefont {Forman}, \citenamefont {Orellana}, \citenamefont {Zhang},
  \citenamefont {Maynor}, \citenamefont {Betts}, \citenamefont {DeSimone},\
  and\ \citenamefont {Jaeger}}]{brown2010generality}%
  \BibitemOpen
  \bibfield  {author} {\bibinfo {author} {\bibfnamefont {E.}~\bibnamefont
  {Brown}}, \bibinfo {author} {\bibfnamefont {N.~A.}\ \bibnamefont {Forman}},
  \bibinfo {author} {\bibfnamefont {C.~S.}\ \bibnamefont {Orellana}}, \bibinfo
  {author} {\bibfnamefont {H.}~\bibnamefont {Zhang}}, \bibinfo {author}
  {\bibfnamefont {B.~W.}\ \bibnamefont {Maynor}}, \bibinfo {author}
  {\bibfnamefont {D.~E.}\ \bibnamefont {Betts}}, \bibinfo {author}
  {\bibfnamefont {J.~M.}\ \bibnamefont {DeSimone}}, \ and\ \bibinfo {author}
  {\bibfnamefont {H.~M.}\ \bibnamefont {Jaeger}},\ }\bibfield  {title}
  {\enquote {\bibinfo {title} {Generality of shear thickening in dense
  suspensions},}\ }\href@noop {} {\bibfield  {journal} {\bibinfo  {journal}
  {Nature materials}\ }\textbf {\bibinfo {volume} {9}},\ \bibinfo {pages}
  {220--224} (\bibinfo {year} {2010})}\BibitemShut {NoStop}%
\bibitem [{\citenamefont {Wagner}\ and\ \citenamefont
  {Brady}(2009)}]{wagner2009shear}%
  \BibitemOpen
  \bibfield  {author} {\bibinfo {author} {\bibfnamefont {N.~J.}\ \bibnamefont
  {Wagner}}\ and\ \bibinfo {author} {\bibfnamefont {J.~F.}\ \bibnamefont
  {Brady}},\ }\bibfield  {title} {\enquote {\bibinfo {title} {Shear thickening
  in colloidal dispersions},}\ }\href@noop {} {\bibfield  {journal} {\bibinfo
  {journal} {Physics Today}\ }\textbf {\bibinfo {volume} {62}},\ \bibinfo
  {pages} {27--32} (\bibinfo {year} {2009})}\BibitemShut {NoStop}%
\bibitem [{\citenamefont {Brown}\ and\ \citenamefont
  {Jaeger}(2014)}]{brown2014shear}%
  \BibitemOpen
  \bibfield  {author} {\bibinfo {author} {\bibfnamefont {E.}~\bibnamefont
  {Brown}}\ and\ \bibinfo {author} {\bibfnamefont {H.~M.}\ \bibnamefont
  {Jaeger}},\ }\bibfield  {title} {\enquote {\bibinfo {title} {Shear thickening
  in concentrated suspensions: phenomenology, mechanisms and relations to
  jamming},}\ }\href@noop {} {\bibfield  {journal} {\bibinfo  {journal}
  {Reports on Progress in Physics}\ }\textbf {\bibinfo {volume} {77}},\
  \bibinfo {pages} {046602} (\bibinfo {year} {2014})}\BibitemShut {NoStop}%
\bibitem [{Note1()}]{Note1}%
  \BibitemOpen
  \bibinfo {note} {Search www.youtube.com for walking on
  cornstarch}\BibitemShut {NoStop}%
\bibitem [{\citenamefont {Lin}\ \emph {et~al.}(2015)\citenamefont {Lin},
  \citenamefont {Guy}, \citenamefont {Hermes}, \citenamefont {Ness},
  \citenamefont {Sun}, \citenamefont {Poon},\ and\ \citenamefont
  {Cohen}}]{lin2015hydrodynamic}%
  \BibitemOpen
  \bibfield  {author} {\bibinfo {author} {\bibfnamefont {N.~Y.}\ \bibnamefont
  {Lin}}, \bibinfo {author} {\bibfnamefont {B.~M.}\ \bibnamefont {Guy}},
  \bibinfo {author} {\bibfnamefont {M.}~\bibnamefont {Hermes}}, \bibinfo
  {author} {\bibfnamefont {C.}~\bibnamefont {Ness}}, \bibinfo {author}
  {\bibfnamefont {J.}~\bibnamefont {Sun}}, \bibinfo {author} {\bibfnamefont
  {W.~C.}\ \bibnamefont {Poon}}, \ and\ \bibinfo {author} {\bibfnamefont
  {I.}~\bibnamefont {Cohen}},\ }\bibfield  {title} {\enquote {\bibinfo {title}
  {Hydrodynamic and contact contributions to continuous shear thickening in
  colloidal suspensions},}\ }\href@noop {} {\bibfield  {journal} {\bibinfo
  {journal} {Phys. Rev. Lett.}\ }\textbf {\bibinfo {volume} {115}},\ \bibinfo
  {pages} {228304} (\bibinfo {year} {2015})}\BibitemShut {NoStop}%
\bibitem [{\citenamefont {Wyart}\ and\ \citenamefont
  {Cates}(2014)}]{wyart2014discontinuous}%
  \BibitemOpen
  \bibfield  {author} {\bibinfo {author} {\bibfnamefont {M.}~\bibnamefont
  {Wyart}}\ and\ \bibinfo {author} {\bibfnamefont {M.}~\bibnamefont {Cates}},\
  }\bibfield  {title} {\enquote {\bibinfo {title} {Discontinuous shear
  thickening without inertia in dense non-brownian suspensions},}\ }\href@noop
  {} {\bibfield  {journal} {\bibinfo  {journal} {Phys. Rev. Lett.}\ }\textbf
  {\bibinfo {volume} {112}},\ \bibinfo {pages} {098302} (\bibinfo {year}
  {2014})}\BibitemShut {NoStop}%
\bibitem [{\citenamefont {Cates}\ and\ \citenamefont
  {Wyart}(2014)}]{cates2014granulation}%
  \BibitemOpen
  \bibfield  {author} {\bibinfo {author} {\bibfnamefont {M.~E.}\ \bibnamefont
  {Cates}}\ and\ \bibinfo {author} {\bibfnamefont {M.}~\bibnamefont {Wyart}},\
  }\bibfield  {title} {\enquote {\bibinfo {title} {Granulation and bistability
  in non-brownian suspensions},}\ }\href@noop {} {\bibfield  {journal}
  {\bibinfo  {journal} {Rheologica Acta}\ }\textbf {\bibinfo {volume} {53}},\
  \bibinfo {pages} {755--764} (\bibinfo {year} {2014})}\BibitemShut {NoStop}%
\bibitem [{\citenamefont {Lootens}\ \emph {et~al.}(2005)\citenamefont
  {Lootens}, \citenamefont {Van~Damme}, \citenamefont {H{\'e}mar},\ and\
  \citenamefont {H{\'e}braud}}]{lootens2005dilatant}%
  \BibitemOpen
  \bibfield  {author} {\bibinfo {author} {\bibfnamefont {D.}~\bibnamefont
  {Lootens}}, \bibinfo {author} {\bibfnamefont {H.}~\bibnamefont {Van~Damme}},
  \bibinfo {author} {\bibfnamefont {Y.}~\bibnamefont {H{\'e}mar}}, \ and\
  \bibinfo {author} {\bibfnamefont {P.}~\bibnamefont {H{\'e}braud}},\
  }\bibfield  {title} {\enquote {\bibinfo {title} {Dilatant flow of
  concentrated suspensions of rough particles},}\ }\href@noop {} {\bibfield
  {journal} {\bibinfo  {journal} {Phys. Rev. Lett.}\ }\textbf {\bibinfo
  {volume} {95}},\ \bibinfo {pages} {268302} (\bibinfo {year}
  {2005})}\BibitemShut {NoStop}%
\bibitem [{\citenamefont {Seto}\ \emph {et~al.}(2013)\citenamefont {Seto},
  \citenamefont {Mari}, \citenamefont {Morris},\ and\ \citenamefont
  {Denn}}]{seto2013discontinuous}%
  \BibitemOpen
  \bibfield  {author} {\bibinfo {author} {\bibfnamefont {R.}~\bibnamefont
  {Seto}}, \bibinfo {author} {\bibfnamefont {R.}~\bibnamefont {Mari}}, \bibinfo
  {author} {\bibfnamefont {J.~F.}\ \bibnamefont {Morris}}, \ and\ \bibinfo
  {author} {\bibfnamefont {M.~M.}\ \bibnamefont {Denn}},\ }\bibfield  {title}
  {\enquote {\bibinfo {title} {Discontinuous shear thickening of frictional
  hard-sphere suspensions},}\ }\href@noop {} {\bibfield  {journal} {\bibinfo
  {journal} {Phys. Rev. Lett.}\ }\textbf {\bibinfo {volume} {111}},\ \bibinfo
  {pages} {218301} (\bibinfo {year} {2013})}\BibitemShut {NoStop}%
\bibitem [{\citenamefont {Brown}\ and\ \citenamefont
  {Jaeger}(2009)}]{brown2009dynamic}%
  \BibitemOpen
  \bibfield  {author} {\bibinfo {author} {\bibfnamefont {E.}~\bibnamefont
  {Brown}}\ and\ \bibinfo {author} {\bibfnamefont {H.~M.}\ \bibnamefont
  {Jaeger}},\ }\bibfield  {title} {\enquote {\bibinfo {title} {Dynamic jamming
  point for shear thickening suspensions},}\ }\href@noop {} {\bibfield
  {journal} {\bibinfo  {journal} {Physical review letters}\ }\textbf {\bibinfo
  {volume} {103}},\ \bibinfo {pages} {086001} (\bibinfo {year}
  {2009})}\BibitemShut {NoStop}%
\bibitem [{\citenamefont {Hsiao}\ \emph {et~al.}(2017)\citenamefont {Hsiao},
  \citenamefont {Jamali}, \citenamefont {Glynos}, \citenamefont {Green},
  \citenamefont {Larson},\ and\ \citenamefont
  {Solomon}}]{hsiao2017rheological}%
  \BibitemOpen
  \bibfield  {author} {\bibinfo {author} {\bibfnamefont {L.~C.}\ \bibnamefont
  {Hsiao}}, \bibinfo {author} {\bibfnamefont {S.}~\bibnamefont {Jamali}},
  \bibinfo {author} {\bibfnamefont {E.}~\bibnamefont {Glynos}}, \bibinfo
  {author} {\bibfnamefont {P.~F.}\ \bibnamefont {Green}}, \bibinfo {author}
  {\bibfnamefont {R.~G.}\ \bibnamefont {Larson}}, \ and\ \bibinfo {author}
  {\bibfnamefont {M.~J.}\ \bibnamefont {Solomon}},\ }\bibfield  {title}
  {\enquote {\bibinfo {title} {Rheological state diagrams for rough colloids in
  shear flow},}\ }\href@noop {} {\bibfield  {journal} {\bibinfo  {journal}
  {Physical review letters}\ }\textbf {\bibinfo {volume} {119}},\ \bibinfo
  {pages} {158001} (\bibinfo {year} {2017})}\BibitemShut {NoStop}%
\bibitem [{\citenamefont {Hsu}\ \emph {et~al.}(2018)\citenamefont {Hsu},
  \citenamefont {Ramakrishna}, \citenamefont {Zanini}, \citenamefont
  {Spencer},\ and\ \citenamefont {Isa}}]{hsu2018roughness}%
  \BibitemOpen
  \bibfield  {author} {\bibinfo {author} {\bibfnamefont {C.-P.}\ \bibnamefont
  {Hsu}}, \bibinfo {author} {\bibfnamefont {S.~N.}\ \bibnamefont
  {Ramakrishna}}, \bibinfo {author} {\bibfnamefont {M.}~\bibnamefont {Zanini}},
  \bibinfo {author} {\bibfnamefont {N.~D.}\ \bibnamefont {Spencer}}, \ and\
  \bibinfo {author} {\bibfnamefont {L.}~\bibnamefont {Isa}},\ }\bibfield
  {title} {\enquote {\bibinfo {title} {Roughness-dependent tribology effects on
  discontinuous shear thickening},}\ }\href@noop {} {\bibfield  {journal}
  {\bibinfo  {journal} {Proceedings of the National Academy of Sciences}\
  }\textbf {\bibinfo {volume} {115}},\ \bibinfo {pages} {5117--5122} (\bibinfo
  {year} {2018})}\BibitemShut {NoStop}%
\bibitem [{\citenamefont {Hsu}\ \emph {et~al.}(2021)\citenamefont {Hsu},
  \citenamefont {Mandal}, \citenamefont {Ramakrishna}, \citenamefont
  {Spencer},\ and\ \citenamefont {Isa}}]{hsu2021exploring}%
  \BibitemOpen
  \bibfield  {author} {\bibinfo {author} {\bibfnamefont {C.-P.}\ \bibnamefont
  {Hsu}}, \bibinfo {author} {\bibfnamefont {J.}~\bibnamefont {Mandal}},
  \bibinfo {author} {\bibfnamefont {S.~N.}\ \bibnamefont {Ramakrishna}},
  \bibinfo {author} {\bibfnamefont {N.~D.}\ \bibnamefont {Spencer}}, \ and\
  \bibinfo {author} {\bibfnamefont {L.}~\bibnamefont {Isa}},\ }\bibfield
  {title} {\enquote {\bibinfo {title} {Exploring the roles of roughness,
  friction and adhesion in discontinuous shear thickening by means of
  thermo-responsive particles},}\ }\href@noop {} {\bibfield  {journal}
  {\bibinfo  {journal} {Nature communications}\ }\textbf {\bibinfo {volume}
  {12}},\ \bibinfo {pages} {1--10} (\bibinfo {year} {2021})}\BibitemShut
  {NoStop}%
\bibitem [{\citenamefont {Hoffman}(1998)}]{hoffman1998explanations}%
  \BibitemOpen
  \bibfield  {author} {\bibinfo {author} {\bibfnamefont {R.~L.}\ \bibnamefont
  {Hoffman}},\ }\bibfield  {title} {\enquote {\bibinfo {title} {Explanations
  for the cause of shear thickening in concentrated colloidal suspensions},}\
  }\href@noop {} {\bibfield  {journal} {\bibinfo  {journal} {Journal of
  Rheology}\ }\textbf {\bibinfo {volume} {42}},\ \bibinfo {pages} {111--123}
  (\bibinfo {year} {1998})}\BibitemShut {NoStop}%
\bibitem [{\citenamefont {Jamali}\ and\ \citenamefont
  {Brady}(2019)}]{jamali2019alternative}%
  \BibitemOpen
  \bibfield  {author} {\bibinfo {author} {\bibfnamefont {S.}~\bibnamefont
  {Jamali}}\ and\ \bibinfo {author} {\bibfnamefont {J.~F.}\ \bibnamefont
  {Brady}},\ }\bibfield  {title} {\enquote {\bibinfo {title} {Alternative
  frictional model for discontinuous shear thickening of dense suspensions:
  Hydrodynamics},}\ }\href@noop {} {\bibfield  {journal} {\bibinfo  {journal}
  {Physical review letters}\ }\textbf {\bibinfo {volume} {123}},\ \bibinfo
  {pages} {138002} (\bibinfo {year} {2019})}\BibitemShut {NoStop}%
\bibitem [{\citenamefont {More}\ and\ \citenamefont
  {Ardekani}(2020)}]{more2020roughness}%
  \BibitemOpen
  \bibfield  {author} {\bibinfo {author} {\bibfnamefont {R.}~\bibnamefont
  {More}}\ and\ \bibinfo {author} {\bibfnamefont {A.}~\bibnamefont
  {Ardekani}},\ }\bibfield  {title} {\enquote {\bibinfo {title} {Roughness
  induced shear thickening in frictional non-brownian suspensions: A numerical
  study},}\ }\href@noop {} {\bibfield  {journal} {\bibinfo  {journal} {Journal
  of Rheology}\ }\textbf {\bibinfo {volume} {64}},\ \bibinfo {pages} {283--297}
  (\bibinfo {year} {2020})}\BibitemShut {NoStop}%
\bibitem [{\citenamefont {Bourrianne}\ \emph {et~al.}(2022)\citenamefont
  {Bourrianne}, \citenamefont {Niggel}, \citenamefont {Polly}, \citenamefont
  {Divoux},\ and\ \citenamefont {McKinley}}]{bourrianne2022tuning}%
  \BibitemOpen
  \bibfield  {author} {\bibinfo {author} {\bibfnamefont {P.}~\bibnamefont
  {Bourrianne}}, \bibinfo {author} {\bibfnamefont {V.}~\bibnamefont {Niggel}},
  \bibinfo {author} {\bibfnamefont {G.}~\bibnamefont {Polly}}, \bibinfo
  {author} {\bibfnamefont {T.}~\bibnamefont {Divoux}}, \ and\ \bibinfo {author}
  {\bibfnamefont {G.~H.}\ \bibnamefont {McKinley}},\ }\bibfield  {title}
  {\enquote {\bibinfo {title} {Tuning the shear thickening of suspensions
  through surface roughness and physico-chemical interactions},}\ }\href@noop
  {} {\bibfield  {journal} {\bibinfo  {journal} {Physical Review Research}\
  }\textbf {\bibinfo {volume} {4}},\ \bibinfo {pages} {033062} (\bibinfo {year}
  {2022})}\BibitemShut {NoStop}%
\bibitem [{\citenamefont {Jaishankar}\ \emph {et~al.}(2015)\citenamefont
  {Jaishankar}, \citenamefont {Wee}, \citenamefont {Matia-Merino},
  \citenamefont {Goh},\ and\ \citenamefont {McKinley}}]{jaishankar2015probing}%
  \BibitemOpen
  \bibfield  {author} {\bibinfo {author} {\bibfnamefont {A.}~\bibnamefont
  {Jaishankar}}, \bibinfo {author} {\bibfnamefont {M.}~\bibnamefont {Wee}},
  \bibinfo {author} {\bibfnamefont {L.}~\bibnamefont {Matia-Merino}}, \bibinfo
  {author} {\bibfnamefont {K.~K.}\ \bibnamefont {Goh}}, \ and\ \bibinfo
  {author} {\bibfnamefont {G.~H.}\ \bibnamefont {McKinley}},\ }\bibfield
  {title} {\enquote {\bibinfo {title} {Probing hydrogen bond interactions in a
  shear thickening polysaccharide using nonlinear shear and extensional
  rheology},}\ }\href@noop {} {\bibfield  {journal} {\bibinfo  {journal}
  {Carbohydrate polymers}\ }\textbf {\bibinfo {volume} {123}},\ \bibinfo
  {pages} {136--145} (\bibinfo {year} {2015})}\BibitemShut {NoStop}%
\bibitem [{\citenamefont {James}\ \emph {et~al.}(2018)\citenamefont {James},
  \citenamefont {Han}, \citenamefont {de~la Cruz}, \citenamefont {Jureller},\
  and\ \citenamefont {Jaeger}}]{james2018interparticle}%
  \BibitemOpen
  \bibfield  {author} {\bibinfo {author} {\bibfnamefont {N.~M.}\ \bibnamefont
  {James}}, \bibinfo {author} {\bibfnamefont {E.}~\bibnamefont {Han}}, \bibinfo
  {author} {\bibfnamefont {R.~A.~L.}\ \bibnamefont {de~la Cruz}}, \bibinfo
  {author} {\bibfnamefont {J.}~\bibnamefont {Jureller}}, \ and\ \bibinfo
  {author} {\bibfnamefont {H.~M.}\ \bibnamefont {Jaeger}},\ }\bibfield  {title}
  {\enquote {\bibinfo {title} {Interparticle hydrogen bonding can elicit shear
  jamming in dense suspensions},}\ }\href@noop {} {\bibfield  {journal}
  {\bibinfo  {journal} {Nature materials}\ }\textbf {\bibinfo {volume} {17}},\
  \bibinfo {pages} {965--970} (\bibinfo {year} {2018})}\BibitemShut {NoStop}%
\bibitem [{\citenamefont {James}\ \emph
  {et~al.}(2019{\natexlab{a}})\citenamefont {James}, \citenamefont {Hsu},
  \citenamefont {Spencer}, \citenamefont {Jaeger},\ and\ \citenamefont
  {Isa}}]{james2019tuning}%
  \BibitemOpen
  \bibfield  {author} {\bibinfo {author} {\bibfnamefont {N.~M.}\ \bibnamefont
  {James}}, \bibinfo {author} {\bibfnamefont {C.-P.}\ \bibnamefont {Hsu}},
  \bibinfo {author} {\bibfnamefont {N.~D.}\ \bibnamefont {Spencer}}, \bibinfo
  {author} {\bibfnamefont {H.~M.}\ \bibnamefont {Jaeger}}, \ and\ \bibinfo
  {author} {\bibfnamefont {L.}~\bibnamefont {Isa}},\ }\bibfield  {title}
  {\enquote {\bibinfo {title} {Tuning interparticle hydrogen bonding in
  shear-jamming suspensions: Kinetic effects and consequences for tribology and
  rheology},}\ }\href@noop {} {\bibfield  {journal} {\bibinfo  {journal} {The
  journal of physical chemistry letters}\ }\textbf {\bibinfo {volume} {10}},\
  \bibinfo {pages} {1663--1668} (\bibinfo {year}
  {2019}{\natexlab{a}})}\BibitemShut {NoStop}%
\bibitem [{\citenamefont {Brown}\ \emph {et~al.}(2011)\citenamefont {Brown},
  \citenamefont {Zhang}, \citenamefont {Forman}, \citenamefont {Maynor},
  \citenamefont {Betts}, \citenamefont {DeSimone},\ and\ \citenamefont
  {Jaeger}}]{brown2011shear}%
  \BibitemOpen
  \bibfield  {author} {\bibinfo {author} {\bibfnamefont {E.}~\bibnamefont
  {Brown}}, \bibinfo {author} {\bibfnamefont {H.}~\bibnamefont {Zhang}},
  \bibinfo {author} {\bibfnamefont {N.~A.}\ \bibnamefont {Forman}}, \bibinfo
  {author} {\bibfnamefont {B.~W.}\ \bibnamefont {Maynor}}, \bibinfo {author}
  {\bibfnamefont {D.~E.}\ \bibnamefont {Betts}}, \bibinfo {author}
  {\bibfnamefont {J.~M.}\ \bibnamefont {DeSimone}}, \ and\ \bibinfo {author}
  {\bibfnamefont {H.~M.}\ \bibnamefont {Jaeger}},\ }\bibfield  {title}
  {\enquote {\bibinfo {title} {Shear thickening and jamming in densely packed
  suspensions of different particle shapes},}\ }\href@noop {} {\bibfield
  {journal} {\bibinfo  {journal} {Physical Review E}\ }\textbf {\bibinfo
  {volume} {84}},\ \bibinfo {pages} {031408} (\bibinfo {year}
  {2011})}\BibitemShut {NoStop}%
\bibitem [{\citenamefont {James}\ \emph
  {et~al.}(2019{\natexlab{b}})\citenamefont {James}, \citenamefont {Xue},
  \citenamefont {Goyal},\ and\ \citenamefont {Jaeger}}]{james2019controlling}%
  \BibitemOpen
  \bibfield  {author} {\bibinfo {author} {\bibfnamefont {N.~M.}\ \bibnamefont
  {James}}, \bibinfo {author} {\bibfnamefont {H.}~\bibnamefont {Xue}}, \bibinfo
  {author} {\bibfnamefont {M.}~\bibnamefont {Goyal}}, \ and\ \bibinfo {author}
  {\bibfnamefont {H.~M.}\ \bibnamefont {Jaeger}},\ }\bibfield  {title}
  {\enquote {\bibinfo {title} {Controlling shear jamming in dense suspensions
  via the particle aspect ratio},}\ }\href@noop {} {\bibfield  {journal}
  {\bibinfo  {journal} {Soft matter}\ }\textbf {\bibinfo {volume} {15}},\
  \bibinfo {pages} {3649--3654} (\bibinfo {year}
  {2019}{\natexlab{b}})}\BibitemShut {NoStop}%
\bibitem [{\citenamefont {Rathee}\ \emph {et~al.}(2020)\citenamefont {Rathee},
  \citenamefont {Arora}, \citenamefont {Blair}, \citenamefont {Urbach},
  \citenamefont {Sood},\ and\ \citenamefont {Ganapathy}}]{rathee2020role}%
  \BibitemOpen
  \bibfield  {author} {\bibinfo {author} {\bibfnamefont {V.}~\bibnamefont
  {Rathee}}, \bibinfo {author} {\bibfnamefont {S.}~\bibnamefont {Arora}},
  \bibinfo {author} {\bibfnamefont {D.~L.}\ \bibnamefont {Blair}}, \bibinfo
  {author} {\bibfnamefont {J.~S.}\ \bibnamefont {Urbach}}, \bibinfo {author}
  {\bibfnamefont {A.}~\bibnamefont {Sood}}, \ and\ \bibinfo {author}
  {\bibfnamefont {R.}~\bibnamefont {Ganapathy}},\ }\bibfield  {title} {\enquote
  {\bibinfo {title} {Role of particle orientational order during shear
  thickening in suspensions of colloidal rods},}\ }\href@noop {} {\bibfield
  {journal} {\bibinfo  {journal} {Physical Review E}\ }\textbf {\bibinfo
  {volume} {101}},\ \bibinfo {pages} {040601} (\bibinfo {year}
  {2020})}\BibitemShut {NoStop}%
\bibitem [{\citenamefont {Guy}\ \emph {et~al.}(2020)\citenamefont {Guy},
  \citenamefont {Ness}, \citenamefont {Hermes}, \citenamefont {Sawiak},
  \citenamefont {Sun},\ and\ \citenamefont {Poon}}]{guy2020testing}%
  \BibitemOpen
  \bibfield  {author} {\bibinfo {author} {\bibfnamefont {B.~M.}\ \bibnamefont
  {Guy}}, \bibinfo {author} {\bibfnamefont {C.}~\bibnamefont {Ness}}, \bibinfo
  {author} {\bibfnamefont {M.}~\bibnamefont {Hermes}}, \bibinfo {author}
  {\bibfnamefont {L.~J.}\ \bibnamefont {Sawiak}}, \bibinfo {author}
  {\bibfnamefont {J.}~\bibnamefont {Sun}}, \ and\ \bibinfo {author}
  {\bibfnamefont {W.~C.}\ \bibnamefont {Poon}},\ }\bibfield  {title} {\enquote
  {\bibinfo {title} {Testing the wyart--cates model for non-brownian shear
  thickening using bidisperse suspensions},}\ }\href@noop {} {\bibfield
  {journal} {\bibinfo  {journal} {Soft matter}\ }\textbf {\bibinfo {volume}
  {16}},\ \bibinfo {pages} {229--237} (\bibinfo {year} {2020})}\BibitemShut
  {NoStop}%
\bibitem [{\citenamefont {Bender}\ and\ \citenamefont
  {Wagner}(1996)}]{bender1996reversible}%
  \BibitemOpen
  \bibfield  {author} {\bibinfo {author} {\bibfnamefont {J.}~\bibnamefont
  {Bender}}\ and\ \bibinfo {author} {\bibfnamefont {N.~J.}\ \bibnamefont
  {Wagner}},\ }\bibfield  {title} {\enquote {\bibinfo {title} {Reversible shear
  thickening in monodisperse and bidisperse colloidal dispersions},}\
  }\href@noop {} {\bibfield  {journal} {\bibinfo  {journal} {Journal of
  Rheology}\ }\textbf {\bibinfo {volume} {40}},\ \bibinfo {pages} {899--916}
  (\bibinfo {year} {1996})}\BibitemShut {NoStop}%
\bibitem [{\citenamefont {Ness}, \citenamefont {Mari},\ and\ \citenamefont
  {Cates}(2018)}]{ness2018shaken}%
  \BibitemOpen
  \bibfield  {author} {\bibinfo {author} {\bibfnamefont {C.}~\bibnamefont
  {Ness}}, \bibinfo {author} {\bibfnamefont {R.}~\bibnamefont {Mari}}, \ and\
  \bibinfo {author} {\bibfnamefont {M.~E.}\ \bibnamefont {Cates}},\ }\bibfield
  {title} {\enquote {\bibinfo {title} {Shaken and stirred: Random organization
  reduces viscosity and dissipation in granular suspensions},}\ }\href@noop {}
  {\bibfield  {journal} {\bibinfo  {journal} {Science advances}\ }\textbf
  {\bibinfo {volume} {4}},\ \bibinfo {pages} {eaar3296} (\bibinfo {year}
  {2018})}\BibitemShut {NoStop}%
\bibitem [{\citenamefont {Niu}\ \emph {et~al.}(2020)\citenamefont {Niu},
  \citenamefont {Ramaswamy}, \citenamefont {Ness}, \citenamefont {Shetty},\
  and\ \citenamefont {Cohen}}]{niu2020tunable}%
  \BibitemOpen
  \bibfield  {author} {\bibinfo {author} {\bibfnamefont {R.}~\bibnamefont
  {Niu}}, \bibinfo {author} {\bibfnamefont {M.}~\bibnamefont {Ramaswamy}},
  \bibinfo {author} {\bibfnamefont {C.}~\bibnamefont {Ness}}, \bibinfo {author}
  {\bibfnamefont {A.}~\bibnamefont {Shetty}}, \ and\ \bibinfo {author}
  {\bibfnamefont {I.}~\bibnamefont {Cohen}},\ }\bibfield  {title} {\enquote
  {\bibinfo {title} {Tunable solidification of cornstarch under impact: How to
  make someone walking on cornstarch sink},}\ }\href@noop {} {\bibfield
  {journal} {\bibinfo  {journal} {Science advances}\ }\textbf {\bibinfo
  {volume} {6}},\ \bibinfo {pages} {eaay6661} (\bibinfo {year}
  {2020})}\BibitemShut {NoStop}%
\bibitem [{\citenamefont {Lin}\ \emph {et~al.}(2016)\citenamefont {Lin},
  \citenamefont {Ness}, \citenamefont {Cates}, \citenamefont {Sun},\ and\
  \citenamefont {Cohen}}]{lin2016tunable}%
  \BibitemOpen
  \bibfield  {author} {\bibinfo {author} {\bibfnamefont {N.~Y.}\ \bibnamefont
  {Lin}}, \bibinfo {author} {\bibfnamefont {C.}~\bibnamefont {Ness}}, \bibinfo
  {author} {\bibfnamefont {M.~E.}\ \bibnamefont {Cates}}, \bibinfo {author}
  {\bibfnamefont {J.}~\bibnamefont {Sun}}, \ and\ \bibinfo {author}
  {\bibfnamefont {I.}~\bibnamefont {Cohen}},\ }\bibfield  {title} {\enquote
  {\bibinfo {title} {Tunable shear thickening in suspensions},}\ }\href@noop {}
  {\bibfield  {journal} {\bibinfo  {journal} {Proceedings of the National
  Academy of Sciences}\ }\textbf {\bibinfo {volume} {113}},\ \bibinfo {pages}
  {10774--10778} (\bibinfo {year} {2016})}\BibitemShut {NoStop}%
\bibitem [{\citenamefont {Sehgal}\ \emph {et~al.}(2019)\citenamefont {Sehgal},
  \citenamefont {Ramaswamy}, \citenamefont {Cohen},\ and\ \citenamefont
  {Kirby}}]{sehgal2019using}%
  \BibitemOpen
  \bibfield  {author} {\bibinfo {author} {\bibfnamefont {P.}~\bibnamefont
  {Sehgal}}, \bibinfo {author} {\bibfnamefont {M.}~\bibnamefont {Ramaswamy}},
  \bibinfo {author} {\bibfnamefont {I.}~\bibnamefont {Cohen}}, \ and\ \bibinfo
  {author} {\bibfnamefont {B.~J.}\ \bibnamefont {Kirby}},\ }\bibfield  {title}
  {\enquote {\bibinfo {title} {Using acoustic perturbations to dynamically tune
  shear thickening in colloidal suspensions},}\ }\href@noop {} {\bibfield
  {journal} {\bibinfo  {journal} {Physical review letters}\ }\textbf {\bibinfo
  {volume} {123}},\ \bibinfo {pages} {128001} (\bibinfo {year}
  {2019})}\BibitemShut {NoStop}%
\bibitem [{\citenamefont {Gillissen}\ \emph {et~al.}(2020)\citenamefont
  {Gillissen}, \citenamefont {Ness}, \citenamefont {Peterson}, \citenamefont
  {Wilson},\ and\ \citenamefont {Cates}}]{gillissen2020constitutive}%
  \BibitemOpen
  \bibfield  {author} {\bibinfo {author} {\bibfnamefont {J.}~\bibnamefont
  {Gillissen}}, \bibinfo {author} {\bibfnamefont {C.}~\bibnamefont {Ness}},
  \bibinfo {author} {\bibfnamefont {J.}~\bibnamefont {Peterson}}, \bibinfo
  {author} {\bibfnamefont {H.}~\bibnamefont {Wilson}}, \ and\ \bibinfo {author}
  {\bibfnamefont {M.}~\bibnamefont {Cates}},\ }\bibfield  {title} {\enquote
  {\bibinfo {title} {Constitutive model for shear-thickening suspensions:
  Predictions for steady shear with superposed transverse oscillations},}\
  }\href@noop {} {\bibfield  {journal} {\bibinfo  {journal} {Journal of
  Rheology}\ }\textbf {\bibinfo {volume} {64}},\ \bibinfo {pages} {353--365}
  (\bibinfo {year} {2020})}\BibitemShut {NoStop}%
\bibitem [{\citenamefont {Chen}\ \emph {et~al.}(2022)\citenamefont {Chen},
  \citenamefont {van~de Naald}, \citenamefont {Singh}, \citenamefont
  {Dolinski}, \citenamefont {Jackson}, \citenamefont {Jaeger}, \citenamefont
  {Rowan},\ and\ \citenamefont {de~Pablo}}]{chen2022leveraging}%
  \BibitemOpen
  \bibfield  {author} {\bibinfo {author} {\bibfnamefont {C.}~\bibnamefont
  {Chen}}, \bibinfo {author} {\bibfnamefont {M.}~\bibnamefont {van~de Naald}},
  \bibinfo {author} {\bibfnamefont {A.}~\bibnamefont {Singh}}, \bibinfo
  {author} {\bibfnamefont {N.}~\bibnamefont {Dolinski}}, \bibinfo {author}
  {\bibfnamefont {G.}~\bibnamefont {Jackson}}, \bibinfo {author} {\bibfnamefont
  {H.}~\bibnamefont {Jaeger}}, \bibinfo {author} {\bibfnamefont
  {S.}~\bibnamefont {Rowan}}, \ and\ \bibinfo {author} {\bibfnamefont
  {J.}~\bibnamefont {de~Pablo}},\ }\bibfield  {title} {\enquote {\bibinfo
  {title} {Leveraging the polymer glass transition to access
  thermally-switchable shear jamming suspensions},}\ }\href@noop {} {\
  (\bibinfo {year} {2022})}\BibitemShut {NoStop}%
\bibitem [{\citenamefont {Jackson}\ \emph {et~al.}(2022)\citenamefont
  {Jackson}, \citenamefont {Dennis}, \citenamefont {Dolinski}, \citenamefont
  {van~der Naald}, \citenamefont {Kim}, \citenamefont {Eom}, \citenamefont
  {Rowan},\ and\ \citenamefont {Jaeger}}]{jackson2022designing}%
  \BibitemOpen
  \bibfield  {author} {\bibinfo {author} {\bibfnamefont {G.~L.}\ \bibnamefont
  {Jackson}}, \bibinfo {author} {\bibfnamefont {J.~M.}\ \bibnamefont {Dennis}},
  \bibinfo {author} {\bibfnamefont {N.~D.}\ \bibnamefont {Dolinski}}, \bibinfo
  {author} {\bibfnamefont {M.}~\bibnamefont {van~der Naald}}, \bibinfo {author}
  {\bibfnamefont {H.}~\bibnamefont {Kim}}, \bibinfo {author} {\bibfnamefont
  {C.}~\bibnamefont {Eom}}, \bibinfo {author} {\bibfnamefont {S.~J.}\
  \bibnamefont {Rowan}}, \ and\ \bibinfo {author} {\bibfnamefont {H.~M.}\
  \bibnamefont {Jaeger}},\ }\bibfield  {title} {\enquote {\bibinfo {title}
  {Designing stress-adaptive dense suspensions using dynamic covalent
  chemistry},}\ }\href@noop {} {\bibfield  {journal} {\bibinfo  {journal}
  {Macromolecules}\ }\textbf {\bibinfo {volume} {55}},\ \bibinfo {pages}
  {6453--6461} (\bibinfo {year} {2022})}\BibitemShut {NoStop}%
\bibitem [{\citenamefont {Singh}\ \emph {et~al.}(2018)\citenamefont {Singh},
  \citenamefont {Mari}, \citenamefont {Denn},\ and\ \citenamefont
  {Morris}}]{singh2018constitutive}%
  \BibitemOpen
  \bibfield  {author} {\bibinfo {author} {\bibfnamefont {A.}~\bibnamefont
  {Singh}}, \bibinfo {author} {\bibfnamefont {R.}~\bibnamefont {Mari}},
  \bibinfo {author} {\bibfnamefont {M.~M.}\ \bibnamefont {Denn}}, \ and\
  \bibinfo {author} {\bibfnamefont {J.~F.}\ \bibnamefont {Morris}},\ }\bibfield
   {title} {\enquote {\bibinfo {title} {A constitutive model for simple shear
  of dense frictional suspensions},}\ }\href@noop {} {\bibfield  {journal}
  {\bibinfo  {journal} {Journal of Rheology}\ }\textbf {\bibinfo {volume}
  {62}},\ \bibinfo {pages} {457--468} (\bibinfo {year} {2018})}\BibitemShut
  {NoStop}%
\bibitem [{\citenamefont {Royer}, \citenamefont {Blair},\ and\ \citenamefont
  {Hudson}(2016)}]{royer2016rheological}%
  \BibitemOpen
  \bibfield  {author} {\bibinfo {author} {\bibfnamefont {J.~R.}\ \bibnamefont
  {Royer}}, \bibinfo {author} {\bibfnamefont {D.~L.}\ \bibnamefont {Blair}}, \
  and\ \bibinfo {author} {\bibfnamefont {S.~D.}\ \bibnamefont {Hudson}},\
  }\bibfield  {title} {\enquote {\bibinfo {title} {Rheological signature of
  frictional interactions in shear thickening suspensions},}\ }\href@noop {}
  {\bibfield  {journal} {\bibinfo  {journal} {Phys. Rev. Lett.}\ }\textbf
  {\bibinfo {volume} {116}},\ \bibinfo {pages} {188301} (\bibinfo {year}
  {2016})}\BibitemShut {NoStop}%
\bibitem [{\citenamefont {Lee}\ \emph {et~al.}(2020)\citenamefont {Lee},
  \citenamefont {Luo}, \citenamefont {Brown},\ and\ \citenamefont
  {Wagner}}]{lee2020experimental}%
  \BibitemOpen
  \bibfield  {author} {\bibinfo {author} {\bibfnamefont {Y.-F.}\ \bibnamefont
  {Lee}}, \bibinfo {author} {\bibfnamefont {Y.}~\bibnamefont {Luo}}, \bibinfo
  {author} {\bibfnamefont {S.~C.}\ \bibnamefont {Brown}}, \ and\ \bibinfo
  {author} {\bibfnamefont {N.~J.}\ \bibnamefont {Wagner}},\ }\bibfield  {title}
  {\enquote {\bibinfo {title} {Experimental test of a frictional contact model
  for shear thickening in concentrated colloidal suspensions},}\ }\href@noop {}
  {\bibfield  {journal} {\bibinfo  {journal} {Journal of Rheology}\ }\textbf
  {\bibinfo {volume} {64}},\ \bibinfo {pages} {267--282} (\bibinfo {year}
  {2020})}\BibitemShut {NoStop}%
\bibitem [{\citenamefont {Gillissen}\ \emph {et~al.}(2019)\citenamefont
  {Gillissen}, \citenamefont {Ness}, \citenamefont {Peterson}, \citenamefont
  {Wilson},\ and\ \citenamefont {Cates}}]{gillissen2019constitutive}%
  \BibitemOpen
  \bibfield  {author} {\bibinfo {author} {\bibfnamefont {J.~J.}\ \bibnamefont
  {Gillissen}}, \bibinfo {author} {\bibfnamefont {C.}~\bibnamefont {Ness}},
  \bibinfo {author} {\bibfnamefont {J.~D.}\ \bibnamefont {Peterson}}, \bibinfo
  {author} {\bibfnamefont {H.~J.}\ \bibnamefont {Wilson}}, \ and\ \bibinfo
  {author} {\bibfnamefont {M.~E.}\ \bibnamefont {Cates}},\ }\bibfield  {title}
  {\enquote {\bibinfo {title} {Constitutive model for time-dependent flows of
  shear-thickening suspensions},}\ }\href@noop {} {\bibfield  {journal}
  {\bibinfo  {journal} {Physical review letters}\ }\textbf {\bibinfo {volume}
  {123}},\ \bibinfo {pages} {214504} (\bibinfo {year} {2019})}\BibitemShut
  {NoStop}%
\bibitem [{\citenamefont {Pradeep}, \citenamefont {Jacob},\ and\ \citenamefont
  {Hsiao}(2020)}]{pradeep2020jamming}%
  \BibitemOpen
  \bibfield  {author} {\bibinfo {author} {\bibfnamefont {S.}~\bibnamefont
  {Pradeep}}, \bibinfo {author} {\bibfnamefont {A.~R.}\ \bibnamefont {Jacob}},
  \ and\ \bibinfo {author} {\bibfnamefont {L.~C.}\ \bibnamefont {Hsiao}},\
  }\bibfield  {title} {\enquote {\bibinfo {title} {Jamming distance dictates
  colloidal shear thickening},}\ }\href@noop {} {\bibfield  {journal} {\bibinfo
   {journal} {arXiv preprint arXiv:2007.01825}\ } (\bibinfo {year}
  {2020})}\BibitemShut {NoStop}%
\bibitem [{\citenamefont {Cardy}(1996)}]{cardy1996scaling}%
  \BibitemOpen
  \bibfield  {author} {\bibinfo {author} {\bibfnamefont {J.}~\bibnamefont
  {Cardy}},\ }\href@noop {} {\emph {\bibinfo {title} {Scaling and
  renormalization in statistical physics}}},\ Vol.~\bibinfo {volume} {5}\
  (\bibinfo  {publisher} {Cambridge university press},\ \bibinfo {year}
  {1996})\BibitemShut {NoStop}%
\bibitem [{\citenamefont {Sachdev}(2011)}]{sachdev2011quantum}%
  \BibitemOpen
  \bibfield  {author} {\bibinfo {author} {\bibfnamefont {S.}~\bibnamefont
  {Sachdev}},\ }\href@noop {} {\emph {\bibinfo {title} {Quantum phase
  transitions}}}\ (\bibinfo  {publisher} {Cambridge university press},\
  \bibinfo {year} {2011})\BibitemShut {NoStop}%
\bibitem [{\citenamefont {Sachdev}(1997)}]{sachdev1997theory}%
  \BibitemOpen
  \bibfield  {author} {\bibinfo {author} {\bibfnamefont {S.}~\bibnamefont
  {Sachdev}},\ }\bibfield  {title} {\enquote {\bibinfo {title} {Theory of
  finite-temperature crossovers near quantum critical points close to, or
  above, their upper-critical dimension},}\ }\href@noop {} {\bibfield
  {journal} {\bibinfo  {journal} {Physical Review B}\ }\textbf {\bibinfo
  {volume} {55}},\ \bibinfo {pages} {142} (\bibinfo {year} {1997})}\BibitemShut
  {NoStop}%
\bibitem [{\citenamefont {Sachdev}(1999)}]{sachdev1999universal}%
  \BibitemOpen
  \bibfield  {author} {\bibinfo {author} {\bibfnamefont {S.}~\bibnamefont
  {Sachdev}},\ }\bibfield  {title} {\enquote {\bibinfo {title} {Universal
  relaxational dynamics near two-dimensional quantum critical points},}\
  }\href@noop {} {\bibfield  {journal} {\bibinfo  {journal} {Physical Review
  B}\ }\textbf {\bibinfo {volume} {59}},\ \bibinfo {pages} {14054} (\bibinfo
  {year} {1999})}\BibitemShut {NoStop}%
\bibitem [{\citenamefont {Adam}\ \emph {et~al.}(2002)\citenamefont {Adam},
  \citenamefont {Brouwer}, \citenamefont {Sethna},\ and\ \citenamefont
  {Waintal}}]{adam2002enhanced}%
  \BibitemOpen
  \bibfield  {author} {\bibinfo {author} {\bibfnamefont {S.}~\bibnamefont
  {Adam}}, \bibinfo {author} {\bibfnamefont {P.~W.}\ \bibnamefont {Brouwer}},
  \bibinfo {author} {\bibfnamefont {J.~P.}\ \bibnamefont {Sethna}}, \ and\
  \bibinfo {author} {\bibfnamefont {X.}~\bibnamefont {Waintal}},\ }\bibfield
  {title} {\enquote {\bibinfo {title} {Enhanced mesoscopic fluctuations in the
  crossover between random-matrix ensembles},}\ }\href@noop {} {\bibfield
  {journal} {\bibinfo  {journal} {Physical Review B}\ }\textbf {\bibinfo
  {volume} {66}},\ \bibinfo {pages} {165310} (\bibinfo {year}
  {2002})}\BibitemShut {NoStop}%
\bibitem [{\citenamefont {Chen}, \citenamefont {Zapperi},\ and\ \citenamefont
  {Sethna}(2015)}]{chen2015crossover}%
  \BibitemOpen
  \bibfield  {author} {\bibinfo {author} {\bibfnamefont {Y.}~\bibnamefont
  {Chen}}, \bibinfo {author} {\bibfnamefont {S.}~\bibnamefont {Zapperi}}, \
  and\ \bibinfo {author} {\bibfnamefont {J.~P.}\ \bibnamefont {Sethna}},\
  }\bibfield  {title} {\enquote {\bibinfo {title} {Crossover behavior in
  interface depinning},}\ }\href@noop {} {\bibfield  {journal} {\bibinfo
  {journal} {Physical Review E}\ }\textbf {\bibinfo {volume} {92}},\ \bibinfo
  {pages} {022146} (\bibinfo {year} {2015})}\BibitemShut {NoStop}%
\bibitem [{\citenamefont {Guy}, \citenamefont {Hermes},\ and\ \citenamefont
  {Poon}(2015)}]{guy2015towards}%
  \BibitemOpen
  \bibfield  {author} {\bibinfo {author} {\bibfnamefont {B.}~\bibnamefont
  {Guy}}, \bibinfo {author} {\bibfnamefont {M.}~\bibnamefont {Hermes}}, \ and\
  \bibinfo {author} {\bibfnamefont {W.~C.}\ \bibnamefont {Poon}},\ }\bibfield
  {title} {\enquote {\bibinfo {title} {Towards a unified description of the
  rheology of hard-particle suspensions},}\ }\href@noop {} {\bibfield
  {journal} {\bibinfo  {journal} {Physical review letters}\ }\textbf {\bibinfo
  {volume} {115}},\ \bibinfo {pages} {088304} (\bibinfo {year}
  {2015})}\BibitemShut {NoStop}%
\bibitem [{Note2()}]{Note2}%
  \BibitemOpen
  \bibinfo {note} {We find that the scaling functions for cornstarch silica
  differ by a multiplicative factor of $\sim $2, which simply reflects a
  different solvent viscosity.}\BibitemShut {Stop}%
\bibitem [{\citenamefont {Guazzelli}\ and\ \citenamefont
  {Pouliquen}(2018)}]{guazzelli2018rheology}%
  \BibitemOpen
  \bibfield  {author} {\bibinfo {author} {\bibfnamefont {{\'E}.}~\bibnamefont
  {Guazzelli}}\ and\ \bibinfo {author} {\bibfnamefont {O.}~\bibnamefont
  {Pouliquen}},\ }\bibfield  {title} {\enquote {\bibinfo {title} {Rheology of
  dense granular suspensions},}\ }\href@noop {} {\bibfield  {journal} {\bibinfo
   {journal} {Journal of Fluid Mechanics}\ }\textbf {\bibinfo {volume} {852}}
  (\bibinfo {year} {2018})}\BibitemShut {NoStop}%
\bibitem [{\citenamefont {Morris}\ and\ \citenamefont
  {Boulay}(1999)}]{morris1999curvilinear}%
  \BibitemOpen
  \bibfield  {author} {\bibinfo {author} {\bibfnamefont {J.~F.}\ \bibnamefont
  {Morris}}\ and\ \bibinfo {author} {\bibfnamefont {F.}~\bibnamefont
  {Boulay}},\ }\bibfield  {title} {\enquote {\bibinfo {title} {Curvilinear
  flows of noncolloidal suspensions: The role of normal stresses},}\
  }\href@noop {} {\bibfield  {journal} {\bibinfo  {journal} {Journal of
  rheology}\ }\textbf {\bibinfo {volume} {43}},\ \bibinfo {pages} {1213--1237}
  (\bibinfo {year} {1999})}\BibitemShut {NoStop}%
\bibitem [{Note3()}]{Note3}%
  \BibitemOpen
  \bibinfo {note} {To construct this phase diagram we use an extended data set
  with data closer to $x_c$. See SI for more details}\BibitemShut {NoStop}%
\bibitem [{\citenamefont {Peters}, \citenamefont {Majumdar},\ and\
  \citenamefont {Jaeger}(2016)}]{peters2016direct}%
  \BibitemOpen
  \bibfield  {author} {\bibinfo {author} {\bibfnamefont {I.~R.}\ \bibnamefont
  {Peters}}, \bibinfo {author} {\bibfnamefont {S.}~\bibnamefont {Majumdar}}, \
  and\ \bibinfo {author} {\bibfnamefont {H.~M.}\ \bibnamefont {Jaeger}},\
  }\bibfield  {title} {\enquote {\bibinfo {title} {Direct observation of
  dynamic shear jamming in dense suspensions},}\ }\href@noop {} {\bibfield
  {journal} {\bibinfo  {journal} {Nature}\ }\textbf {\bibinfo {volume} {532}},\
  \bibinfo {pages} {214--217} (\bibinfo {year} {2016})}\BibitemShut {NoStop}%
\bibitem [{\citenamefont {Bi}\ \emph {et~al.}(2011)\citenamefont {Bi},
  \citenamefont {Zhang}, \citenamefont {Chakraborty},\ and\ \citenamefont
  {Behringer}}]{bi2011jamming}%
  \BibitemOpen
  \bibfield  {author} {\bibinfo {author} {\bibfnamefont {D.}~\bibnamefont
  {Bi}}, \bibinfo {author} {\bibfnamefont {J.}~\bibnamefont {Zhang}}, \bibinfo
  {author} {\bibfnamefont {B.}~\bibnamefont {Chakraborty}}, \ and\ \bibinfo
  {author} {\bibfnamefont {R.~P.}\ \bibnamefont {Behringer}},\ }\bibfield
  {title} {\enquote {\bibinfo {title} {Jamming by shear},}\ }\href@noop {}
  {\bibfield  {journal} {\bibinfo  {journal} {Nature}\ }\textbf {\bibinfo
  {volume} {480}},\ \bibinfo {pages} {355} (\bibinfo {year}
  {2011})}\BibitemShut {NoStop}%
\bibitem [{\citenamefont {Gameiro}\ \emph {et~al.}(2020)\citenamefont
  {Gameiro}, \citenamefont {Singh}, \citenamefont {Kondic}, \citenamefont
  {Mischaikow},\ and\ \citenamefont {Morris}}]{gameiro2020interaction}%
  \BibitemOpen
  \bibfield  {author} {\bibinfo {author} {\bibfnamefont {M.}~\bibnamefont
  {Gameiro}}, \bibinfo {author} {\bibfnamefont {A.}~\bibnamefont {Singh}},
  \bibinfo {author} {\bibfnamefont {L.}~\bibnamefont {Kondic}}, \bibinfo
  {author} {\bibfnamefont {K.}~\bibnamefont {Mischaikow}}, \ and\ \bibinfo
  {author} {\bibfnamefont {J.~F.}\ \bibnamefont {Morris}},\ }\bibfield  {title}
  {\enquote {\bibinfo {title} {Interaction network analysis in shear thickening
  suspensions},}\ }\href@noop {} {\bibfield  {journal} {\bibinfo  {journal}
  {Physical Review Fluids}\ }\textbf {\bibinfo {volume} {5}},\ \bibinfo {pages}
  {034307} (\bibinfo {year} {2020})}\BibitemShut {NoStop}%
\bibitem [{\citenamefont {Singh}\ \emph {et~al.}(2020)\citenamefont {Singh},
  \citenamefont {Ness}, \citenamefont {Seto}, \citenamefont {de~Pablo},\ and\
  \citenamefont {Jaeger}}]{singh2020shear}%
  \BibitemOpen
  \bibfield  {author} {\bibinfo {author} {\bibfnamefont {A.}~\bibnamefont
  {Singh}}, \bibinfo {author} {\bibfnamefont {C.}~\bibnamefont {Ness}},
  \bibinfo {author} {\bibfnamefont {R.}~\bibnamefont {Seto}}, \bibinfo {author}
  {\bibfnamefont {J.~J.}\ \bibnamefont {de~Pablo}}, \ and\ \bibinfo {author}
  {\bibfnamefont {H.~M.}\ \bibnamefont {Jaeger}},\ }\bibfield  {title}
  {\enquote {\bibinfo {title} {Shear thickening and jamming of dense
  suspensions: the “roll” of friction},}\ }\href@noop {} {\bibfield
  {journal} {\bibinfo  {journal} {Physical Review Letters}\ }\textbf {\bibinfo
  {volume} {124}},\ \bibinfo {pages} {248005} (\bibinfo {year}
  {2020})}\BibitemShut {NoStop}%
\bibitem [{\citenamefont {Baumgarten}\ and\ \citenamefont
  {Kamrin}(2019)}]{baumgarten2019general}%
  \BibitemOpen
  \bibfield  {author} {\bibinfo {author} {\bibfnamefont {A.~S.}\ \bibnamefont
  {Baumgarten}}\ and\ \bibinfo {author} {\bibfnamefont {K.}~\bibnamefont
  {Kamrin}},\ }\bibfield  {title} {\enquote {\bibinfo {title} {A general
  constitutive model for dense, fine-particle suspensions validated in many
  geometries},}\ }\href@noop {} {\bibfield  {journal} {\bibinfo  {journal}
  {Proceedings of the National Academy of Sciences}\ }\textbf {\bibinfo
  {volume} {116}},\ \bibinfo {pages} {20828--20836} (\bibinfo {year}
  {2019})}\BibitemShut {NoStop}%
\bibitem [{\citenamefont {van~der Naald}\ \emph {et~al.}(2021)\citenamefont
  {van~der Naald}, \citenamefont {Zhao}, \citenamefont {Jackson},\ and\
  \citenamefont {Jaeger}}]{van2021role}%
  \BibitemOpen
  \bibfield  {author} {\bibinfo {author} {\bibfnamefont {M.}~\bibnamefont
  {van~der Naald}}, \bibinfo {author} {\bibfnamefont {L.}~\bibnamefont {Zhao}},
  \bibinfo {author} {\bibfnamefont {G.~L.}\ \bibnamefont {Jackson}}, \ and\
  \bibinfo {author} {\bibfnamefont {H.~M.}\ \bibnamefont {Jaeger}},\ }\bibfield
   {title} {\enquote {\bibinfo {title} {The role of solvent molecular weight in
  shear thickening and shear jamming},}\ }\href@noop {} {\bibfield  {journal}
  {\bibinfo  {journal} {Soft Matter}\ }\textbf {\bibinfo {volume} {17}},\
  \bibinfo {pages} {3144--3152} (\bibinfo {year} {2021})}\BibitemShut {NoStop}%
\bibitem [{\citenamefont {Townsend}\ and\ \citenamefont
  {Wilson}(2017)}]{townsend2017frictional}%
  \BibitemOpen
  \bibfield  {author} {\bibinfo {author} {\bibfnamefont {A.~K.}\ \bibnamefont
  {Townsend}}\ and\ \bibinfo {author} {\bibfnamefont {H.~J.}\ \bibnamefont
  {Wilson}},\ }\bibfield  {title} {\enquote {\bibinfo {title} {Frictional shear
  thickening in suspensions: The effect of rigid asperities},}\ }\href@noop {}
  {\bibfield  {journal} {\bibinfo  {journal} {Physics of Fluids}\ }\textbf
  {\bibinfo {volume} {29}},\ \bibinfo {pages} {121607} (\bibinfo {year}
  {2017})}\BibitemShut {NoStop}%
\bibitem [{\citenamefont {O’Neill}, \citenamefont {Royer},\ and\
  \citenamefont {Poon}(2019)}]{o2019liquid}%
  \BibitemOpen
  \bibfield  {author} {\bibinfo {author} {\bibfnamefont {R.~E.}\ \bibnamefont
  {O’Neill}}, \bibinfo {author} {\bibfnamefont {J.~R.}\ \bibnamefont
  {Royer}}, \ and\ \bibinfo {author} {\bibfnamefont {W.~C.}\ \bibnamefont
  {Poon}},\ }\bibfield  {title} {\enquote {\bibinfo {title} {Liquid migration
  in shear thickening suspensions flowing through constrictions},}\ }\href@noop
  {} {\bibfield  {journal} {\bibinfo  {journal} {Physical review letters}\
  }\textbf {\bibinfo {volume} {123}},\ \bibinfo {pages} {128002} (\bibinfo
  {year} {2019})}\BibitemShut {NoStop}%
\bibitem [{\citenamefont {White}, \citenamefont {Chellamuthu},\ and\
  \citenamefont {Rothstein}(2010)}]{white2010extensional}%
  \BibitemOpen
  \bibfield  {author} {\bibinfo {author} {\bibfnamefont {E.~E.~B.}\
  \bibnamefont {White}}, \bibinfo {author} {\bibfnamefont {M.}~\bibnamefont
  {Chellamuthu}}, \ and\ \bibinfo {author} {\bibfnamefont {J.~P.}\ \bibnamefont
  {Rothstein}},\ }\bibfield  {title} {\enquote {\bibinfo {title} {Extensional
  rheology of a shear-thickening cornstarch and water suspension},}\
  }\href@noop {} {\bibfield  {journal} {\bibinfo  {journal} {Rheologica acta}\
  }\textbf {\bibinfo {volume} {49}},\ \bibinfo {pages} {119--129} (\bibinfo
  {year} {2010})}\BibitemShut {NoStop}%
\bibitem [{\citenamefont {Han}\ \emph {et~al.}(2018)\citenamefont {Han},
  \citenamefont {Wyart}, \citenamefont {Peters},\ and\ \citenamefont
  {Jaeger}}]{han2018shear}%
  \BibitemOpen
  \bibfield  {author} {\bibinfo {author} {\bibfnamefont {E.}~\bibnamefont
  {Han}}, \bibinfo {author} {\bibfnamefont {M.}~\bibnamefont {Wyart}}, \bibinfo
  {author} {\bibfnamefont {I.~R.}\ \bibnamefont {Peters}}, \ and\ \bibinfo
  {author} {\bibfnamefont {H.~M.}\ \bibnamefont {Jaeger}},\ }\bibfield  {title}
  {\enquote {\bibinfo {title} {Shear fronts in shear-thickening suspensions},}\
  }\href@noop {} {\bibfield  {journal} {\bibinfo  {journal} {Physical Review
  Fluids}\ }\textbf {\bibinfo {volume} {3}},\ \bibinfo {pages} {073301}
  (\bibinfo {year} {2018})}\BibitemShut {NoStop}%
\bibitem [{\citenamefont {DeGiuli}\ \emph {et~al.}(2015)\citenamefont
  {DeGiuli}, \citenamefont {D{\"u}ring}, \citenamefont {Lerner},\ and\
  \citenamefont {Wyart}}]{degiuli2015unified}%
  \BibitemOpen
  \bibfield  {author} {\bibinfo {author} {\bibfnamefont {E.}~\bibnamefont
  {DeGiuli}}, \bibinfo {author} {\bibfnamefont {G.}~\bibnamefont {D{\"u}ring}},
  \bibinfo {author} {\bibfnamefont {E.}~\bibnamefont {Lerner}}, \ and\ \bibinfo
  {author} {\bibfnamefont {M.}~\bibnamefont {Wyart}},\ }\bibfield  {title}
  {\enquote {\bibinfo {title} {Unified theory of inertial granular flows and
  non-brownian suspensions},}\ }\href@noop {} {\bibfield  {journal} {\bibinfo
  {journal} {Physical Review E}\ }\textbf {\bibinfo {volume} {91}},\ \bibinfo
  {pages} {062206} (\bibinfo {year} {2015})}\BibitemShut {NoStop}%
\bibitem [{\citenamefont {Boyer}, \citenamefont {Guazzelli},\ and\
  \citenamefont {Pouliquen}(2011)}]{boyer2011unifying}%
  \BibitemOpen
  \bibfield  {author} {\bibinfo {author} {\bibfnamefont {F.}~\bibnamefont
  {Boyer}}, \bibinfo {author} {\bibfnamefont {{\'E}.}~\bibnamefont
  {Guazzelli}}, \ and\ \bibinfo {author} {\bibfnamefont {O.}~\bibnamefont
  {Pouliquen}},\ }\bibfield  {title} {\enquote {\bibinfo {title} {Unifying
  suspension and granular rheology},}\ }\href@noop {} {\bibfield  {journal}
  {\bibinfo  {journal} {Physical review letters}\ }\textbf {\bibinfo {volume}
  {107}},\ \bibinfo {pages} {188301} (\bibinfo {year} {2011})}\BibitemShut
  {NoStop}%
\bibitem [{\citenamefont {Perrin}\ \emph {et~al.}(2019)\citenamefont {Perrin},
  \citenamefont {Clavaud}, \citenamefont {Wyart}, \citenamefont {Metzger},\
  and\ \citenamefont {Forterre}}]{perrin2019interparticle}%
  \BibitemOpen
  \bibfield  {author} {\bibinfo {author} {\bibfnamefont {H.}~\bibnamefont
  {Perrin}}, \bibinfo {author} {\bibfnamefont {C.}~\bibnamefont {Clavaud}},
  \bibinfo {author} {\bibfnamefont {M.}~\bibnamefont {Wyart}}, \bibinfo
  {author} {\bibfnamefont {B.}~\bibnamefont {Metzger}}, \ and\ \bibinfo
  {author} {\bibfnamefont {Y.}~\bibnamefont {Forterre}},\ }\bibfield  {title}
  {\enquote {\bibinfo {title} {Interparticle friction leads to nonmonotonic
  flow curves and hysteresis in viscous suspensions},}\ }\href@noop {}
  {\bibfield  {journal} {\bibinfo  {journal} {Physical Review X}\ }\textbf
  {\bibinfo {volume} {9}},\ \bibinfo {pages} {031027} (\bibinfo {year}
  {2019})}\BibitemShut {NoStop}%
\bibitem [{\citenamefont {Perrin}\ \emph {et~al.}(2021)\citenamefont {Perrin},
  \citenamefont {Wyart}, \citenamefont {Metzger},\ and\ \citenamefont
  {Forterre}}]{perrin2021nonlocal}%
  \BibitemOpen
  \bibfield  {author} {\bibinfo {author} {\bibfnamefont {H.}~\bibnamefont
  {Perrin}}, \bibinfo {author} {\bibfnamefont {M.}~\bibnamefont {Wyart}},
  \bibinfo {author} {\bibfnamefont {B.}~\bibnamefont {Metzger}}, \ and\
  \bibinfo {author} {\bibfnamefont {Y.}~\bibnamefont {Forterre}},\ }\bibfield
  {title} {\enquote {\bibinfo {title} {Nonlocal effects reflect the jamming
  criticality in frictionless granular flows down inclines},}\ }\href@noop {}
  {\bibfield  {journal} {\bibinfo  {journal} {Physical Review Letters}\
  }\textbf {\bibinfo {volume} {126}},\ \bibinfo {pages} {228002} (\bibinfo
  {year} {2021})}\BibitemShut {NoStop}%
\bibitem [{\citenamefont {Malbranche}\ \emph {et~al.}(2022)\citenamefont
  {Malbranche}, \citenamefont {Santra}, \citenamefont {Chakraborty},\ and\
  \citenamefont {Morris}}]{malbranche2022scaling}%
  \BibitemOpen
  \bibfield  {author} {\bibinfo {author} {\bibfnamefont {N.}~\bibnamefont
  {Malbranche}}, \bibinfo {author} {\bibfnamefont {A.}~\bibnamefont {Santra}},
  \bibinfo {author} {\bibfnamefont {B.}~\bibnamefont {Chakraborty}}, \ and\
  \bibinfo {author} {\bibfnamefont {J.~F.}\ \bibnamefont {Morris}},\ }\bibfield
   {title} {\enquote {\bibinfo {title} {Scaling analysis of shear thickening
  suspensions},}\ }\href@noop {} {\bibfield  {journal} {\bibinfo  {journal}
  {Frontiers in physics}\ }\textbf {\bibinfo {volume} {10}},\ \bibinfo {pages}
  {946221} (\bibinfo {year} {2022})}\BibitemShut {NoStop}%
\bibitem [{\citenamefont {Malbranche}, \citenamefont {Chakraborty},\ and\
  \citenamefont {Morris}(2023)}]{malbranche2023shear}%
  \BibitemOpen
  \bibfield  {author} {\bibinfo {author} {\bibfnamefont {N.}~\bibnamefont
  {Malbranche}}, \bibinfo {author} {\bibfnamefont {B.}~\bibnamefont
  {Chakraborty}}, \ and\ \bibinfo {author} {\bibfnamefont {J.~F.}\ \bibnamefont
  {Morris}},\ }\bibfield  {title} {\enquote {\bibinfo {title} {Shear thickening
  in dense bidisperse suspensions},}\ }\href@noop {} {\bibfield  {journal}
  {\bibinfo  {journal} {Journal of Rheology}\ }\textbf {\bibinfo {volume}
  {67}},\ \bibinfo {pages} {91--104} (\bibinfo {year} {2023})}\BibitemShut
  {NoStop}%
\bibitem [{\citenamefont {Mari}\ \emph {et~al.}(2014)\citenamefont {Mari},
  \citenamefont {Seto}, \citenamefont {Morris},\ and\ \citenamefont
  {Denn}}]{mari2014shear}%
  \BibitemOpen
  \bibfield  {author} {\bibinfo {author} {\bibfnamefont {R.}~\bibnamefont
  {Mari}}, \bibinfo {author} {\bibfnamefont {R.}~\bibnamefont {Seto}}, \bibinfo
  {author} {\bibfnamefont {J.~F.}\ \bibnamefont {Morris}}, \ and\ \bibinfo
  {author} {\bibfnamefont {M.~M.}\ \bibnamefont {Denn}},\ }\bibfield  {title}
  {\enquote {\bibinfo {title} {Shear thickening, frictionless and frictional
  rheologies in non-brownian suspensions},}\ }\href@noop {} {\bibfield
  {journal} {\bibinfo  {journal} {Journal of Rheology}\ }\textbf {\bibinfo
  {volume} {58}},\ \bibinfo {pages} {1693--1724} (\bibinfo {year}
  {2014})}\BibitemShut {NoStop}%
\bibitem [{Note4()}]{Note4}%
  \BibitemOpen
  \bibinfo {note} {Just as one defines $u_h(P,T)$ as the magnetic-field like
  perturbation away from the liquid-gas critical point~\cite
  {cardy1996scaling}. Usually, analytic corrections to scaling are treated
  perturbatively near criticality. Here we use them to extend our theory to the
  entire shear-thickening regime.}\BibitemShut {Stop}%
\bibitem [{\citenamefont {Goodrich}, \citenamefont {Liu},\ and\ \citenamefont
  {Sethna}(2016)}]{goodrich2016scaling}%
  \BibitemOpen
  \bibfield  {author} {\bibinfo {author} {\bibfnamefont {C.~P.}\ \bibnamefont
  {Goodrich}}, \bibinfo {author} {\bibfnamefont {A.~J.}\ \bibnamefont {Liu}}, \
  and\ \bibinfo {author} {\bibfnamefont {J.~P.}\ \bibnamefont {Sethna}},\
  }\bibfield  {title} {\enquote {\bibinfo {title} {Scaling ansatz for the
  jamming transition},}\ }\href@noop {} {\bibfield  {journal} {\bibinfo
  {journal} {Proceedings of the National Academy of Sciences}\ }\textbf
  {\bibinfo {volume} {113}},\ \bibinfo {pages} {9745--9750} (\bibinfo {year}
  {2016})}\BibitemShut {NoStop}%
\bibitem [{\citenamefont {O’hern}\ \emph {et~al.}(2003)\citenamefont
  {O’hern}, \citenamefont {Silbert}, \citenamefont {Liu},\ and\ \citenamefont
  {Nagel}}]{o2003jamming}%
  \BibitemOpen
  \bibfield  {author} {\bibinfo {author} {\bibfnamefont {C.~S.}\ \bibnamefont
  {O’hern}}, \bibinfo {author} {\bibfnamefont {L.~E.}\ \bibnamefont
  {Silbert}}, \bibinfo {author} {\bibfnamefont {A.~J.}\ \bibnamefont {Liu}}, \
  and\ \bibinfo {author} {\bibfnamefont {S.~R.}\ \bibnamefont {Nagel}},\
  }\bibfield  {title} {\enquote {\bibinfo {title} {Jamming at zero temperature
  and zero applied stress: The epitome of disorder},}\ }\href@noop {}
  {\bibfield  {journal} {\bibinfo  {journal} {Physical Review E}\ }\textbf
  {\bibinfo {volume} {68}},\ \bibinfo {pages} {011306} (\bibinfo {year}
  {2003})}\BibitemShut {NoStop}%
\bibitem [{\citenamefont {Goodrich}\ \emph {et~al.}(2014)\citenamefont
  {Goodrich}, \citenamefont {Dagois-Bohy}, \citenamefont {Tighe}, \citenamefont
  {Van~Hecke}, \citenamefont {Liu},\ and\ \citenamefont
  {Nagel}}]{goodrich2014jamming}%
  \BibitemOpen
  \bibfield  {author} {\bibinfo {author} {\bibfnamefont {C.~P.}\ \bibnamefont
  {Goodrich}}, \bibinfo {author} {\bibfnamefont {S.}~\bibnamefont
  {Dagois-Bohy}}, \bibinfo {author} {\bibfnamefont {B.~P.}\ \bibnamefont
  {Tighe}}, \bibinfo {author} {\bibfnamefont {M.}~\bibnamefont {Van~Hecke}},
  \bibinfo {author} {\bibfnamefont {A.~J.}\ \bibnamefont {Liu}}, \ and\
  \bibinfo {author} {\bibfnamefont {S.~R.}\ \bibnamefont {Nagel}},\ }\bibfield
  {title} {\enquote {\bibinfo {title} {Jamming in finite systems: Stability,
  anisotropy, fluctuations, and scaling},}\ }\href@noop {} {\bibfield
  {journal} {\bibinfo  {journal} {Physical Review E}\ }\textbf {\bibinfo
  {volume} {90}},\ \bibinfo {pages} {022138} (\bibinfo {year}
  {2014})}\BibitemShut {NoStop}%
\bibitem [{\citenamefont {Liarte}\ \emph {et~al.}(2019)\citenamefont {Liarte},
  \citenamefont {Mao}, \citenamefont {Stenull},\ and\ \citenamefont
  {Lubensky}}]{liarte2019jamming}%
  \BibitemOpen
  \bibfield  {author} {\bibinfo {author} {\bibfnamefont {D.~B.}\ \bibnamefont
  {Liarte}}, \bibinfo {author} {\bibfnamefont {X.}~\bibnamefont {Mao}},
  \bibinfo {author} {\bibfnamefont {O.}~\bibnamefont {Stenull}}, \ and\
  \bibinfo {author} {\bibfnamefont {T.}~\bibnamefont {Lubensky}},\ }\bibfield
  {title} {\enquote {\bibinfo {title} {Jamming as a multicritical point},}\
  }\href@noop {} {\bibfield  {journal} {\bibinfo  {journal} {Physical review
  letters}\ }\textbf {\bibinfo {volume} {122}},\ \bibinfo {pages} {128006}
  (\bibinfo {year} {2019})}\BibitemShut {NoStop}%
\bibitem [{\citenamefont {Liarte}\ \emph
  {et~al.}(2021{\natexlab{a}})\citenamefont {Liarte}, \citenamefont {Thornton},
  \citenamefont {Schwen}, \citenamefont {Cohen}, \citenamefont {Chowdhury},\
  and\ \citenamefont {Sethna}}]{liarte2021scaling}%
  \BibitemOpen
  \bibfield  {author} {\bibinfo {author} {\bibfnamefont {D.~B.}\ \bibnamefont
  {Liarte}}, \bibinfo {author} {\bibfnamefont {S.~J.}\ \bibnamefont
  {Thornton}}, \bibinfo {author} {\bibfnamefont {E.}~\bibnamefont {Schwen}},
  \bibinfo {author} {\bibfnamefont {I.}~\bibnamefont {Cohen}}, \bibinfo
  {author} {\bibfnamefont {D.}~\bibnamefont {Chowdhury}}, \ and\ \bibinfo
  {author} {\bibfnamefont {J.~P.}\ \bibnamefont {Sethna}},\ }\bibfield  {title}
  {\enquote {\bibinfo {title} {Scaling of dynamical susceptibility at the onset
  of rigidity for disordered viscoelastic matter},}\ }\href@noop {} {\bibfield
  {journal} {\bibinfo  {journal} {arXiv preprint arXiv:2103.07474}\ } (\bibinfo
  {year} {2021}{\natexlab{a}})}\BibitemShut {NoStop}%
\bibitem [{\citenamefont {Liarte}\ \emph
  {et~al.}(2021{\natexlab{b}})\citenamefont {Liarte}, \citenamefont {Thornton},
  \citenamefont {Schwen}, \citenamefont {Cohen}, \citenamefont {Chowdhury},\
  and\ \citenamefont {Sethna}}]{liarte2021universal}%
  \BibitemOpen
  \bibfield  {author} {\bibinfo {author} {\bibfnamefont {D.~B.}\ \bibnamefont
  {Liarte}}, \bibinfo {author} {\bibfnamefont {S.~J.}\ \bibnamefont
  {Thornton}}, \bibinfo {author} {\bibfnamefont {E.}~\bibnamefont {Schwen}},
  \bibinfo {author} {\bibfnamefont {I.}~\bibnamefont {Cohen}}, \bibinfo
  {author} {\bibfnamefont {D.}~\bibnamefont {Chowdhury}}, \ and\ \bibinfo
  {author} {\bibfnamefont {J.~P.}\ \bibnamefont {Sethna}},\ }\bibfield  {title}
  {\enquote {\bibinfo {title} {Universal scaling for disordered viscoelastic
  matter i: Dynamic susceptibility at the onset of rigidity},}\ }\href@noop {}
  {\bibfield  {journal} {\bibinfo  {journal} {arXiv e-prints}\ ,\ \bibinfo
  {pages} {arXiv--2103}} (\bibinfo {year} {2021}{\natexlab{b}})}\BibitemShut
  {NoStop}%
\bibitem [{\citenamefont {Aharony}\ and\ \citenamefont
  {Fisher}(1980)}]{aharony1980universality}%
  \BibitemOpen
  \bibfield  {author} {\bibinfo {author} {\bibfnamefont {A.}~\bibnamefont
  {Aharony}}\ and\ \bibinfo {author} {\bibfnamefont {M.~E.}\ \bibnamefont
  {Fisher}},\ }\bibfield  {title} {\enquote {\bibinfo {title} {Universality in
  analytic corrections to scaling for planar ising models},}\ }\href@noop {}
  {\bibfield  {journal} {\bibinfo  {journal} {Physical Review Letters}\
  }\textbf {\bibinfo {volume} {45}},\ \bibinfo {pages} {1044} (\bibinfo {year}
  {1980})}\BibitemShut {NoStop}%
\bibitem [{\citenamefont {Aharony}\ and\ \citenamefont
  {Fisher}(1983)}]{aharony1983nonlinear}%
  \BibitemOpen
  \bibfield  {author} {\bibinfo {author} {\bibfnamefont {A.}~\bibnamefont
  {Aharony}}\ and\ \bibinfo {author} {\bibfnamefont {M.~E.}\ \bibnamefont
  {Fisher}},\ }\bibfield  {title} {\enquote {\bibinfo {title} {Nonlinear
  scaling fields and corrections to scaling near criticality},}\ }\href@noop {}
  {\bibfield  {journal} {\bibinfo  {journal} {Physical Review B}\ }\textbf
  {\bibinfo {volume} {27}},\ \bibinfo {pages} {4394} (\bibinfo {year}
  {1983})}\BibitemShut {NoStop}%
\bibitem [{\citenamefont {Barnes}(1989)}]{barnes1989shear}%
  \BibitemOpen
  \bibfield  {author} {\bibinfo {author} {\bibfnamefont {H.}~\bibnamefont
  {Barnes}},\ }\bibfield  {title} {\enquote {\bibinfo {title} {Shear-thickening
  (“dilatancy”) in suspensions of nonaggregating solid particles dispersed
  in newtonian liquids},}\ }\href@noop {} {\bibfield  {journal} {\bibinfo
  {journal} {Journal of Rheology}\ }\textbf {\bibinfo {volume} {33}},\ \bibinfo
  {pages} {329--366} (\bibinfo {year} {1989})}\BibitemShut {NoStop}%
\bibitem [{\citenamefont {Brady}(1993)}]{brady1993rheological}%
  \BibitemOpen
  \bibfield  {author} {\bibinfo {author} {\bibfnamefont {J.~F.}\ \bibnamefont
  {Brady}},\ }\bibfield  {title} {\enquote {\bibinfo {title} {The rheological
  behavior of concentrated colloidal dispersions},}\ }\href@noop {} {\bibfield
  {journal} {\bibinfo  {journal} {The Journal of Chemical Physics}\ }\textbf
  {\bibinfo {volume} {99}},\ \bibinfo {pages} {567--581} (\bibinfo {year}
  {1993})}\BibitemShut {NoStop}%
\bibitem [{\citenamefont {Brady}\ and\ \citenamefont
  {Bossis}(1985)}]{brady1985rheology}%
  \BibitemOpen
  \bibfield  {author} {\bibinfo {author} {\bibfnamefont {J.~F.}\ \bibnamefont
  {Brady}}\ and\ \bibinfo {author} {\bibfnamefont {G.}~\bibnamefont {Bossis}},\
  }\bibfield  {title} {\enquote {\bibinfo {title} {The rheology of concentrated
  suspensions of spheres in simple shear flow by numerical simulation},}\
  }\href@noop {} {\bibfield  {journal} {\bibinfo  {journal} {Journal of Fluid
  Mechanics}\ }\textbf {\bibinfo {volume} {155}},\ \bibinfo {pages} {105--129}
  (\bibinfo {year} {1985})}\BibitemShut {NoStop}%
\bibitem [{\citenamefont {Cwalina}\ and\ \citenamefont
  {Wagner}(2014)}]{cwalina2014material}%
  \BibitemOpen
  \bibfield  {author} {\bibinfo {author} {\bibfnamefont {C.~D.}\ \bibnamefont
  {Cwalina}}\ and\ \bibinfo {author} {\bibfnamefont {N.~J.}\ \bibnamefont
  {Wagner}},\ }\bibfield  {title} {\enquote {\bibinfo {title} {Material
  properties of the shear-thickened state in concentrated near hard-sphere
  colloidal dispersions},}\ }\href@noop {} {\bibfield  {journal} {\bibinfo
  {journal} {Journal of Rheology}\ }\textbf {\bibinfo {volume} {58}},\ \bibinfo
  {pages} {949--967} (\bibinfo {year} {2014})}\BibitemShut {NoStop}%
\bibitem [{\citenamefont {Dong}\ and\ \citenamefont
  {Trulsson}(2020)}]{dong2020unifying}%
  \BibitemOpen
  \bibfield  {author} {\bibinfo {author} {\bibfnamefont {J.}~\bibnamefont
  {Dong}}\ and\ \bibinfo {author} {\bibfnamefont {M.}~\bibnamefont
  {Trulsson}},\ }\bibfield  {title} {\enquote {\bibinfo {title} {Unifying
  viscous and inertial regimes of discontinuous shear thickening
  suspensions},}\ }\href@noop {} {\bibfield  {journal} {\bibinfo  {journal}
  {Journal of Rheology}\ }\textbf {\bibinfo {volume} {64}},\ \bibinfo {pages}
  {255--266} (\bibinfo {year} {2020})}\BibitemShut {NoStop}%
\bibitem [{\citenamefont {Dogic}\ \emph {et~al.}(2000)\citenamefont {Dogic},
  \citenamefont {Philipse}, \citenamefont {Fraden},\ and\ \citenamefont
  {Dhont}}]{dogic2000concentration}%
  \BibitemOpen
  \bibfield  {author} {\bibinfo {author} {\bibfnamefont {Z.}~\bibnamefont
  {Dogic}}, \bibinfo {author} {\bibfnamefont {A.}~\bibnamefont {Philipse}},
  \bibinfo {author} {\bibfnamefont {S.}~\bibnamefont {Fraden}}, \ and\ \bibinfo
  {author} {\bibfnamefont {J.}~\bibnamefont {Dhont}},\ }\bibfield  {title}
  {\enquote {\bibinfo {title} {Concentration-dependent sedimentation of
  colloidal rods},}\ }\href@noop {} {\bibfield  {journal} {\bibinfo  {journal}
  {The Journal of Chemical Physics}\ }\textbf {\bibinfo {volume} {113}},\
  \bibinfo {pages} {8368--8380} (\bibinfo {year} {2000})}\BibitemShut {NoStop}%
\bibitem [{\citenamefont {Chacko}\ \emph {et~al.}(2018)\citenamefont {Chacko},
  \citenamefont {Mari}, \citenamefont {Fielding},\ and\ \citenamefont
  {Cates}}]{chacko2018shear}%
  \BibitemOpen
  \bibfield  {author} {\bibinfo {author} {\bibfnamefont {R.~N.}\ \bibnamefont
  {Chacko}}, \bibinfo {author} {\bibfnamefont {R.}~\bibnamefont {Mari}},
  \bibinfo {author} {\bibfnamefont {S.~M.}\ \bibnamefont {Fielding}}, \ and\
  \bibinfo {author} {\bibfnamefont {M.~E.}\ \bibnamefont {Cates}},\ }\bibfield
  {title} {\enquote {\bibinfo {title} {Shear reversal in dense suspensions: The
  challenge to fabric evolution models from simulation data},}\ }\href@noop {}
  {\bibfield  {journal} {\bibinfo  {journal} {Journal of Fluid Mechanics}\
  }\textbf {\bibinfo {volume} {847}},\ \bibinfo {pages} {700--734} (\bibinfo
  {year} {2018})}\BibitemShut {NoStop}%
\bibitem [{\citenamefont {Isa}, \citenamefont {Besseling},\ and\ \citenamefont
  {Poon}(2007)}]{isa2007shear}%
  \BibitemOpen
  \bibfield  {author} {\bibinfo {author} {\bibfnamefont {L.}~\bibnamefont
  {Isa}}, \bibinfo {author} {\bibfnamefont {R.}~\bibnamefont {Besseling}}, \
  and\ \bibinfo {author} {\bibfnamefont {W.~C.}\ \bibnamefont {Poon}},\
  }\bibfield  {title} {\enquote {\bibinfo {title} {Shear zones and wall slip in
  the capillary flow of concentrated colloidal suspensions},}\ }\href@noop {}
  {\bibfield  {journal} {\bibinfo  {journal} {Physical Review Letters}\
  }\textbf {\bibinfo {volume} {98}},\ \bibinfo {pages} {198305} (\bibinfo
  {year} {2007})}\BibitemShut {NoStop}%
\bibitem [{\citenamefont {Lee}\ and\ \citenamefont
  {Wagner}(2006)}]{lee2006rheological}%
  \BibitemOpen
  \bibfield  {author} {\bibinfo {author} {\bibfnamefont {Y.~S.}\ \bibnamefont
  {Lee}}\ and\ \bibinfo {author} {\bibfnamefont {N.~J.}\ \bibnamefont
  {Wagner}},\ }\bibfield  {title} {\enquote {\bibinfo {title} {Rheological
  properties and small-angle neutron scattering of a shear thickening,
  nanoparticle dispersion at high shear rates},}\ }\href@noop {} {\bibfield
  {journal} {\bibinfo  {journal} {Industrial \& engineering chemistry
  research}\ }\textbf {\bibinfo {volume} {45}},\ \bibinfo {pages} {7015--7024}
  (\bibinfo {year} {2006})}\BibitemShut {NoStop}%
\bibitem [{\citenamefont {Liu}, \citenamefont {Henkes},\ and\ \citenamefont
  {Schwarz}(2019)}]{liu2019frictional}%
  \BibitemOpen
  \bibfield  {author} {\bibinfo {author} {\bibfnamefont {K.}~\bibnamefont
  {Liu}}, \bibinfo {author} {\bibfnamefont {S.}~\bibnamefont {Henkes}}, \ and\
  \bibinfo {author} {\bibfnamefont {J.}~\bibnamefont {Schwarz}},\ }\bibfield
  {title} {\enquote {\bibinfo {title} {Frictional rigidity percolation: A new
  universality class and its superuniversal connections through minimal
  rigidity proliferation},}\ }\href@noop {} {\bibfield  {journal} {\bibinfo
  {journal} {Physical Review X}\ }\textbf {\bibinfo {volume} {9}},\ \bibinfo
  {pages} {021006} (\bibinfo {year} {2019})}\BibitemShut {NoStop}%
\bibitem [{\citenamefont {Liu}\ \emph {et~al.}(2021)\citenamefont {Liu},
  \citenamefont {Kollmer}, \citenamefont {Daniels}, \citenamefont {Schwarz},\
  and\ \citenamefont {Henkes}}]{liu2021spongelike}%
  \BibitemOpen
  \bibfield  {author} {\bibinfo {author} {\bibfnamefont {K.}~\bibnamefont
  {Liu}}, \bibinfo {author} {\bibfnamefont {J.~E.}\ \bibnamefont {Kollmer}},
  \bibinfo {author} {\bibfnamefont {K.~E.}\ \bibnamefont {Daniels}}, \bibinfo
  {author} {\bibfnamefont {J.}~\bibnamefont {Schwarz}}, \ and\ \bibinfo
  {author} {\bibfnamefont {S.}~\bibnamefont {Henkes}},\ }\bibfield  {title}
  {\enquote {\bibinfo {title} {Spongelike rigid structures in frictional
  granular packings},}\ }\href@noop {} {\bibfield  {journal} {\bibinfo
  {journal} {Physical Review Letters}\ }\textbf {\bibinfo {volume} {126}},\
  \bibinfo {pages} {088002} (\bibinfo {year} {2021})}\BibitemShut {NoStop}%
\bibitem [{\citenamefont {Olsson}\ and\ \citenamefont
  {Teitel}(2020)}]{olsson2020dynamic}%
  \BibitemOpen
  \bibfield  {author} {\bibinfo {author} {\bibfnamefont {P.}~\bibnamefont
  {Olsson}}\ and\ \bibinfo {author} {\bibfnamefont {S.}~\bibnamefont
  {Teitel}},\ }\bibfield  {title} {\enquote {\bibinfo {title} {Dynamic length
  scales in athermal, shear-driven jamming of frictionless disks in two
  dimensions},}\ }\href@noop {} {\bibfield  {journal} {\bibinfo  {journal}
  {Physical Review E}\ }\textbf {\bibinfo {volume} {102}},\ \bibinfo {pages}
  {042906} (\bibinfo {year} {2020})}\BibitemShut {NoStop}%
\bibitem [{\citenamefont {Olsson}\ and\ \citenamefont
  {Teitel}(2011)}]{olsson2011critical}%
  \BibitemOpen
  \bibfield  {author} {\bibinfo {author} {\bibfnamefont {P.}~\bibnamefont
  {Olsson}}\ and\ \bibinfo {author} {\bibfnamefont {S.}~\bibnamefont
  {Teitel}},\ }\bibfield  {title} {\enquote {\bibinfo {title} {Critical scaling
  of shearing rheology at the jamming transition of soft-core frictionless
  disks},}\ }\href@noop {} {\bibfield  {journal} {\bibinfo  {journal} {Physical
  Review E}\ }\textbf {\bibinfo {volume} {83}},\ \bibinfo {pages} {030302}
  (\bibinfo {year} {2011})}\BibitemShut {NoStop}%
\bibitem [{\citenamefont {Royall}\ \emph {et~al.}(2007)\citenamefont {Royall},
  \citenamefont {Dzubiella}, \citenamefont {Schmidt},\ and\ \citenamefont {van
  Blaaderen}}]{royall2007nonequilibrium}%
  \BibitemOpen
  \bibfield  {author} {\bibinfo {author} {\bibfnamefont {C.~P.}\ \bibnamefont
  {Royall}}, \bibinfo {author} {\bibfnamefont {J.}~\bibnamefont {Dzubiella}},
  \bibinfo {author} {\bibfnamefont {M.}~\bibnamefont {Schmidt}}, \ and\
  \bibinfo {author} {\bibfnamefont {A.}~\bibnamefont {van Blaaderen}},\
  }\bibfield  {title} {\enquote {\bibinfo {title} {Nonequilibrium sedimentation
  of colloids on the particle scale},}\ }\href@noop {} {\bibfield  {journal}
  {\bibinfo  {journal} {Physical review letters}\ }\textbf {\bibinfo {volume}
  {98}},\ \bibinfo {pages} {188304} (\bibinfo {year} {2007})}\BibitemShut
  {NoStop}%
\bibitem [{\citenamefont {Thies-Weesie}\ \emph {et~al.}(1995)\citenamefont
  {Thies-Weesie}, \citenamefont {Philipse}, \citenamefont {N{\"a}gele},
  \citenamefont {Mandl},\ and\ \citenamefont {Klein}}]{thies1995nonanalytical}%
  \BibitemOpen
  \bibfield  {author} {\bibinfo {author} {\bibfnamefont {D.~M.}\ \bibnamefont
  {Thies-Weesie}}, \bibinfo {author} {\bibfnamefont {A.~P.}\ \bibnamefont
  {Philipse}}, \bibinfo {author} {\bibfnamefont {G.}~\bibnamefont
  {N{\"a}gele}}, \bibinfo {author} {\bibfnamefont {B.}~\bibnamefont {Mandl}}, \
  and\ \bibinfo {author} {\bibfnamefont {R.}~\bibnamefont {Klein}},\ }\bibfield
   {title} {\enquote {\bibinfo {title} {Nonanalytical concentration dependence
  of sedimentation of charged silica spheres in an organic solvent: experiments
  and calculations},}\ }\href@noop {} {\bibfield  {journal} {\bibinfo
  {journal} {Journal of colloid and interface science}\ }\textbf {\bibinfo
  {volume} {176}},\ \bibinfo {pages} {43--54} (\bibinfo {year}
  {1995})}\BibitemShut {NoStop}%
\end{thebibliography}%


\begin{thebibliography}{1}
\expandafter\ifx\csname url\endcsname\relax
  \def\url#1{\texttt{#1}}\fi
\expandafter\ifx\csname urlprefix\endcsname\relax\def\urlprefix{URL }\fi
\providecommand{\bibinfo}[2]{#2}
\providecommand{\eprint}[2][]{\url{#2}}

\bibitem{aharony1980universality}
\bibinfo{author}{Aharony, A.} \& \bibinfo{author}{Fisher, M.~E.}
\newblock \bibinfo{title}{Universality in analytic corrections to scaling for
  planar ising models}.
\newblock \emph{\bibinfo{journal}{Physical Review Letters}}
  \textbf{\bibinfo{volume}{45}}, \bibinfo{pages}{1044} (\bibinfo{year}{1980}).

\bibitem{aharony1983nonlinear}
\bibinfo{author}{Aharony, A.} \& \bibinfo{author}{Fisher, M.~E.}
\newblock \bibinfo{title}{Nonlinear scaling fields and corrections to scaling
  near criticality}.
\newblock \emph{\bibinfo{journal}{Physical Review B}}
  \textbf{\bibinfo{volume}{27}}, \bibinfo{pages}{4394} (\bibinfo{year}{1983}).

\bibitem{brady1993rheological}
\bibinfo{author}{Brady, J.~F.}
\newblock \bibinfo{title}{The rheological behavior of concentrated colloidal
  dispersions}.
\newblock \emph{\bibinfo{journal}{The Journal of Chemical Physics}}
  \textbf{\bibinfo{volume}{99}}, \bibinfo{pages}{567--581}
  (\bibinfo{year}{1993}).

\bibitem{cardy1996scaling}
\bibinfo{author}{Cardy, J.}
\newblock \emph{\bibinfo{title}{Scaling and renormalization in statistical
  physics}}, vol.~\bibinfo{volume}{5} (\bibinfo{publisher}{Cambridge university
  press}, \bibinfo{year}{1996}).

\bibitem{goodrich2016scaling}
\bibinfo{author}{Goodrich, C.~P.}, \bibinfo{author}{Liu, A.~J.} \&
  \bibinfo{author}{Sethna, J.~P.}
\newblock \bibinfo{title}{Scaling ansatz for the jamming transition}.
\newblock \emph{\bibinfo{journal}{Proceedings of the National Academy of
  Sciences}} \textbf{\bibinfo{volume}{113}}, \bibinfo{pages}{9745--9750}
  (\bibinfo{year}{2016}).

\bibitem{o2003jamming}
\bibinfo{author}{O’hern, C.~S.}, \bibinfo{author}{Silbert, L.~E.},
  \bibinfo{author}{Liu, A.~J.} \& \bibinfo{author}{Nagel, S.~R.}
\newblock \bibinfo{title}{Jamming at zero temperature and zero applied stress:
  The epitome of disorder}.
\newblock \emph{\bibinfo{journal}{Physical Review E}}
  \textbf{\bibinfo{volume}{68}}, \bibinfo{pages}{011306}
  (\bibinfo{year}{2003}).

\bibitem{goodrich2014jamming}
\bibinfo{author}{Goodrich, C.~P.} \emph{et~al.}
\newblock \bibinfo{title}{Jamming in finite systems: Stability, anisotropy,
  fluctuations, and scaling}.
\newblock \emph{\bibinfo{journal}{Physical Review E}}
  \textbf{\bibinfo{volume}{90}}, \bibinfo{pages}{022138}
  (\bibinfo{year}{2014}).

\bibitem{liarte2019jamming}
\bibinfo{author}{Liarte, D.~B.}, \bibinfo{author}{Mao, X.},
  \bibinfo{author}{Stenull, O.} \& \bibinfo{author}{Lubensky, T.}
\newblock \bibinfo{title}{Jamming as a multicritical point}.
\newblock \emph{\bibinfo{journal}{Physical review letters}}
  \textbf{\bibinfo{volume}{122}}, \bibinfo{pages}{128006}
  (\bibinfo{year}{2019}).

\bibitem{liarte2021scaling}
\bibinfo{author}{Liarte, D.~B.} \emph{et~al.}
\newblock \bibinfo{title}{Scaling of dynamical susceptibility at the onset of
  rigidity for disordered viscoelastic matter}.
\newblock \emph{\bibinfo{journal}{arXiv preprint arXiv:2103.07474}}
  (\bibinfo{year}{2021}).

\end{thebibliography}

\end{document}


\preprint{APS/123-QED}

\title{Universal scaling of shear thickening transitions}
\renewcommand{\theequation}{S\arabic{equation}}
\renewcommand{\thefigure}{S\arabic{figure}}
\renewcommand{\thesection}{S\arabic{section}}

\begin{center}
   \large{\textbf{Universal scaling of shear thickening transitions}}
    
    \textbf{Supplementary Materials}
\end{center}

\section{\label{sec:Phi0Calc} Estimation of $\phi_0$}
\begin{figure}[b]
 \begin{center}
\includegraphics[width=0.8\linewidth]{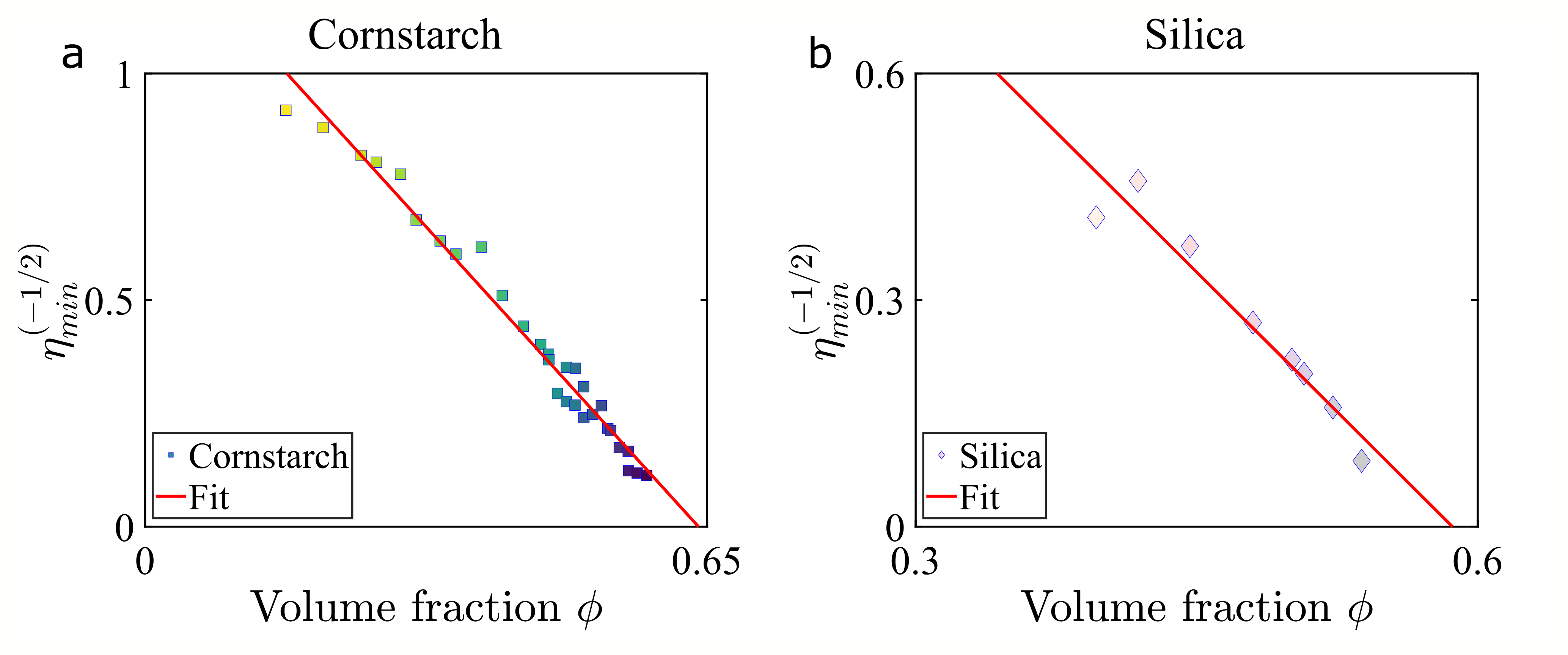}
\caption{\textbf{Estimation of $\phi_0$.} Plot of $\eta_{min}^{-1/2}$ as a function of the volume fraction $\phi$, for \textbf{a} cornstarch and \textbf{b} silica, where $\eta_{min}$ is the viscosity prior to shear thickening. The red line is a linear fit to the data and the x intercept is the estimated value of $\phi_0$.}
\label{fig:Phi0Calc}
\end{center}
\end{figure}

We use the low stress viscosity across all volume fractions to estimate the value of $\phi_0$. It has been shown previously that the low stress viscosity prior to shear thickening diverges at $\phi_0$ as $\eta \sim 1/(\phi_0 - \phi)^2$. Thus, we plot $1/\sqrt{\eta_{min}}$ as a function of $\phi$, and use the $x$ intercept of the graph as the estimate of $\phi_0$ (Fig.~\ref{fig:Phi0Calc}). As seen in Fig.1C, this estimate of $\phi_0$ works well to collapse the data at small $x$.

\section{Scaling collapse of the high volume fraction data}
 \begin{figure}
\includegraphics[width=1\textwidth]{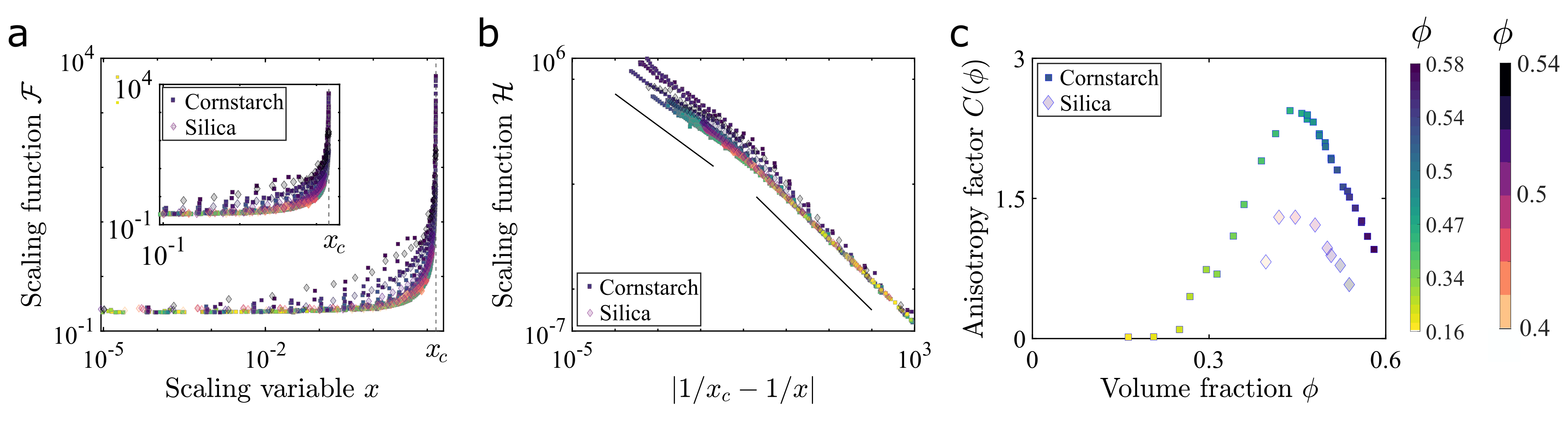}
\caption{\textbf{Scaling collapse at high volume fractions} \textbf{a.} The scaling function $\mathcal{F} = \eta(\phi_0 - \phi)^2$ as a function of the scaling variable $x = e^{-\sigma^{*}/\sigma}C(\phi)/(\phi_0 - \phi)$ for all the cornstarch (squares) and silica (diamond) suspensions data. We find that the data largely collapse onto a single curve that diverges at $x = x_c$. The higher volume fractions show small deviations at intermediate values of $x$. Inset shows a zoom-in of the data. \textbf{b}. The scaling function $\mathcal{H} = \eta(g(\sigma, \phi))^2$ versus $|1/x_c - 1/x|$ for all the cornstarch and silica suspensions data. This way of scaling the data clearly illustrates two distinct regimes - a regime characterized by a power law of -2 at small $x$ and a power law of $\sim -3/2$ at $x \approx x_c$. The solid black lines indicate power laws of $-2$ and $-1.5$ respectively. \textbf{c}. The anisotropy factor $C(\phi)$ as a function of the volume fraction for both silica and cornstarch.}
\label{fig:HighPhiCollapse}
\end{figure}
At higher volume fractions, the measured flow curves are noisier, particularly at the intermediate stresses. This deviation is likely because of the varying strain rate across the parallel plate geometry coupled with the discontinuous nature of the transition. Nonetheless, we attempt a collapse of the data as shown in Fig~\ref{fig:HighPhiCollapse}. We observe excellent collapse at low and high stresses with some deviations at intermediate values of the scaling variable. 

\section{\label{x_cx} Determining the scaling exponent}
 \begin{figure}
 \begin{center}
\includegraphics[width = \textwidth]{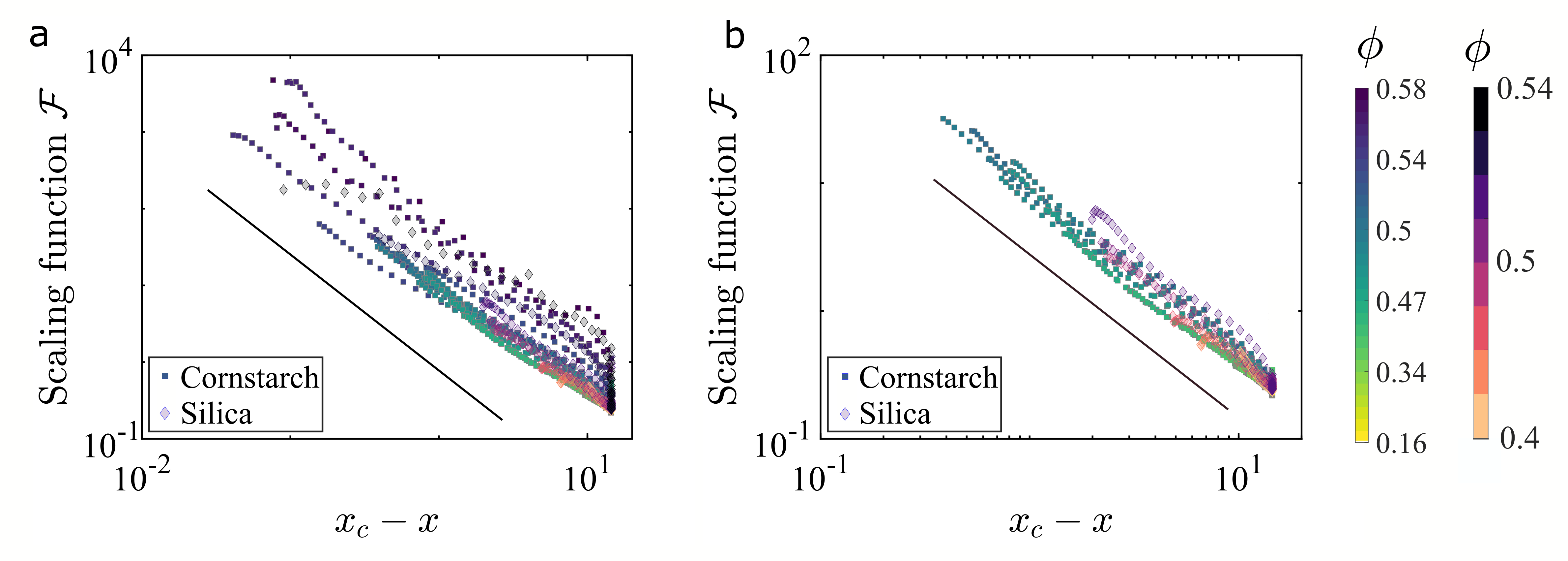}
\caption{\textbf{Power law scaling of cornstarch and silica viscosity data.} \textbf{a} The scaling function $\mathcal{F} =  \eta(\phi_0 - \phi)^2$ as a function of the scaling variable $x = e^{-\sigma^{*}/\sigma}C(\phi)/(\phi_0 - \phi)$ for the cornstarch (squares) and silica (diamond) suspensions data. We find that the data shows power law scaling with an exponent of $\sim -3/2$. The black solid line is a line with power law -1.5. \textbf{b} Only the low volume fraction data -- the scaling function $\mathcal{F} = \eta(\phi_0 - \phi)^2$ as a function of the scaling variable $x = e^{-\sigma^{*}/\sigma}C(\phi)/(\phi_0 - \phi)$ for continuously shear thickening cornstarch (squares) and silica (diamond) suspensions data.The black solid line is a line with power law -1.5.}
\label{fig:xc_xPhi0}
\end{center}
\end{figure}

From the scaling collapse of the data shown in Fig. 1C, the scaling function $\mathcal{F}$ appears to diverge at $x_c$: 
\begin{equation}
    \mathcal{F} = (x_c - x)^{-\delta}
\end{equation}
To determine $\delta$, we plot $\eta(\phi_0 - \phi)^2$ as a function of $x_c - x$. We find power law scaling with an exponent $\delta \sim 3/2$. 

\section{\label{SJLime} Shear Jamming Line Calculation}
The shear jamming line is given by the divergence of the scaling function $\mathcal{F}$ at $x = x_c$. Using 
\begin{equation}
    x = \frac{e^{-\sigma^*/\sigma} C(\phi)}{(\phi_0 - \phi)} = x_c,
\end{equation}
and a linear fit to the high volume fraction $C(\phi) = c_1 \phi + c_2$, we can solve for the shear jamming line:
\begin{equation}
    \phi = \frac{x_c \phi_0 - c_2 e^{-\sigma^*/\sigma}}{c_1 e^{-\sigma^*/\sigma} + x_c}
\end{equation}
For both the cornstarch and silica data, $x_c = 14.5$. The values of $c_1$ and $c_2$ are suspension dependent and the exact values used to generate the phase diagram are listed in Table 1.

\section{\label{DSTLime} DST Line Calculation}
\begin{table}
\begin{center}
\begin{tabular}{ |c|c| } 
 \hline
 \textbf{Parameters} & \textbf{Values} \\ 
 \hline
 Cornstarch $C(\phi)$ & -12.47$\phi$ + 8.23\\ 
 \hline
 Silica $C(\phi)$ & -10.5$\phi$ + 6.2\\ 
 \hline
 Cornstarch $\sigma^*$ & 3.9\\ 
 \hline
 Silica $\sigma^*$ & 48.2\\ 
 \hline
  Cornstarch $\phi_0$ & 0.65\\ 
 \hline
 Silica $\phi_0$ & 0.58\\ 
 \hline
 Fit $\mathcal{H}$ - $a$ & -1.5 $\pm$ 0.3 \\
 \hline
 Fit $\mathcal{H}$ - $b$ & -1 $\pm$ 1 \\
 \hline
 Fit $\mathcal{H}$ - $c$ & -2.00 $\pm$ 0.04\\
 \hline
 Fit $\mathcal{H}$ - $d$ & -1.3 $\pm$ 0.3\\
 \hline
 Fit $\mathcal{H}$ - $e$ & 0.3 $\pm$ 0.2\\
 \hline
 Fit $\mathcal{H}$ - $f$ & -3 $\pm$ 2\\
 \hline
 \end{tabular}
\label{table:ParametersPhi0}
\end{center}
\caption{Parameters to determine the shear jamming and DST lines for the phase diagram with only the $C(\phi)$ modification.}
\end{table}
The DST line is calculated as 
\begin{equation}
    \frac{d \log \eta}{d \log \sigma} = 1
\end{equation}
or, 
\begin{equation}
    \frac{\sigma}{\eta}\frac{d\eta}{d\sigma} = 1
\end{equation}
To calculate the DST line, we thus need the full functional form of the  scaling function. From Fig. 1D in the main text, we know that
\begin{equation}
    \eta (C(\phi)f)^2 = \mathcal{H}(1/x - 1/x_c)
\end{equation}
or 
\begin{equation}
\label{eq:CardyScalingManipulated}
    \eta = \frac{1}{(C(\phi)f)^{2}} \mathcal{H}(1/x - 1/x_c)
\end{equation}
where $x = fC(\phi)/(\phi_0 - \phi)$, and $f = e^{-\sigma^*/\sigma}$
\begin{equation}
    \frac{\sigma}{\eta}\frac{d\eta}{d\sigma} = \frac{\sigma}{\eta}\frac{1}{(C(\phi))^2}\left( \frac{-2}{f^3}\mathcal{H}\frac{df}{d\sigma} + \frac{1}{f^2}\frac{d\mathcal{H}}{d\sigma}\right) = 1
\end{equation}
\begin{equation}
    \frac{\sigma}{\eta}\frac{1}{(C(\phi))^2}\left( \frac{-2}{f^3}\mathcal{H}e^{-\sigma^*/\sigma}\frac{\sigma^*}{\sigma^2} + \frac{1}{f^2}\frac{d\mathcal{H}\left(\frac{1}{x} - \frac{1}{x_c} \right)}{d\left(\frac{1}{x}- \frac{1}{x_c}\right)}\frac{d \left(\frac{1}{x} - \frac{1}{x_c}\right)}{d \sigma}\right) = 1
\end{equation}
\begin{equation}
    \frac{\sigma}{\eta}\frac{1}{(C(\phi))^2}\left( \frac{-2}{f^2}\mathcal{H}\frac{\sigma^*}{\sigma^2} + \frac{1}{f^2}\frac{d\mathcal{H}\left(\frac{1}{x} - \frac{1}{x_c} \right)}{d\left(\frac{1}{x}- \frac{1}{x_c}\right)}\frac{-1}{x^2}\frac{d x}{d \sigma}\right) = 1
\end{equation}
\begin{equation}
    \frac{\sigma}{\eta}\frac{1}{(C(\phi))^2}\left( \frac{-2}{f^2}\mathcal{H}\frac{\sigma^*}{\sigma^2} + \frac{1}{f^2}\frac{d\mathcal{H}\left(t\right)}{dt}\left(\frac{-1}{x^2}\right)\frac{e^{-\sigma^*/\sigma}C(\phi)}{\phi_0 - \phi}\left(\frac{\sigma^*}{\sigma^2}\right)\right) = 1
\end{equation}
where $t = 1/x - 1/x_c$. Further simplifying, 
\begin{equation}
    \frac{1}{\eta}\frac{1}{(C(\phi))^2}\frac{\sigma^*}{\sigma}\frac{1}{f^2}\left( -2\mathcal{H} + \frac{d\mathcal{H}\left(t\right)}{dt}\left(\frac{-1}{x^2}\right)x\right) = 1
\end{equation}
We then substitute Eq.~\ref{eq:CardyScalingManipulated} above, we get
\begin{equation}
    \frac{(C(\phi)f)^2}{\mathcal{H}}\frac{1}{(C(\phi))^2}\frac{\sigma^*}{\sigma}\frac{1}{f^2}\left( -2\mathcal{H} + \frac{d\mathcal{H}\left(t\right)}{dt}\left(\frac{-1}{x}\right)\right) = 1
\end{equation}
\begin{equation}
    \left( -2\frac{\sigma^*}{\sigma} - \frac{1}{t}\frac{d\log\mathcal{H}\left( t\right)}{d \log t}\left(\frac{1}{x}\right)\left(\frac{\sigma^*}{\sigma}\right)\right) = 1
\end{equation}
Simplifying:
\begin{equation}
\label{eq:DSTCond}
    \left( -2\frac{\sigma^*}{\sigma} - \left(\frac{x_c}{x_c - x}\right)\left(\frac{\sigma^*}{\sigma}\right)\frac{d\log\mathcal{H}\left( t\right)}{d \log t}\right) = 1
\end{equation}
\newline
We fit the function in Fig. 1D on a log scale to a crossover between two straight lines. 
\begin{equation}
\label{eq:HFit}
    \log \mathcal{H}(y) = ay + b + \left( \frac{1 + \tanh(e(y - f))}{2}\right)\left( (c-a)y + d -b\right)
\end{equation}
where $y = \log(1/x - 1/x_c)$, and $a$, $b$, $c$, $d$, $e$ and $f$ are fitting parameters. Using this expression for $\mathcal{H}$ it is very easy to take the derivative. 
\begin{equation}
    \label{eq:DH}
    \frac{d\log \mathcal{H}}{dy} = a + ((c - a)y + d - b)\frac{1}{2\cosh^2(e(y - f))}e + \left(\frac{1 +  \tanh(e(y - f))}{2}\right)(c - a)
\end{equation}
From Eq.~\ref{eq:DH} and Eq.~\ref{eq:DSTCond}, and the fits for $a$, $b$, $c$, $d$, $e$ and $f$, we can solve for the DST line. Finally, we note that these parameters are required to obtain a good analytical fit to the scaling function $\mathcal{H}$ so that we can easily take derivatives. One could have generated a similar phase diagram by simply taking numerical derivatives of the data in Fig. 1D without any fitting parameters.        

More specifically, the exact parameters used to solve for the shear jamming and the DST line in silica and cornstarch respectively are shown in Table 1. 

\section{Effect of the fraction of frictional contacts}

\begin{figure}[t]
\begin{center}
\includegraphics[width=0.8\linewidth]{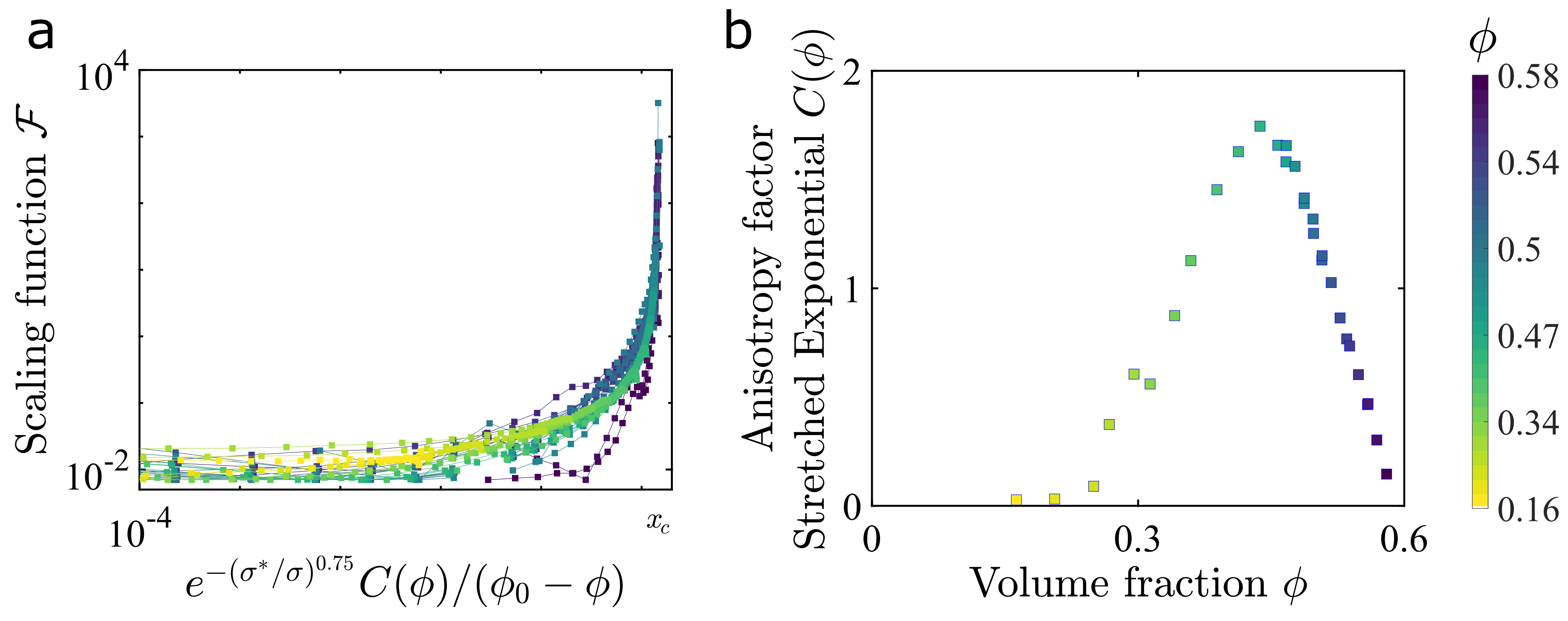}
\caption{\label{fig:Varyingf} \textbf{Scaling collapse and $C(\phi)$ for the stretched exponential form of $f$}. \textbf{a} The scaling collapse for the cornstarch data assuming a fraction of frictional contacts  $f(\sigma) = e^{-(\sigma^*/\sigma)^{3/4}}$. The scaling function $\mathcal{F} = (\eta - \beta_\eta(\phi))(\phi^*_0 - \phi)^2$. \textbf{b} The $C(\phi)$ corresponding to the scaling collapse in \textbf{c}. }
\end{center}
\end{figure}
The fraction of frictional contacts in the systems, $f(\sigma)$, depicts the increase in the frictional contacts with increasing stress. At small stresses, $f \rightarrow 0$, and large stress $f \rightarrow 1$. The onset stress for the formation of frictional contacts is $\sigma^*$. Several studies attempting to fit experimental and simulation data to the Wyart and Cates model use various functional forms for $f(\sigma)$ including $f(\sigma) = 1 - e^{-\sigma/\sigma^*}$, $f(\sigma) =  e^{-\sigma^*/\sigma}$  and  $f(\sigma) = e^{-(\sigma^*/\sigma)^\beta}$. As a comparison, we attempt the scaling collapse using the stretched exponential form for $f(\sigma)$,  $f(\sigma) = e^{-(\sigma^*/\sigma)^{3/4}}$ and the results are shown in Fig~\ref{fig:Varyingf}. We note that while quantitatively, the results vary, with slightly different values of $\phi_0$, $\sigma^*$, and $C(\phi)$, qualitatively the results remain the same.

\section{Adding additional corrections to scaling to improve the fit}

\begin{figure*}[t]
\includegraphics[width=\linewidth]{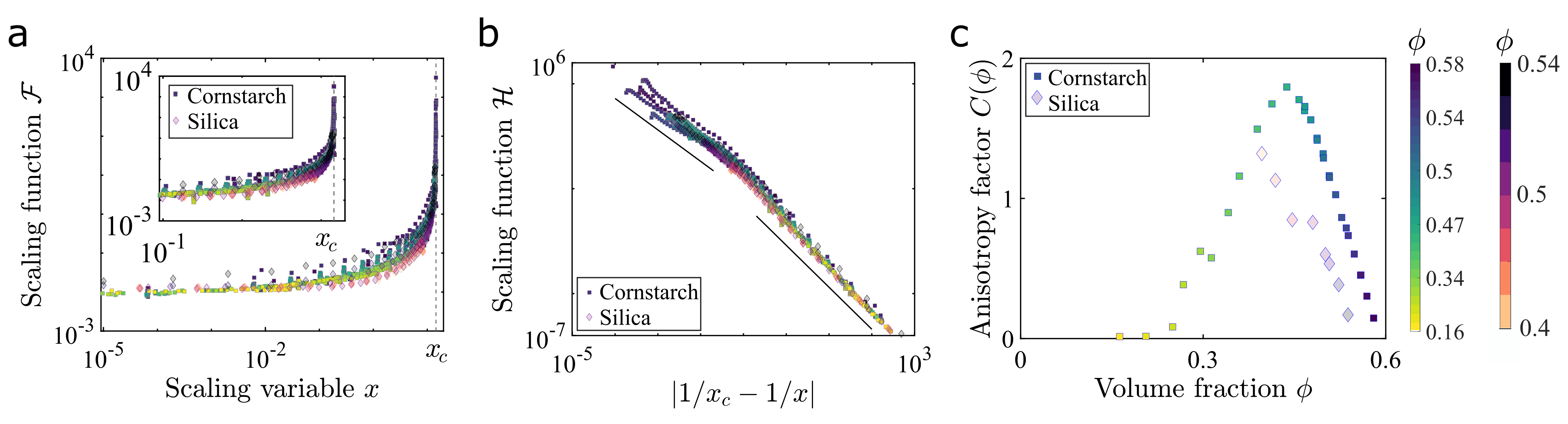}
\caption{\label{fig:ScalingPhi0Str} \textbf{Universal scaling of the suspension viscosity with additional corrections to scaling}. \textbf{a}. The scaling function $\mathcal{F} = (\eta - \beta_\eta(\phi))(\phi^*_0 - \phi)^2$ as a function of the scaling variable $x = e^{-\sigma^*/\sigma}C(\phi)/(\phi^*_0 - \phi)$ for all the cornstarch (squares) and silica (diamond) suspensions data. We find that all the data collapse onto a single universal curve that diverges at $x = x_c$. \textbf{b}. The scaling function $
\mathcal{H} = (\eta - \beta_\eta(\phi))(g(\sigma, \phi))^2$ versus $|1/x_c - 1/x|$ for all the cornstarch and silica suspensions data. This way of scaling the data clearly illustrates two distinct regimes - a regime characterized by a power law of -2 at small $x$ and a power law of $\sim -3/2$ at $x \approx x_c$. The solid black lines indicate power laws of $-2$ and $-1.5$ respectively. \textbf{c}. The anisotropy factor $C(\phi)$ as a function of the volume fraction for both silica and cornstarch. We note that $C(\phi)$ is a smooth analytic function as required by theories for scaling variables such as $g(\sigma,\phi)=f(\sigma)C(\phi)$ far from the critical point.} 
\end{figure*}
\begin{figure}[t]
\begin{center}
\includegraphics[width=0.8\linewidth]{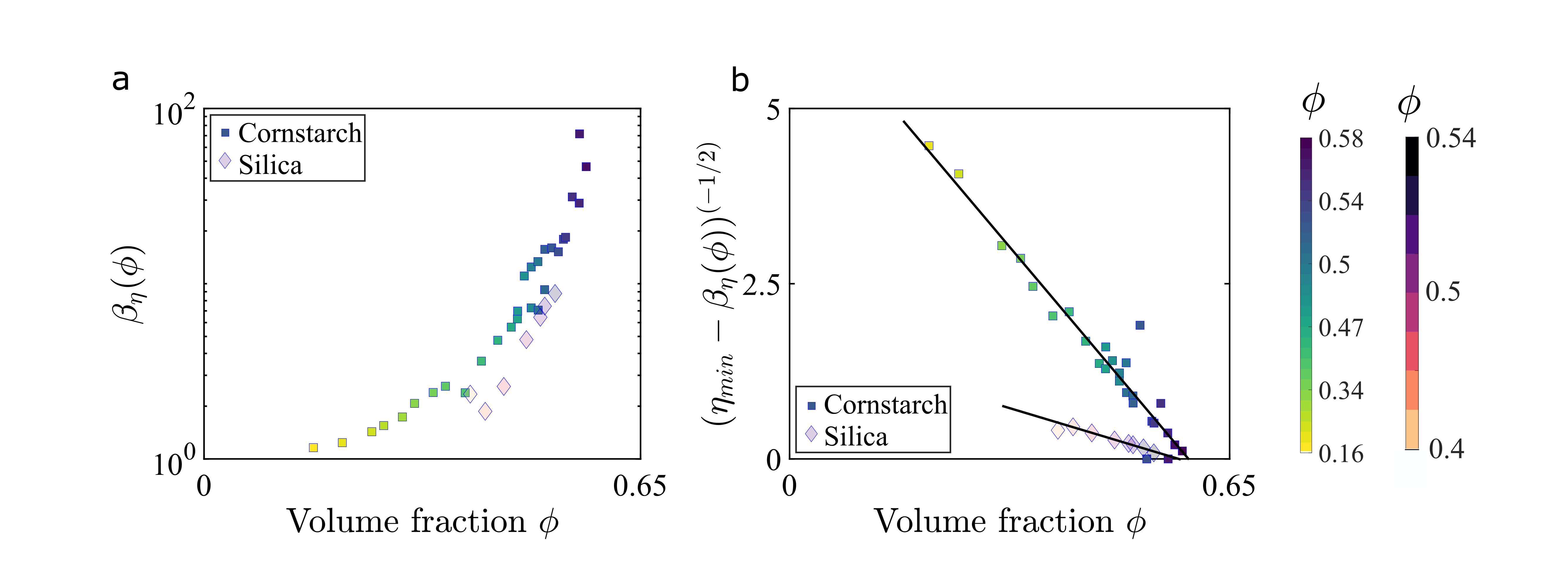}
\caption{\label{fig:BetaEta} \textbf{Low Volume fraction subtraction parameter $\beta_\eta(\phi)$}. \textbf{a} $\beta_\eta(\phi)$ as a function of the viscosity for cornstarch (square) and silica (diamond) suspensions. $\beta_\eta(\phi)$ largely increases with the volume fraction except at the highest volume fraction where it is set to zero. \textbf{b} $(\eta_{min} - \beta_\eta(\phi))^{-1/2}$ as a function of volume fraction, where $\eta_{min}$ is the low stress viscosity. The x intercept of this graph is the divergence of $\eta_{min} - \beta_\eta(\phi)$ at low stresses. Black lines are straight line guides to the eye. }
\end{center}
\end{figure}

We find that additional corrections to scaling can improve the scaling collapse of the data, particularly at intermediate values of the scaling variables (Fig.~\ref{fig:ScalingPhi0Str}) \cite{aharony1980universality, aharony1983nonlinear}. To determine these corrections to scaling, we modify the value of $\phi_0$ to $\phi^*_0$. In particular, we reduce the spread of the scaling function $\mathcal{F}$, plotted as a function  $x_c - x$, attempting to get good collapse. The $C(\phi)$ values are multiplied by $(\phi^*_0 - \phi)/(\phi_0 - \phi)$, to obtain the $C(\phi)$ values corresponding to $\phi^*_0$ (Fig.~\ref{fig:ScalingPhi0Str}\textbf{c}). As required by scaling theories far from the critical point, $C(\phi)$ remains a smooth analytic function. We find that to ensure collapse of the low stress regime it is necessary to subtract a volume fraction dependent term for all packing fractions $\beta_\eta(\phi)$. This parameter may indicate that a fraction of the viscosity is not governed by the crossover scaling. We interpret this term as accounting for the high frequency dynamic viscosity associated with lubrication forces rather than the developing contact network \cite{brady1993rheological}. The values of $\beta_\eta(\phi)$ used in the scaling collapse for cornstarch and silica are shown in Fig.~\ref{fig:BetaEta}\textbf{a}. Importantly, even though $\beta_\eta(\phi)$ increases with the volume fraction, subtracting this parameter from the viscosity does not remove the divergence at small $x$. Rather, it shifts the  divergence at small stresses from $\phi_0$ to $\phi^*_0$ (Fig~\ref{fig:BetaEta}\textbf{b}). 

\begin{figure}
 \begin{center}
\includegraphics[width=0.6\linewidth]{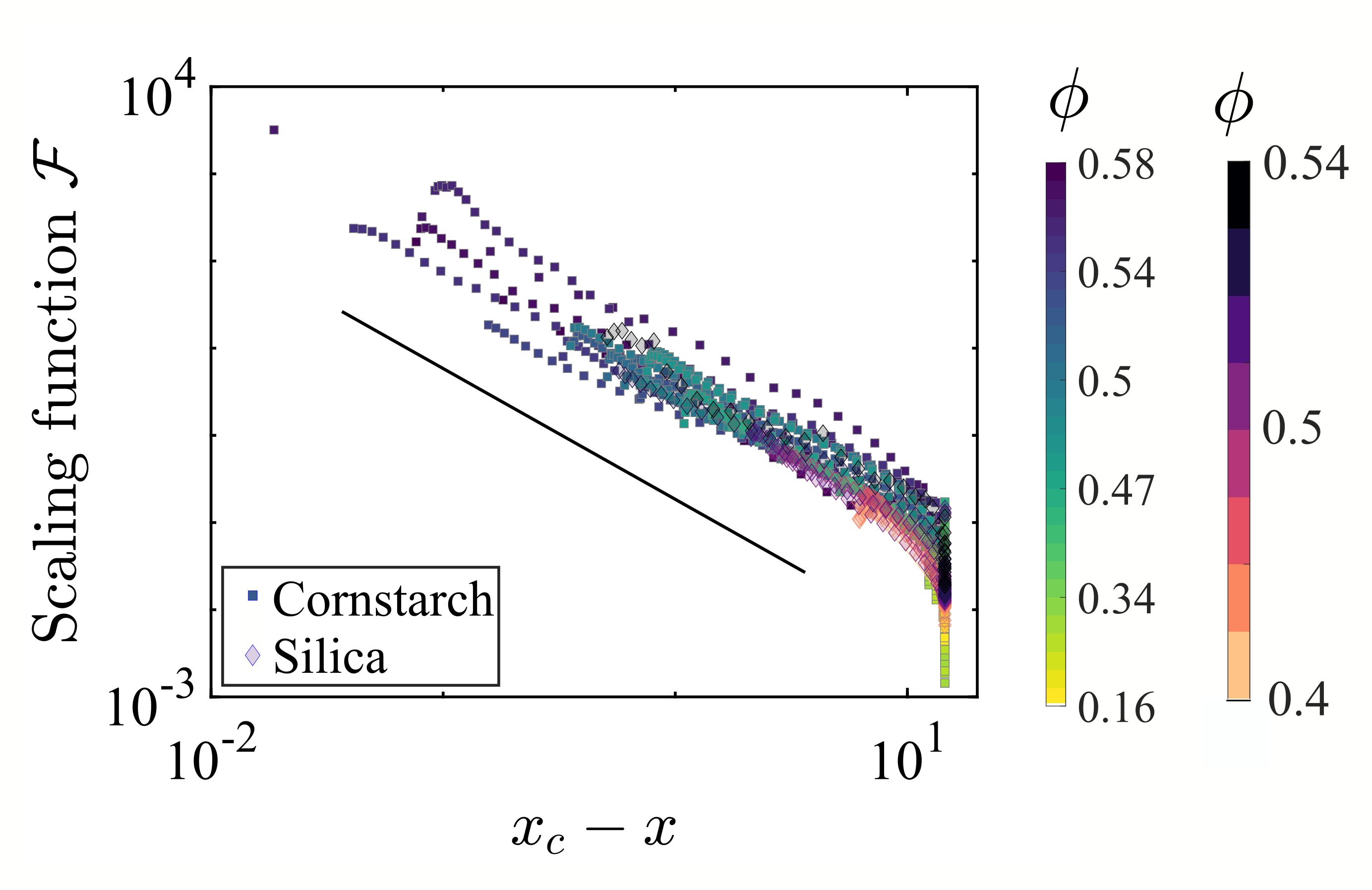}
\caption{\textbf{Power law scaling of cornstarch and silica viscosity data with corrections to scaling.} The scaling function $\mathcal{F} = (\eta - \beta_\eta(\phi))(\phi^*_0 - \phi)^2$ as a function of the scaling variable $x = e^{-\sigma^{*}/\sigma}C(\phi)/(\phi^*_0 - \phi)$ for the cornstarch (squares) and silica (diamond) suspensions data. We find that the data shows power law scaling with an exponent of $\sim -1.5$. The black solid line is a line with power law -1.5.}
\label{fig:x_c_xPhi0Str}
\end{center}
\end{figure}

These additional corrections to the scaling do not alter the divergence at shear jamming. We expect the scaling function to diverge as  $\mathcal{F} \sim 1/(x_c - x)^\delta$ near shear jamming. To determine the scaling exponent $\delta$, we plot the scaling function $\mathcal{F} = (\eta - \beta_\eta(\phi))(\phi^*_0 - \phi)^2$ as a function of $x_c - x$ (Fig.~\ref{fig:x_c_xPhi0Str}). We find that the divergence is indeed a power law, with $\delta \sim 1.5$. This value is distinct from $\delta = 2$, predicted by the reformulated Wyart and Cates model (Eq.~5). 

To better visualize the change in scaling exponent, we can replot the data in a manner analogous to Fig. 1D : 
\begin{equation}
(\eta   - \beta_\eta(\phi))(g(\sigma, \phi))^2 = \mathcal{H}\left( |1/x_c - 1/x | \right)    
\end{equation}
where $x = g(\sigma, \phi)/(\phi^*_0 - \phi)$. We find excellent scaling collapse over six orders of magnitude in the scaling variable with two distinct regimes, each characterized by clearly different power-law exponents (Fig.~\ref{fig:ScalingPhi0Str}\textbf{b}). At small $x$, far from $x_c$, the behaviour is governed by the frictionless jamming point $\phi_0$ and $\mathcal{H} \sim  |1/x_c - 1/x|^{-2}$. As $x$ approaches $x_c$, the exponent changes to $\approx -3/2$, with a crossover between the two regimes at $x/x_c \sim$ 0.1.

\begin{figure*}[t]
\begin{center}
\includegraphics[width=0.8\linewidth]{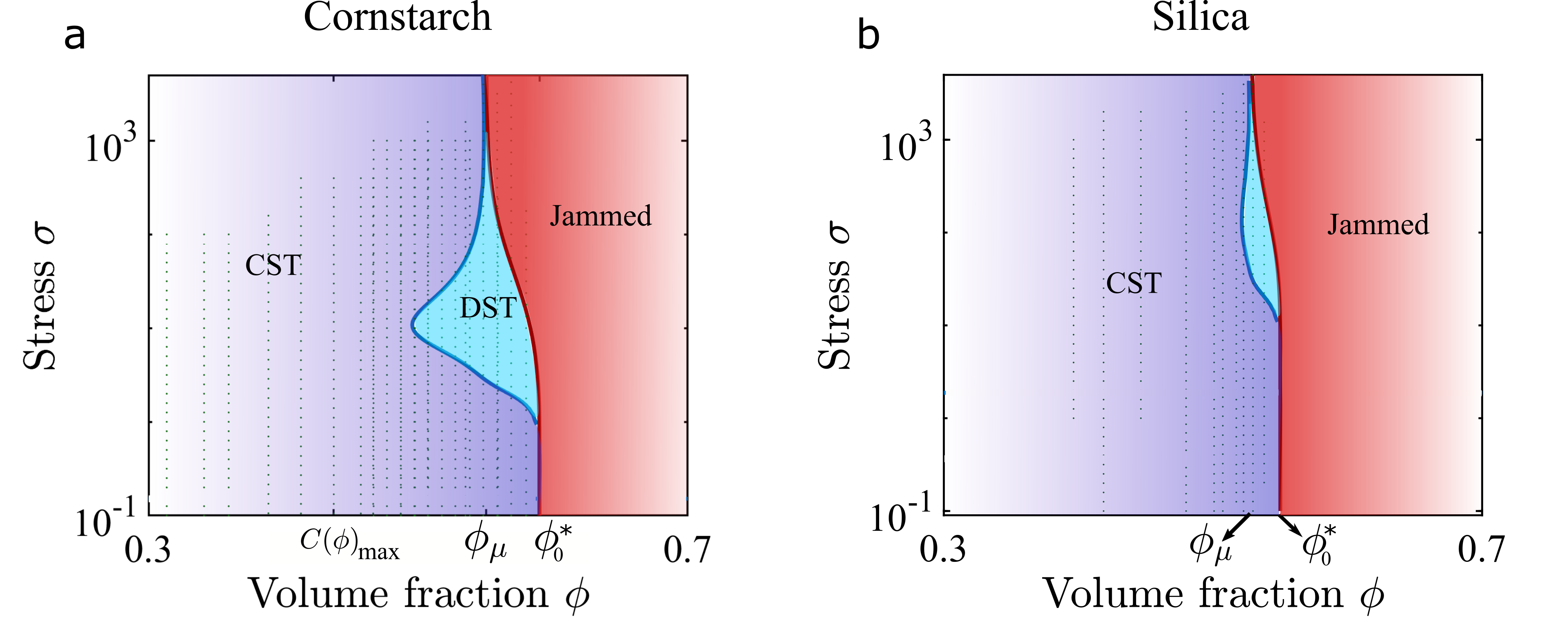}
\caption{\label{fig:PhaseDiagramPhi0Str}  \textbf{Phase diagrams for cornstarch and silica suspensions as derived from the scaling analysis with analytical corrections to scaling}. Three distinct regions are seen in the phase diagrams for the cornstarch \textbf{a} and silica \textbf{b} systems- continuous shear thickening (CST) in purple, discontinuous shear thickening (DST) in blue and a jammed region in red. The shear jamming line (maroon) is determined by $x = x_c$, where $x = e^{-\sigma^*/\sigma}C(\phi)/(\phi^*_0 - \phi)$ and the DST line (blue) is determined by the condition $d\log\eta/d\log\sigma = 1$. The vertical dotted lines indicate the values of frictionless jamming point, $\phi^*_0$, and shear jammed point, $\phi_{\mu}$, and the volume fraction at which $C(\phi)$ is maximum.}
\end{center}
\end{figure*}

From this scaling collapse, we can again draw the phase diagrams for the cornstarch and silica suspensions. We fit the scaling function $\mathcal{H}$ to a crossover between two different power laws :
\begin{equation}
    \log \mathcal{H}(y) = ay + b + \left( \frac{1 + \tanh(e(y - f))}{2}\right)\left( (c-a)y + d -b\right)
\end{equation}
and use the fit to determine the DST line:
\begin{equation}
    \frac{d \log \eta}{d \log \sigma} = 1
\end{equation}
We can also determine the shear jamming line as $x = x_c$. The resultant phase diagram for silica and cornstarch are shown in Fig.~\ref{fig:PhaseDiagramPhi0Str}. The exact parameters used to generate the phase diagram is shown in Table 2. We find that the phase diagrams are qualitatively similar to that shown in the main text, however, there are small differences in $\phi_\mu$, the shear jamming line and the DST line. The key difference is that the shear jamming line and DST line converge at $\phi^*_0$ and not $\phi_0$ at small stresses. 

\begin{table}
\begin{center}
\begin{tabular}{ |c|c| } 
 \hline
 \textbf{Parameters} & \textbf{Values} \\ 
 \hline
 Cornstarch $C(\phi)$ & -13.11$\phi$ + 7.79\\ 
 \hline
 Silica $C(\phi)$ & -11.14$\phi$ + 6.2\\ 
 \hline
 Cornstarch $\sigma^*$ & 3.9\\ 
 \hline
 Silica $\sigma^*$ & 48.2\\ 
 \hline
  Cornstarch $\phi^*_0$ & 0.59\\ 
 \hline
 Silica $\phi^*_0$ & 0.55\\ 
 \hline
 Fit $\mathcal{H}$ - $a$ & -1.51 $\pm$ 0.06 \\
 \hline
 Fit $\mathcal{H}$ - $b$ & -1.4 $\pm$ 0.4 \\
 \hline
 Fit $\mathcal{H}$ - $c$ & -2.04 $\pm$ 0.08\\
 \hline
 Fit $\mathcal{H}$ - $d$ & -3.1 $\pm$ 0.3\\
 \hline
 Fit $\mathcal{H}$ - $e$ & 0.5 $\pm$ 0.2\\
 \hline
 Fit $\mathcal{H}$ - $f$ & -2.1 $\pm$ 1\\
 \hline
 \end{tabular}
\label{table:ParametersPhi0Str}
\end{center}
\caption{Parameters to determine the shear jamming and DST lines for the phase diagram with the additional corrections to scaling.}
\end{table}

\section{Easy Visualization of the scaling exponent}

\begin{figure}[t]
\begin{center}
\includegraphics[width=0.6\linewidth]{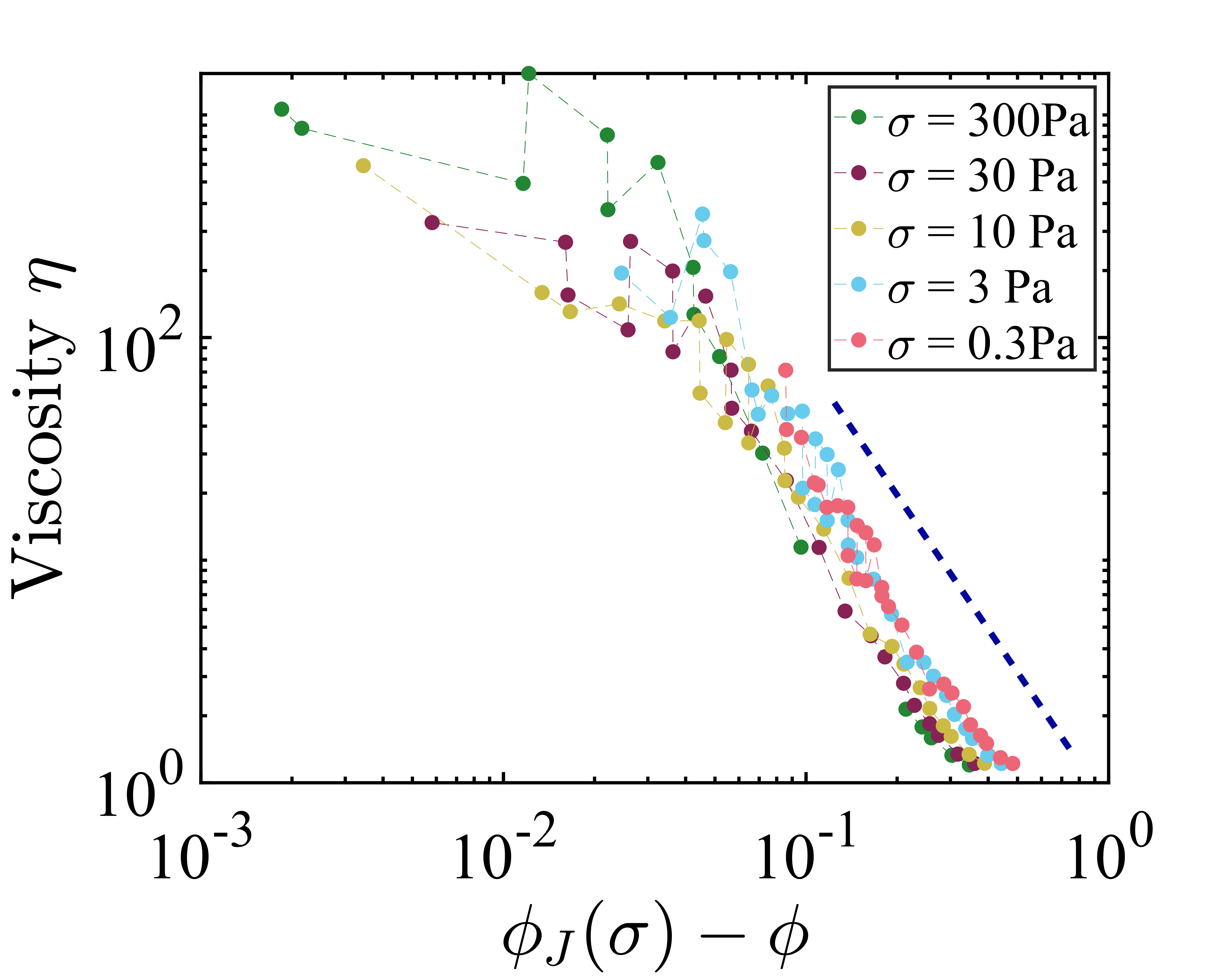}
\caption{\label{fig:VaryingPhiJ} \textbf{Visualizing the change in the scaling exponent}. The viscosity at constant stress for different packing fractions is plotted as a function of $\phi_J(\sigma) - \phi$, where $\phi_J(\sigma) = \phi_0(1 - f(\sigma)) + \phi_\mu(f(\sigma))$. While the data is noisy, it appears that at small stresses, the data appears to be a power law with a single exponent (pink data). At higher stresses, however, there appears to be a second exponent at higher packing fractions (green data). The blue dashed line is a guide to the eye, indicating a power law of -2, similar to the low stress, low packing fraction data.}
\end{center}
\end{figure}
One way to easily visualize the change in the scaling exponent is to determine $\phi_J(\sigma) = \phi_0(1 - f(\sigma)) + \phi_\mu(f(\sigma))$. We can then plot the viscosity at different stresses as a function $\phi_J(\sigma) - \phi$ (Fig.~\ref{fig:VaryingPhiJ}). While the data is noisy, at small stresses, the viscosity appears to diverge with a single exponent $\sim 2$ (pink data). At larger stresses (green data) there appears to be two distinct regimes, one with an exponent of $\sim 2$ at small $\phi$, and one with a slope $<2$ at larger $\phi$. This behaviour suggests that there exists a change in the scaling exponent that occurs at a constant value of a variable that is combination of both the stress and the packing fraction, and is consistent with the scaling collapse and the crossover scaling picture that we have presented. In our picture, the scaling variable, which is a combination of both the stress and the packing fraction, reaches a particular value ($x \sim 0.1$), at which point the scaling exponent changes.

\section{\label{Sedimentation } Sedimentation of the sample}

We expect the effects of sedimentation in these suspensions to be minimal despite the density mismatch. The Shield's number, $\tau^*$, the ratio of stress to gravitational forces:
\begin{equation}
    \tau^* = \frac{\sigma}{\Delta \rho g D}
\end{equation}
where $\sigma$ is the stress, $\Delta \rho$ is the density mismatch, $g$ is the acceleration due to gravity, and $D$ is the particle diameter, ranges from $10$ to $10^4$ for silica and $1$ and $10^4$ for the cornstarch suspensions.

\section{\label{CphiSensitivty} Sensitivity of phase diagrams}
\begin{figure}[t]
\begin{center}
\includegraphics{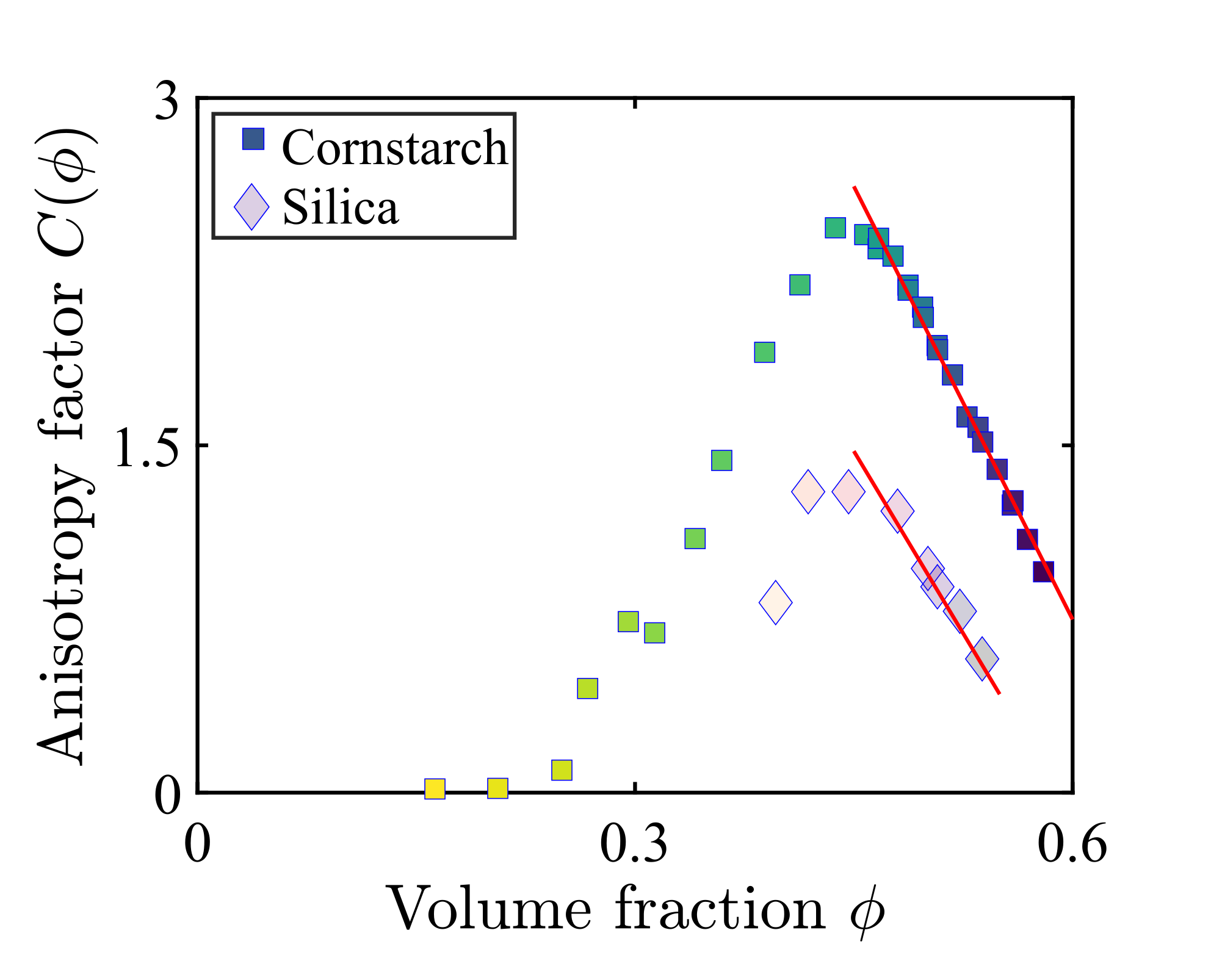}
\caption{\label{fig:CPhiFit} \textbf{Fits to the anisotropy $C(\phi)$}. The anisotropy factor $C(\phi)$ for cornstarch (square) and silica (diamond) suspensions and the fits (red) used to generate the phase diagram. We find that the data is fit fairly well by a straight line.}
\end{center}
\end{figure}

\begin{figure}[t]
\begin{center}
\includegraphics[width=0.8\linewidth]{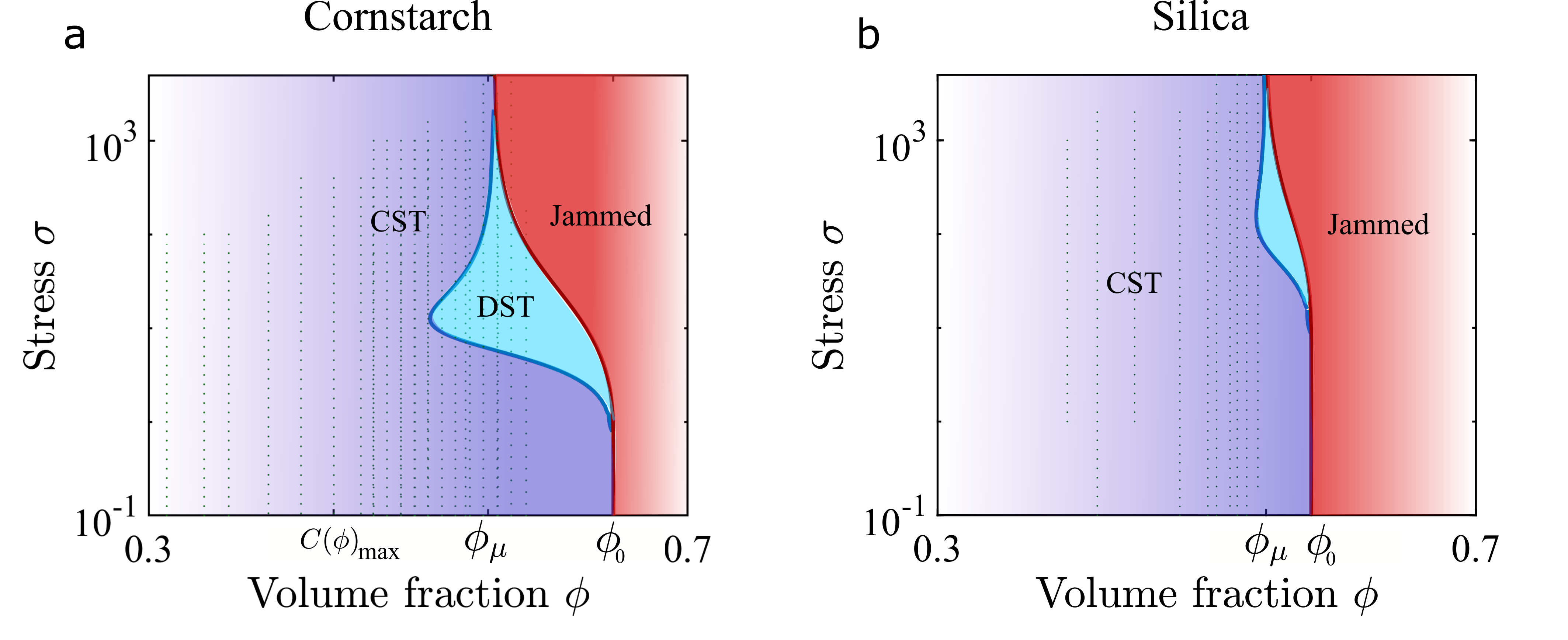}
\caption{\label{fig:PhaseDiagramPhi0Dots} \textbf{Phase diagrams for cornstarch and silica suspensions with experimental data points}. Three distinct regions are seen in the phase diagrams for the cornstarch \textbf{a} and silica \textbf{b} systems- continuous shear thickening (CST) in purple, discontinuous shear thickening (DST) in blue and a jammed region in red. The small green dots indicate the experimental stresses and volume fractions used for the scaling analysis. The shear jamming line (maroon) is determined by $x = x_c$, where $x = e^{-\sigma^*/\sigma}C(\phi)/(\phi_0 - \phi)$ and the DST line (blue) is determined by the condition $d\log\eta/d\log\sigma = 1$.}
\end{center}
\end{figure}

To generate the phase diagram, we use fits to both $C(\phi)$ and the scaling function $\mathcal{H}$. We fit a line to the high volume fraction $C(\phi)$ data as shown in Fig.~\ref{fig:CPhiFit}, and $\mathcal{H}$ is fit by crossover between two power laws as described in Eq.~\ref{eq:HFit}. The lines in the phase diagram (Fig. 3) are sensitive to these fits. More specifically, as the scaling variable  $x = fC(\phi)/(\phi_0 - \phi)$, and $\phi \rightarrow \phi_0$, the shear jamming line is very sensitive to the fit of $C(\phi)$. This sensitivity is illustrated by the overlay of experimental points on the phase diagram (Fig.~\ref{fig:PhaseDiagramPhi0Dots}), where data are projected to be on the jammed side of the phase diagram via the fit, while the actual values of $C(\phi)$ places the data in the DST region. The DST line is especially sensitive to the power law exponents $a$ and $c$. These parameters change the extent of the DST region, moving the onset volume fraction for DST.  

\section{Predictive powers of scaling functions}Universal scaling theories and scaling functions such as those presented in this paper can be used to predict a number of physical properties of the system. In the well studied system of Heisenberg and Ising magnets for instance, from the crossover scaling function describing the free energy one can determine the specific heat, correlation functions and many other system properties \cite{cardy1996scaling}. Similarly, recent studies have investigated the scaling relations for the jamming transition for frictionless particles \cite{goodrich2016scaling, o2003jamming, goodrich2014jamming, liarte2019jamming, liarte2021scaling}, demonstrating the both static and dynamic viscoelastic properties can be predicted by such a scaling formulation. For instance, \cite{goodrich2016scaling} proposes a scaling \textit{ansatz} for the elastic energy:
\begin{equation}
    E(\Delta Z, \Delta\phi, \epsilon, N) = \Delta Z^\zeta \mathcal{E}_0\left( \frac{\Delta\phi}{\Delta Z^{\beta_\phi}}, \frac{\epsilon}{\Delta Z^{\beta_\epsilon}}, N\Delta Z^\psi\right)  
\end{equation} 
where $\Delta \phi = \phi - \phi_{c, \Lambda}$ is the excess volume fraction ($\phi_{c, \
\Lambda}$ is the volume fraction at jamming, $\Delta Z$ is the excess contact number, N is the system size, and $\epsilon$ is the strain. From this scaling anstaz, by taking derivatives of the energy, they predict scaling forms for the pressure, shear stress, bulk modulus, and shear modulus that are verified by simulations. Similarly, \cite{liarte2021scaling} introduces a scaling \textit{ansatz} for the longitudinal response function near jamming, and determines the exact scaling function for the response function through the effective medium theory 
\cite{liarte2019jamming}. From this scaling function, they predict the dynamic (frequency dependent) shear and bulk moduli, the density response and the correlation function near jamming. These studies demonstrate the power of such scaling theories and universal scaling functions, indicating that our work can be the first step in generating similar predictions and theories for shear thickening and frictional jamming.
    



\bibliography{aipsamp.bib}
\bibliographystyle{naturemag}